\documentclass[11pt]{article}
\usepackage[utf8]{inputenc}
\usepackage[left=25mm,right=25mm,top=25mm,bottom=25mm]{geometry}
\usepackage[T1]{fontenc}
\usepackage{stmaryrd}

\newcommand{\blue}[1]{\textcolor{black}{#1}}

\newcommand{\A}{{\boldsymbol{A}}}
\newcommand{\F}{{\boldsymbol{F}}}

\newcommand{\R}{\boldsymbol{{R}}}
\newcommand{\U}{\boldsymbol{{U}}}

\newcommand{\ba}{\boldsymbol{a}}
\newcommand{\bb}{\boldsymbol{b}}
\newcommand{\bd}{\boldsymbol{d}}
\newcommand{\be}{\boldsymbol{e}}

\newcommand{\bp}{\boldsymbol{p}}
\newcommand{\bA}{\boldsymbol{A}}
\newcommand{\bB}{\boldsymbol{B}}
\newcommand{\bC}{\boldsymbol{C}}

\newcommand{\bF}{\boldsymbol{F}}
\newcommand{\bG}{\boldsymbol{G}}
\newcommand{\bH}{\boldsymbol{H}}
\newcommand{\bI}{\boldsymbol{I}}

\newcommand{\bP}{\boldsymbol{P}}
\newcommand{\bQ}{\boldsymbol{Q}}

\newcommand{\bW}{\boldsymbol{W}}
\newcommand{\bX}{\boldsymbol{X}}
\newcommand{\bY}{\boldsymbol{Y}}
\newcommand{\bnull}{\boldsymbol{0}}

\newcommand{\uuI}{\mathbbm{1}}

\newcommand{\GL}{\text{GL}}
\newcommand{\SO}{\text{SO}}

\newcommand{\bnabla}{\vect{\nabla}}

\newcommand{\cB}{{\mathcal{B}}}
\newcommand{\cC}{{\mathcal{C}}}

\newcommand{\cG}{{\mathcal{G}}}

\newcommand{\cI}{{\mathcal{I}}}

\newcommand{\cO}{{\mathcal{O}}}
\newcommand{\cP}{{\mathcal{P}}}

\newcommand{\cS}{{\mathcal{S}}}

\newcommand{\cU}{{\mathcal{U}}}
\newcommand{\cV}{{\mathcal{V}}}

\newcommand{\bbG}{{\mathbb{G}}}

\newcommand{\bbR}{{\mathbb{R}}}

\usepackage{multirow}
\usepackage{booktabs}

\usepackage{enumitem}

\usepackage{amssymb}
\usepackage{amsmath}
\usepackage{amsthm}
\usepackage{nicefrac}
\usepackage{bbm}

\theoremstyle{definition}   
\usepackage{chngcntr}
\usepackage{apptools}

\usepackage[usenames,dvipsnames,svgnames]{xcolor}
\usepackage{tikz}
\usepackage{pgfplots}
\usepgfplotslibrary{groupplots}
\usepackage{pgfplotstable}
\usepackage{subcaption}

\usepackage[font=small]{caption}

\usepackage{wrapfig}

\pgfplotsset{compat=newest}
\usepgfplotslibrary{fillbetween}
\usetikzlibrary{patterns}
\usepgfplotslibrary{colormaps}

\definecolor{color11a}{RGB}{236, 188, 0}
\definecolor{color12a}{RGB}{50, 67, 121}
\definecolor{color13}{RGB}{222, 222, 222}
\definecolor{color21a}{RGB}{92, 134, 196}
\definecolor{color22a}{RGB}{249, 156, 0}
\definecolor{color23}{RGB}{222, 222, 222}
\definecolor{color31}{RGB}{222, 222, 222}
\definecolor{color32}{RGB}{222, 222, 222}
\definecolor{color33a}{RGB}{191, 60, 60}

\colorlet{color11}{color11a!90!color13}
\colorlet{color12}{color12a!90!color13}
\colorlet{color21}{color21a!90!color13}
\colorlet{color22}{color22a!90!color13}
\colorlet{color33}{color33a!90!color13}

\usetikzlibrary{shapes}
\usetikzlibrary{arrows.meta} 

\tikzset{>=latex} 
\tikzstyle{node}=[thick,circle,draw=black,minimum size=22,inner sep=0.5,outer sep=0.6]
\tikzstyle{node icnn}=[color=color22!10!black,node,draw=black,fill=color22!25, text = black]
\tikzstyle{node in_out}=[node,color21!10!black,draw=black,fill=color21!25, text=black]
\tikzstyle{node inv_pot}=[rectangle, node,color33!10!black,draw=black,fill=color33!25, text=black]
\tikzstyle{connect}=[thick,black] 
\tikzstyle{connect arrow}=[-{Latex[length=4,width=3.5]},thick,black,shorten <=0.5,shorten >=1]

\pgfplotscreateplotcyclelist{mycolorlist_mixed}{%
only marks,solid, mark options={solid,fill=color12,fill opacity=1,mark size=4pt},mark=square*,color12\\
only marks,solid, very thick, mark options={solid,fill=color12,fill opacity=1,mark size=4pt},mark=triangle*,color12\\
only marks,solid, very thick, mark options={solid,fill=color33,fill opacity=1,mark size=4pt},mark=triangle*,color33\\
only marks,solid, mark options={solid,fill=color12,fill opacity=1,mark size=4pt},mark=*,color12\\
only marks,solid, very thick, mark options={solid,fill=color12,fill opacity=1,mark size=4pt},mark=diamond*,color12\\
only marks,solid, mark options={solid,fill=color33,fill opacity=1,mark size=4pt},mark=square*,color33\\
only marks,solid, very thick, mark options={solid,fill=color22,fill opacity=1,mark size=4pt},mark=triangle*,color22\\
only marks,solid, mark options={solid,fill=color33,fill opacity=1,mark size=4pt},mark=*,color33\\
only marks,solid, very thick, mark options={solid,fill=color33,fill opacity=1,mark size=4pt},mark=diamond*,color33\\
only marks,solid, very thick, mark options={solid,fill=color22,fill opacity=1,mark size=4pt},mark=diamond*,color22\\
only marks,solid, mark options={solid,fill=color22,fill opacity=1,mark size=4pt},mark=square*,color22\\
only marks,solid, mark options={solid,fill=color22,fill opacity=1,mark size=4pt},mark=*,color22\\
color12, only marks, very thick, mark options = {solid}, mark=square*, mark repeat = 15\\
color12, only marks, very thick ,mark=triangle*, mark options={solid}, mark repeat = 15\\
color33, only marks, very thick ,mark=triangle*, mark options={solid}, mark repeat = 15\\
color12, only marks, very thick ,mark=diamond*, mark options ={solid}, mark repeat = 15\\
color33, only marks, very thick ,mark=square*, mark options={solid}, mark repeat = 15\\
color22, only marks, very thick ,mark=triangle*, mark options={solid}, mark repeat = 15\\
color33, only marks, very thick,mark=diamond*, mark options={solid} , mark repeat = 15\\
color22, only marks, very thick ,mark=diamond*, mark options={solid}, mark repeat = 15\\
color22, only marks, very thick,mark=square*, mark options={solid} , mark repeat = 15\\
color12, very thick \\
color12, very thick \\
color33, very thick \\
color12, very thick \\
color33, very thick \\
color22, very thick \\
color33, very thick \\
color22, very thick \\ 
color22, very thick \\
color12, only marks, very thick ,mark=*, mark options={solid}, mark repeat = 15\\
color33, only marks, very thick ,mark=*, mark options={solid}, mark repeat = 15\\
color22, only marks, very thick ,mark=*, mark options={solid}, mark repeat = 15\\
color12, very thick \\
color33, very thick \\
color22, very thick \\
}

\pgfplotscreateplotcyclelist{mycolorlist_mixed_2}{%
only marks,solid, mark options={solid,fill=color13,fill opacity=1,mark size=4pt},mark=square*,color13\\
only marks,solid, very thick, mark options={solid,fill=color12,fill opacity=1,mark size=4pt},mark=triangle*,color12\\
only marks,solid, very thick, mark options={solid,fill=color13,fill opacity=1,mark size=4pt},mark=triangle*,color13\\
only marks,solid, mark options={solid,fill=color12,fill opacity=1,mark size=4pt},mark=*,color12\\
only marks,solid, very thick, mark options={fill=color13,fill opacity=1,mark size=4pt},mark=triangle*,color13\\
only marks,solid, mark options={solid,fill=color33,fill opacity=1,mark size=4pt},mark=square*,color33\\
only marks,solid, very thick, mark options={solid,fill=color13,fill opacity=1,mark size=4pt},mark=triangle*,color13\\
only marks,solid, mark options={solid,fill=color33,fill opacity=1,mark size=4pt},mark=*,color33\\
only marks,solid, very thick, mark options={fill=color21,fill opacity=1,mark size=4pt},mark=triangle*,color21\\
only marks,solid, very thick, mark options={fill=color13,fill opacity=1,mark size=4pt},mark=triangle*,color13\\
only marks,solid, mark options={solid,fill=color22,fill opacity=1,mark size=4pt},mark=square*,color22\\
only marks,solid, mark options={solid,fill=color22,fill opacity=1,mark size=4pt},mark=*,color22\\
color13, only marks, very thick ,mark=square*, mark options = {fill opacity = 1}, mark repeat* = 5\\
color13, only marks, very thick ,mark=triangle*, mark options = {fill opacity = 1}, mark repeat* = 5\\
color13, only marks, very thick ,mark=triangle*, mark options = {fill opacity = 1}, mark repeat* = 5\\
color13, only marks, very thick ,mark=triangle*, mark options = {fill opacity = 1}, mark repeat* = 5\\
color13, only marks, very thick ,mark=triangle*, mark options = {fill opacity = 1}, mark repeat* = 5\\
color12, only marks, very thick ,mark=triangle*, mark options = {fill opacity = 1}, mark repeat* = 5\\
color33, only marks, very thick ,mark=square*, mark options = {fill opacity = 1}, mark repeat* = 5\\
color21, only marks, very thick ,mark=triangle*, mark options = {fill opacity = 1}, mark repeat* = 5\\
color22, only marks, very thick ,mark=square*, mark options = {fill opacity = 1}, mark repeat* = 5\\
color13,  thick \\
color13,  thick \\
color13,  thick \\
color13,  thick \\
color13,  thick \\
color12,  ultra thick \\
color33, ultra thick \\
color21, ultra thick \\
color22, ultra thick \\
color12, only marks, very thick ,mark=*, mark options={solid}, mark repeat* = 5\\
color33, only marks, very thick ,mark=*, mark options={solid}, mark repeat* = 5\\
color22, only marks, very thick ,mark=*, mark options={solid}, mark repeat* = 5\\
color12, ultra thick \\
color33, ultra thick \\
color22, ultra thick \\
}

\pgfplotscreateplotcyclelist{mycolorlist_TI}{%
only marks,solid, mark options={solid,fill=color33,fill opacity=1,mark size=4pt},mark=square*,color33\\
only marks,solid, very thick, mark options={solid,fill=color13,fill opacity=1,mark size=4pt},mark=triangle*,color13\\
only marks,solid, very thick, mark options={solid,fill=color13,fill opacity=1,mark size=4pt},mark=triangle*,color13\\
only marks,solid, mark options={solid,fill=color12,fill opacity=1,mark size=4pt},mark=*,color12\\
only marks,solid, very thick, mark options={fill=color13,fill opacity=1,mark size=4pt},mark=triangle*,color13\\
only marks,solid, mark options={solid,fill=color13,fill opacity=1,mark size=4pt},mark=square*,color13\\
only marks,solid, very thick, mark options={solid,fill=color12,fill opacity=1,mark size=4pt},mark=triangle*,color12\\
only marks,solid, mark options={solid,fill=color33,fill opacity=1,mark size=4pt},mark=*,color33\\
only marks,solid, very thick, mark options={fill=color21,fill opacity=1,mark size=4pt},mark=triangle*,color21\\
only marks,solid, very thick, mark options={fill=color13,fill opacity=1,mark size=4pt},mark=triangle*,color13\\
only marks,solid, mark options={solid,fill=color22,fill opacity=1,mark size=4pt},mark=square*,color22\\
only marks,solid, mark options={solid,fill=color22,fill opacity=1,mark size=4pt},mark=*,color22\\
color13, dashed,  thick, mark options = {solid}, mark=triangle*,mark repeat* = 5\\
color13, dashed,  thick ,mark=triangle*, mark options={solid}, mark repeat* = 5\\
color13, dashed,  thick ,mark=triangle*, mark options ={solid}, mark repeat* = 5\\
color13, dashed,  thick ,mark=square*, mark options={solid}, mark repeat* = 5\\
color13, dashed,  thick,mark=triangle*, mark options={solid} , mark repeat* = 5\\
color33, dashed,  ultra thick ,mark=square*, mark options={solid}, mark repeat* = 5\\
color12, dashed,  ultra thick ,mark=triangle*, mark options={solid}, mark repeat* = 5\\
color21, dashed,  ultra thick ,mark=triangle*, mark options={solid}, mark repeat* = 5\\
color22, dashed, ultra thick,mark=square*, mark options={solid} , mark repeat* = 5\\
color12, dashed, ultra thick ,mark=*, mark options={solid}, mark repeat* = 5\\
color33, dashed, ultra thick ,mark=*, mark options={solid}, mark repeat* = 5\\
color22, dashed, ultra thick ,mark=*, mark options={solid}, mark repeat* = 5\\
color13, only marks, very thick ,mark=square*, mark options = {fill opacity = 1}, mark repeat* = 5\\
color13, only marks, very thick ,mark=triangle*, mark options = {fill opacity = 1}, mark repeat* = 5\\
color13, only marks, very thick ,mark=triangle*, mark options = {fill opacity = 1}, mark repeat* = 5\\
color13, only marks, very thick ,mark=triangle*, mark options = {fill opacity = 1}, mark repeat* = 5\\
color13, only marks, very thick ,mark=triangle*, mark options = {fill opacity = 1}, mark repeat* = 5\\
color33, only marks, very thick ,mark=square*, mark options = {fill opacity = 1}, mark repeat* = 5\\
color12, only marks, very thick ,mark=triangle*, mark options = {fill opacity = 1}, mark repeat* = 5\\
color21, only marks, very thick ,mark=triangle*, mark options = {fill opacity = 1}, mark repeat* = 5\\
color22, only marks, very thick ,mark=square*, mark options = {fill opacity = 1}, mark repeat* = 5\\
color13,  thick \\
color13,  thick \\
color13,  thick \\
color13,  thick \\
color13,  thick \\
color33,  ultra thick \\
color12, ultra thick \\
color21, ultra thick \\
color22, ultra thick \\
color12, only marks, very thick ,mark=*, mark options={solid}, mark repeat* = 5\\
color33, only marks, very thick ,mark=*, mark options={solid}, mark repeat* = 5\\
color22, only marks, very thick ,mark=*, mark options={solid}, mark repeat* = 5\\
color12, ultra thick \\
color33, ultra thick \\
color22, ultra thick \\
}

\pgfplotscreateplotcyclelist{mycolorlist_cub}{%
only marks,solid, mark options={solid,fill=color33,fill opacity=1,mark size=4pt},mark=square*,color33\\
only marks,solid, very thick, mark options={solid,fill=color12,fill opacity=1,mark size=4pt},mark=triangle*,color12\\
only marks,solid, very thick, mark options={solid,fill=color13,fill opacity=1,mark size=4pt},mark=triangle*,color13\\
only marks,solid, mark options={solid,fill=color12,fill opacity=1,mark size=4pt},mark=*,color12\\
only marks,solid, very thick, mark options={fill=color21,fill opacity=1,mark size=4pt},mark=triangle*,color21\\
only marks,solid, mark options={solid,fill=color22,fill opacity=1,mark size=4pt},mark=square*,color22\\
only marks,solid, very thick, mark options={solid,fill=color13,fill opacity=1,mark size=4pt},mark=triangle*,color13\\
only marks,solid, mark options={solid,fill=color33,fill opacity=1,mark size=4pt},mark=*,color33\\
only marks,solid, very thick, mark options={fill=color13,fill opacity=1,mark size=4pt},mark=triangle*,color13\\
only marks,solid, very thick, mark options={fill=color13,fill opacity=1,mark size=4pt},mark=triangle*,color13\\
only marks,solid, mark options={solid,fill=color13,fill opacity=1,mark size=4pt},mark=square*,color13\\
only marks,solid, mark options={solid,fill=color22,fill opacity=1,mark size=4pt},mark=*,color22\\
color13, dashed,  thick, mark options = {solid}, mark=triangle*,mark repeat* = 30\\
color13, dashed,  thick ,mark=triangle*, mark options={solid}, mark repeat* = 30\\
color13, dashed,  thick ,mark=triangle*, mark options ={solid}, mark repeat* = 30\\
color13, dashed,  thick ,mark=triangle*, mark options={solid}, mark repeat* = 30\\
color13, dashed,  thick,mark=square*, mark options={solid} , mark repeat* = 30\\
color33, dashed,  ultra thick ,mark=square*, mark options={solid}, mark repeat* = 30\\
color12, dashed,  ultra thick ,mark=triangle*, mark options={solid}, mark repeat* = 30\\
color21, dashed,  ultra thick ,mark=triangle*, mark options={solid}, mark repeat* = 30\\
color22, dashed, ultra thick,mark=square*, mark options={solid} , mark repeat* = 30\\
color12, dashed, ultra thick ,mark=*, mark options={solid}, mark repeat* = 30\\
color33, dashed, ultra thick ,mark=*, mark options={solid}, mark repeat* = 30\\
color22, dashed, ultra thick ,mark=*, mark options={solid}, mark repeat* = 30\\
color13, only marks, very thick ,mark=triangle*, mark options = {fill opacity = 1}, mark repeat* = 30\\
color13, only marks, very thick ,mark=triangle*, mark options = {fill opacity = 1}, mark repeat* = 30\\
color13, only marks, very thick ,mark=triangle*, mark options = {fill opacity = 1}, mark repeat* = 30\\
color13, only marks, very thick ,mark=triangle*, mark options = {fill opacity = 1}, mark repeat* = 30\\
color13, only marks, very thick ,mark=square*, mark options = {fill opacity = 1}, mark repeat* = 30\\
color33, only marks, very thick ,mark=square*, mark options = {fill opacity = 1}, mark repeat* = 30\\
color12, only marks, very thick ,mark=triangle*, mark options = {fill opacity = 1}, mark repeat* = 30\\
color21, only marks, very thick ,mark=triangle*, mark options = {fill opacity = 1}, mark repeat* = 30\\
color22, only marks, very thick ,mark=square*, mark options = {fill opacity = 1}, mark repeat* = 30\\
color13,  thick \\
color13,  thick \\
color13,  thick \\
color13,  thick \\
color13,  thick \\
color33, ultra thick \\
color12, ultra thick \\
color21, ultra thick \\
color22, ultra thick \\
color12, only marks, very thick ,mark=*, mark options={solid}, mark repeat* = 30\\
color33, only marks, very thick ,mark=*, mark options={solid}, mark repeat* = 30\\
color22, only marks, very thick ,mark=*, mark options={solid}, mark repeat* = 30\\
color12, ultra thick \\
color33, ultra thick \\
color22, ultra thick \\
}

\pgfplotscreateplotcyclelist{mycolorlist_r1_biax}{%
only marks,solid, mark options={solid,fill=color33,fill opacity=1,mark size=4pt},mark=square*,color33\\
only marks,solid, very thick, mark options={solid,fill=color13,fill opacity=1,mark size=4pt},mark=triangle*,color13\\
only marks,solid, very thick, mark options={solid,fill=color13,fill opacity=1,mark size=4pt},mark=triangle*,color13\\
only marks,solid, mark options={solid,fill=color12,fill opacity=1,mark size=4pt},mark=*,color12\\
only marks,solid, very thick, mark options={fill=color13,fill opacity=1,mark size=4pt},mark=triangle*,color13\\
only marks,solid, mark options={solid,fill=color22,fill opacity=1,mark size=4pt},mark=square*,color22\\
only marks,solid, very thick, mark options={solid,fill=color13,fill opacity=1,mark size=4pt},mark=triangle*,color13\\
only marks,solid, mark options={solid,fill=color22,fill opacity=1,mark size=4pt},mark=*,color22\\
only marks,solid, very thick, mark options={fill=color13,fill opacity=1,mark size=4pt},mark=triangle*,color13\\
only marks,solid, very thick, mark options={fill=color13,fill opacity=1,mark size=4pt},mark=triangle*,color13\\
only marks,solid, mark options={solid,fill=color13,fill opacity=1,mark size=4pt},mark=square*,color13\\
only marks,solid, mark options={solid,fill=color33,fill opacity=1,mark size=4pt},mark=*,color33\\
color13, dashed,  thick, mark options = {solid}, mark=triangle*,mark repeat* = 10\\
color13, dashed,  thick ,mark=triangle*, mark options={solid}, mark repeat* = 10\\
color13, dashed,  thick ,mark=triangle*, mark options ={solid}, mark repeat* = 10\\
color13, dashed,  thick ,mark=triangle*, mark options={solid}, mark repeat* = 10\\
color13, dashed,  thick,mark=triangle*, mark options={solid} , mark repeat* = 10\\
color13, dashed,  thick ,mark=triangle*, mark options={solid}, mark repeat* = 10\\
color13, dashed,  thick ,mark=square*, mark options={solid}, mark repeat* = 10\\
color33, dashed,  ultra thick ,mark=square*, mark options={solid}, mark repeat* = 10\\
color22, dashed, ultra thick,mark=square*, mark options={solid} , mark repeat* = 10\\
color12, dashed, ultra thick ,mark=*, mark options={solid}, mark repeat* = 10\\
color22, dashed, ultra thick ,mark=*, mark options={solid}, mark repeat* = 20\\
color33, dashed, ultra thick ,mark=*, mark options={solid}, mark repeat* = 10\\
color13, only marks, very thick ,mark=triangle*, mark options = {fill opacity = 1}, mark repeat* = 10\\
color13, only marks, very thick ,mark=triangle*, mark options = {fill opacity = 1}, mark repeat* = 10\\
color13, only marks, very thick ,mark=triangle*, mark options = {fill opacity = 1}, mark repeat* = 10\\
color13, only marks, very thick ,mark=triangle*, mark options = {fill opacity = 1}, mark repeat* = 10\\
color13, only marks, very thick ,mark=square*, mark options = {fill opacity = 1}, mark repeat* = 10\\
color13, only marks, very thick ,mark=square*, mark options = {fill opacity = 1}, mark repeat* = 10\\
color13, only marks, very thick ,mark=triangle*, mark options = {fill opacity = 1}, mark repeat* = 10\\
color33, only marks, very thick ,mark=square*, mark options = {fill opacity = 1}, mark repeat* = 10\\
color22, only marks, very thick ,mark=square*, mark options = {fill opacity = 1}, mark repeat* = 10\\
color13,  thick \\
color13,  thick \\
color13,  thick \\
color13,  thick \\
color13,  thick \\
color13,  thick \\
color13,  thick \\
color33, ultra thick \\
color22, ultra thick \\
color12, only marks, very thick ,mark=*, mark options={solid}, mark repeat* = 10\\
color22, only marks, very thick ,mark=*, mark options={solid}, mark repeat* = 10\\
color33, only marks, very thick ,mark=*, mark options={solid}, mark repeat* = 10\\
color12, ultra thick \\
color22, ultra thick \\
color33, ultra thick \\
}

\pgfplotscreateplotcyclelist{mycolorlist_r1_mixed}{%
only marks,solid, mark options={solid,fill=color33,fill opacity=1,mark size=4pt},mark=square*,color33\\
only marks,solid, very thick, mark options={solid,fill=color12,fill opacity=1,mark size=4pt},mark=triangle*,color12\\
only marks,solid, very thick, mark options={solid,fill=color13,fill opacity=1,mark size=4pt},mark=triangle*,color13\\
only marks,solid, mark options={solid,fill=color12,fill opacity=1,mark size=4pt},mark=*,color12\\
only marks,solid, very thick, mark options={fill=color21,fill opacity=1,mark size=4pt},mark=triangle*,color21\\
only marks,solid, mark options={solid,fill=color22,fill opacity=1,mark size=4pt},mark=square*,color22\\
only marks,solid, very thick, mark options={solid,fill=color13,fill opacity=1,mark size=4pt},mark=triangle*,color13\\
only marks,solid, mark options={solid,fill=color22,fill opacity=1,mark size=4pt},mark=*,color22\\
only marks,solid, very thick, mark options={fill=color13,fill opacity=1,mark size=4pt},mark=triangle*,color13\\
only marks,solid, very thick, mark options={fill=color13,fill opacity=1,mark size=4pt},mark=triangle*,color13\\
only marks,solid, mark options={solid,fill=color13,fill opacity=1,mark size=4pt},mark=square*,color13\\
only marks,solid, mark options={solid,fill=color33,fill opacity=1,mark size=4pt},mark=*,color33\\
color13, dashed,  thick, mark options = {solid}, mark=triangle*,mark repeat* = 10\\
color13, dashed,  thick ,mark=triangle*, mark options={solid}, mark repeat* = 10\\
color13, dashed,  thick ,mark=triangle*, mark options ={solid}, mark repeat* = 10\\
color13, dashed,  thick ,mark=triangle*, mark options={solid}, mark repeat* = 10\\
color13, dashed,  thick,mark=triangle*, mark options={solid} , mark repeat* = 10\\
color33, dashed,  ultra thick ,mark=square*, mark options={solid}, mark repeat* = 10\\
color12, dashed,  ultra thick ,mark=triangle*, mark options={solid}, mark repeat* = 10\\
color21, dashed,  ultra thick ,mark=triangle*, mark options={solid}, mark repeat* = 10\\
color22, dashed, ultra thick,mark=square*, mark options={solid} , mark repeat* = 10\\
color12, dashed, ultra thick ,mark=*, mark options={solid}, mark repeat* = 10\\
color22, dashed, ultra thick ,mark=*, mark options={solid}, mark repeat* = 20\\
color33, dashed, ultra thick ,mark=*, mark options={solid}, mark repeat* = 10\\
color13, only marks, very thick ,mark=triangle*, mark options = {fill opacity = 1}, mark repeat* = 10\\
color13, only marks, very thick ,mark=triangle*, mark options = {fill opacity = 1}, mark repeat* = 10\\
color13, only marks, very thick ,mark=triangle*, mark options = {fill opacity = 1}, mark repeat* = 10\\
color13, only marks, very thick ,mark=triangle*, mark options = {fill opacity = 1}, mark repeat* = 10\\
color13, only marks, very thick ,mark=square*, mark options = {fill opacity = 1}, mark repeat* = 10\\
color33, only marks, very thick ,mark=square*, mark options = {fill opacity = 1}, mark repeat* = 10\\
color12, only marks, very thick ,mark=triangle*, mark options = {fill opacity = 1}, mark repeat* = 10\\
color21, only marks, very thick ,mark=triangle*, mark options = {fill opacity = 1}, mark repeat* = 10\\
color22, only marks, very thick ,mark=square*, mark options = {fill opacity = 1}, mark repeat* = 10\\
color13,  thick \\
color13,  thick \\
color13,  thick \\
color13,  thick \\
color13,  thick \\
color33, ultra thick \\
color12, ultra thick \\
color21, ultra thick \\
color22, ultra thick \\
color12, only marks, very thick ,mark=*, mark options={solid}, mark repeat* = 10\\
color22, only marks, very thick ,mark=*, mark options={solid}, mark repeat* = 10\\
color33, only marks, very thick ,mark=*, mark options={solid}, mark repeat* = 10\\
color12, ultra thick \\
color22, ultra thick \\
color33, ultra thick \\
}

\pgfplotscreateplotcyclelist{mycolorlist_cub_mixed_t}{%
only marks,solid, mark options={solid,fill=color33,fill opacity=1,mark size=4pt},mark=square*,color33\\
only marks,solid, very thick, mark options={solid,fill=color12,fill opacity=1,mark size=4pt},mark=triangle*,color12\\
only marks,solid, very thick, mark options={solid,fill=color13,fill opacity=1,mark size=4pt},mark=triangle*,color13\\
only marks,solid, mark options={solid,fill=color12,fill opacity=1,mark size=4pt},mark=*,color12\\
only marks,solid, very thick, mark options={fill=color21,fill opacity=1,mark size=4pt},mark=triangle*,color21\\
only marks,solid, mark options={solid,fill=color22,fill opacity=1,mark size=4pt},mark=square*,color22\\
only marks,solid, very thick, mark options={solid,fill=color13,fill opacity=1,mark size=4pt},mark=triangle*,color13\\
only marks,solid, mark options={solid,fill=color33,fill opacity=1,mark size=4pt},mark=*,color33\\
only marks,solid, very thick, mark options={fill=color13,fill opacity=1,mark size=4pt},mark=triangle*,color13\\
only marks,solid, very thick, mark options={fill=color13,fill opacity=1,mark size=4pt},mark=triangle*,color13\\
only marks,solid, mark options={solid,fill=color13,fill opacity=1,mark size=4pt},mark=square*,color13\\
only marks,solid, mark options={solid,fill=color22,fill opacity=1,mark size=4pt},mark=*,color22\\
color13, only marks, very thick ,mark=triangle*, mark options = {fill opacity = 1}, mark repeat* = 10\\
color13, only marks, very thick ,mark=triangle*, mark options = {fill opacity = 1}, mark repeat* = 10\\
color13, only marks, very thick ,mark=triangle*, mark options = {fill opacity = 1}, mark repeat* = 10\\
color13, only marks, very thick ,mark=triangle*, mark options = {fill opacity = 1}, mark repeat* = 10\\
color13, only marks, very thick ,mark=triangle*, mark options = {fill opacity = 1}, mark repeat* = 10\\
color33, only marks, very thick ,mark=square*, mark options = {fill opacity = 1}, mark repeat* = 10\\
color12, only marks, very thick ,mark=triangle*, mark options = {fill opacity = 1}, mark repeat* = 10\\
color21, only marks, very thick ,mark=triangle*, mark options = {fill opacity = 1}, mark repeat* = 10\\
color22, only marks, very thick ,mark=square*, mark options = {fill opacity = 1}, mark repeat* = 10\\
color13,  thick \\
color13,  thick \\
color13,  thick \\
color13,  thick \\
color13,  thick \\
color33, ultra thick \\
color12, ultra thick \\
color21, ultra thick \\
color22, ultra thick \\
color12, only marks, very thick ,mark=*, mark options={solid},  mark repeat* = 10\\
color33, only marks, very thick ,mark=*, mark options={solid},  mark repeat* = 10\\
color22, only marks, very thick ,mark=*, mark options={solid},  mark repeat* = 10\\
color12, ultra thick \\
color33, ultra thick \\
color22, ultra thick \\
}

\pgfplotscreateplotcyclelist{mycolorlist_cub_mech}{%
only marks,solid, mark options={solid,fill=color33,fill opacity=1,mark size=4pt},mark=square*,color33\\
only marks,solid, very thick, mark options={solid,fill=color12,fill opacity=1,mark size=4pt},mark=triangle*,color12\\
only marks,solid, very thick, mark options={solid,fill=color13,fill opacity=1,mark size=4pt},mark=triangle*,color13\\
only marks,solid, very thick, mark options={fill=color21,fill opacity=1,mark size=4pt},mark=triangle*,color21\\
only marks,solid, mark options={solid,fill=color22,fill opacity=1,mark size=4pt},mark=square*,color22\\
only marks,solid, very thick, mark options={solid,fill=color13,fill opacity=1,mark size=4pt},mark=triangle*,color13\\
only marks,solid, very thick, mark options={fill=color13,fill opacity=1,mark size=4pt},mark=triangle*,color13\\
only marks,solid, very thick, mark options={fill=color13,fill opacity=1,mark size=4pt},mark=triangle*,color13\\
only marks,solid, mark options={solid,fill=color13,fill opacity=1,mark size=4pt},mark=square*,color13\\
color13, only marks, very thick ,mark=triangle*, mark options = {fill opacity = 1}, mark repeat* = 10\\
color13, only marks, very thick ,mark=triangle*, mark options = {fill opacity = 1}, mark repeat* = 10\\
color13, only marks, very thick ,mark=triangle*, mark options = {fill opacity = 1}, mark repeat* = 10\\
color13, only marks, very thick ,mark=triangle*, mark options = {fill opacity = 1}, mark repeat* = 10\\
color13, only marks, very thick ,mark=triangle*, mark options = {fill opacity = 1}, mark repeat* = 10\\
color33, only marks, very thick ,mark=square*, mark options = {fill opacity = 1}, mark repeat* = 10\\
color12, only marks, very thick ,mark=triangle*, mark options = {fill opacity = 1}, mark repeat* = 10\\
color21, only marks, very thick ,mark=triangle*, mark options = {fill opacity = 1}, mark repeat* = 10\\
color22, only marks, very thick ,mark=square*, mark options = {fill opacity = 1}, mark repeat* = 10\\
color13,  thick \\
color13,  thick \\
color13,  thick \\
color13,  thick \\
color13,  thick \\
color33, ultra thick \\
color12, ultra thick \\
color21, ultra thick \\
color22, ultra thick \\
}

\pgfplotscreateplotcyclelist{mycolorlist_input}{%
only marks,solid, mark options={solid,fill=color13,fill opacity=1,mark size=4pt},mark=square*,color13\\
only marks,solid, very thick, mark options={solid,fill=color13,fill opacity=1,mark size=4pt},mark=triangle*,color13\\
only marks,solid, very thick, mark options={solid,fill=color12,fill opacity=1,mark size=4pt},mark=triangle*,color12\\
only marks,solid, mark options={solid,fill=color12,fill opacity=1,mark size=4pt},mark=*,color12\\
only marks,solid, very thick, mark options={solid,fill=color13,fill opacity=1,mark size=4pt},mark=triangle*,color13\\
only marks,solid, mark options={solid,fill=color13,fill opacity=1,mark size=4pt},mark=square*,color13\\
only marks,solid, very thick, mark options={solid,fill=color21,fill opacity=1,mark size=4pt},mark=triangle*,color21\\
only marks,solid, mark options={solid,fill=color33,fill opacity=1,mark size=4pt},mark=*,color33\\
only marks,solid, very thick, mark options={solid,fill=color13,fill opacity=1,mark size=4pt},mark=triangle*,color13\\
only marks,solid, very thick, mark options={solid,fill=color13,fill opacity=1,mark size=4pt},mark=triangle*,color13\\
only marks,solid, mark options={solid,fill=color33,fill opacity=1,mark size=4pt},mark=square*,color33\\
only marks,solid, mark options={solid,fill=color22,fill opacity=1,mark size=4pt},mark=*,color22\\
color13, dashed,  thick, mark options = {solid}, mark=square*,mark repeat* = 5\\
color13, dashed,  thick ,mark=triangle*, mark options={solid}, mark repeat* = 5\\
color13, dashed,  thick ,mark=triangle*, mark options ={solid}, mark repeat* = 5\\
color13, dashed,  thick ,mark=square*, mark options={solid}, mark repeat* = 5\\
color13, dashed,  thick,mark=triangle*, mark options={solid} , mark repeat* = 5\\
color13, dashed,  thick ,mark=triangle*, mark options={solid}, mark repeat* = 5\\
color12, dashed,  ultra thick ,mark=triangle*, mark options={solid}, mark repeat* = 5\\
color21, dashed,  ultra thick ,mark=triangle*, mark options={solid}, mark repeat* = 5\\
color33, dashed, ultra thick,mark=square*, mark options={solid} , mark repeat* = 5\\
color12, dashed, ultra thick ,mark=*, mark options={solid}, mark repeat* = 5\\
color33, dashed, ultra thick ,mark=*, mark options={solid}, mark repeat* = 5\\
color22, dashed, ultra thick ,mark=*, mark options={solid}, mark repeat* = 5\\
}

\pgfplotscreateplotcyclelist{mycolorlist_MSE}{%
only marks,solid, mark options={solid,fill=color33,fill opacity=1,mark size=4pt},mark=square*,color33\\
only marks,solid, mark options={solid,fill=color22,fill opacity=1,mark size=4pt},mark=square*,color22\\
only marks,solid, mark options={solid,fill=color12,fill opacity=1,mark size=4pt},mark=square*,color12\\
only marks,solid, mark options={solid,fill=color21,fill opacity=1,mark size=4pt},mark=square*,color21\\
}

\pgfplotscreateplotcyclelist{mycolorlist}{%
only marks,solid, mark options={solid,fill=color12,fill opacity=1,mark size=4pt},mark=square*,color12\\
only marks,solid, mark options={solid,fill=color33,fill opacity=1,mark size=4pt},mark=square*,color33\\
only marks,solid, mark options={solid,fill=color22,fill opacity=1,mark size=4pt},mark=square*,color22\\
color12, dashed, very thick, mark options = {solid}, mark=square*, mark repeat = 5\\
color33, dashed, very thick ,mark=square*, mark options={solid}, mark repeat = 3\\
color22, dashed, very thick,mark=square*, mark options={solid} , mark repeat = 3\\
color12, very thick \\
color33, very thick \\
color22, very thick \\
color12, dashed, very thick, mark options = {solid}, mark=square*, mark repeat = 10\\
color33, dashed, very thick ,mark=square*, mark options={solid}, mark repeat = 10\\
color22, dashed, very thick,mark=square*, mark options={solid} , mark repeat = 10\\
gray,  thick \\
gray,  thick \\
gray,  thick \\
gray,  thick \\
gray,  thick \\
gray,  thick \\
gray,  thick \\
gray,  thick \\
gray,  thick \\
gray,  thick \\
gray,  thick \\
gray,  thick \\
gray,  thick \\
gray,  thick \\
gray,  thick \\
gray,  thick \\
gray,  thick \\
}

\usepackage[maxbibnames=10, maxcitenames=3, giveninits=true,
    natbib=true, bibencoding=utf8, isbn=false, url=false, backend=bibtex, style=numeric-comp]{biblatex}
\AtEveryBibitem{\clearfield{month}}
\AtEveryBibitem{\clearfield{day}}
\AtEveryBibitem{\clearlist{language}}

\usepackage[german, main=english]{babel}
\usepackage[autostyle]{csquotes}

\usepackage[T1]{fontenc}

\usepackage{hyperref}
\usepackage[capitalise]{cleveref}

\usepackage[affil-it]{authblk}
\title{
Finite electro-elasticity with physics-augmented neural networks
}

\author[1,*]{Dominik~K.~Klein}
\author[2]{Rogelio~Ortigosa}
\author[2]{\\Jesús~Martínez-Frutos}
\author[1]{,\;and~Oliver~Weeger}
\affil[1]{\footnotesize Cyber-Physical Simulation Group \& Graduate School of Computational Engineering, \protect \\
Department of Mechanical Engineering \& Centre for Computational Engineering,\protect \\ Technical University of Darmstadt, Dolivostr.~15, 64293 Darmstadt, Germany}
\affil[2]{\footnotesize Technical University of Cartagena, Campus~Muralla~del~Mar, 30202, Cartagena (Murcia), Spain}
\affil[*]{\footnotesize Corresponding author, email: \href{mailto:klein@cps.tu-darmstadt.de}{klein@cps.tu-darmstadt.de}}

\date{July 30, 2022}

\usetikzlibrary{external}
\newtheorem{definition}{Definition}[section]
\newtheorem{remark}[definition]{Remark}
\newcommand{\Cross}{\boldsymbol{\times}}
\newcommand{\vect}[1]{\boldsymbol{#1}}

\pgfplotsset{
    mark repeat*/.style={
        scatter,
        scatter src=x,
        scatter/@pre marker code/.code={
            \pgfmathtruncatemacro\usemark{
                or(mod(\coordindex,#1)==0, (\coordindex==(\numcoords-1))
            }
            \ifnum\usemark=0
                \pgfplotsset{mark=none}
            \fi
        },
        scatter/@post marker code/.code={}
    }
}

\begin{document}

\maketitle

\par\noindent\rule{\textwidth}{0.4pt}
\begin{abstract}
\noindent
In the present work, a machine learning based constitutive model for electro-mechanically coupled material behavior at finite deformations is proposed. Using different sets of invariants as inputs, an internal energy density is formulated as a convex neural network. In this way, the model fulfills the polyconvexity condition which ensures material stability, as well as thermodynamic consistency, objectivity, material symmetry, and growth conditions. Depending on the considered invariants, this physics-augmented machine learning model can either be applied for compressible or nearly incompressible material behavior, as well as for arbitrary material symmetry classes. The applicability and versatility of the approach is demonstrated by calibrating it on transversely isotropic data generated with an analytical potential, as well as for the effective constitutive modeling of an analytically homogenized, transversely isotropic rank-one laminate composite and a numerically homogenized cubic metamaterial. These examinations show the excellent generalization properties that physics-augmented neural networks offer also for multi-physical material modeling such as nonlinear electro-elasticity.
\end{abstract}

\vspace*{0.5ex}
\par\noindent\rule{\textwidth}{0.4pt}\vspace*{2pt}
{\small
Accepted version of manuscript published in \emph{Computer Methods in Applied Mechanics and Engineering}. \\
Date accepted: July 31, 2022. DOI: \href{https://doi.org/10.1016/j.cma.2022.115501}{10.1016/j.cma.2022.115501}. License: \href{https://creativecommons.org/licenses/by-nc-nd/4.0/legalcode}{CC BY-NC-ND 4.0}
}
\vspace*{-1.6mm}
\par\noindent\rule{\textwidth}{0.4pt}

\section{Introduction}
\label{sec:intro}

Electro active polymers (EAPs) belong to a class of multi-functional materials, which can undergo significantly large electrically induced deformations when subjected to an electric field \cite{Pelrine_1998,Pelrine_2003}. They can be broadly classified into ionic EAPs, activated by transport of ions, and electronic EAPs, actuated through the application of an electric field. 
Dielectric elastomers (DEs) are recognised as one of the most popular electronic EAPs \cite{Pelrine_2002,Pelrine_2000} due to their outstanding actuation capabilities (i.e., light weight, fast response time, flexibility, low stiffness properties), which makes them ideal for use as soft robots \cite{DE_review_2008}. 
However, extremely large electric fields are generally required in order to access the large actuation regime in DEs, an aspect that very often places them on the risk of encountering electro-mechanical instabilities or even electrical breakdown \cite{Skov_review_2016,Li_2013_giant_deformation}.

With the aim of reducing the high operational voltage required for actuation in DEs, some authors have advocated for the design of composite-based DEs \cite{Huan_2004,Huan_2004b}.
Typically, these composites combine an ultra-soft and low-permittivity elastomer matrix with a stiffer and high-permittivity inclusion randomly distributed in the form of fibres or particles \cite{Siboni_Casteneda_2014,Siboni_Castaneda_2014_fiber_constrained_DE}. 
Experimental studies have demonstrated a significant enhancement in the electro-mechanical coupling performance of DE composites \cite{Huan_2004}, thus reducing the operational voltage required for actuation. 
Multi-layered laminated DE composites have gained considerable momentum over alternative DE composites. \textcite{Gei_R2L} confirmed that the actuation performance of a DE composite can be amplified with respect to that of a single-phase DE, by optimizing the contrast and volume fraction of the DE composite constituents. 

\medskip

For rank-one laminate composites, it is possible to apply rank-$n$ homogenization principles and to obtain analytical formulas describing their effective material response \cite{Bertoldi_Gei_2011}.
However, depending on the complexity of the constitutive models of the different phases of the laminate, its effective response (i.e., homogenized stress and electric field) might only be available implicitly, requiring the solution of a nonlinear system of equations.
For other types of composites, \textcite{Pamies2019} have used analytical homogenization to derive explicit analytical formulas for the behavior of porous-type materials. 
Unfortunately, analytical expressions might not be available in general, for instance, for metamaterials such as an elastomeric matrix with a stiff spherical or ellipsoidal inclusion. 
In these cases, computational homogenization methods need to be applied in order to obtain the effective response of the composite. However, the use of computational homogenization methods in the context of concurrent multiscale simulations entails extremely high computational costs, as it requires the repeated solution of boundary value problems across scales \cite{Keip_2014}. 
Therefore, the derivation of effective constitutive models that can capture the physics of complex electro-mechanical metamaterials would entail an extraordinary benefit.

Constitutive models should generally well approximate the available data of the material response (whether coming from experiments, analytical or computational homogenization), but also comply with physical requirements, i.e., objectivity, material symmetry, etc. 
For instance, to comply with thermodynamic consistency requirements, constitutive models in (electro-) elasticity are typically formulated in terms of an internal energy density \cite{dorfmann2005}.
Furthermore, the concept of material stability can be related to the mathematical condition of ellipticity \cite{Zee1983} or rank-one convexity of the internal energy density \cite{Marsden_book,Le_Tallec_book}. It prevents the formation of extremely localized deformation modes, guarantees the propagation of real travelling plane waves in the material \cite{Schroder_2005}, and ensures the well-posedness of the underlying boundary value problem \cite{Marsden_book}. 
A sufficient condition complying with the ellipticity condition is the polyconvexity of the internal energy density \cite{Ball1976,Ball1977,Betsch_Energy_Momentum_Polyconvexity,Schroder_2003}. This concept was later extended from elasticity to the field of nonlinear electro-mechanics \cite{gil2016,ortigosa2016,silhavy2018}.

The analytical formulation of constitutive models that can capture the effective response of complex electro-active metamaterials, whilst complying with suitable mathematical requirements, is unfortunately far from trivial, and mostly restricted to certain types of composites \cite{Pamies2019}. However, we envision the use of data-driven methods, such as machine learning techniques, as an extremely adequate and effective tool that can help paving the way to circumvent the current limitations of analytically formulated material models. 

\medskip


Machine learning has reached the world of mechanical constitutive modeling for a while, sparking the development of a variety of models. They have in common that they reduce the \emph{epistemic uncertainty} associated with analytical approaches, in the sense that they reduce their \emph{model uncertainty} \cite{hüllermeier2021}: For analytical models, at some point an explicit choice for some functional relationship has to be made. While some explicit choices have a strong physical motivation, such as the hyperelastic Hencky model \cite{Hencky1928,Hencky1929,neff2016, agn_neff2020axiomatic}, most approaches are of a heuristic nature, e.g., the polynomial form of the hyperelastic Ogden model \cite{ogden2004}. 
The reduced flexibility that often goes along with this human choice of functional relationship has no physical motivation, but purely stems from the necessity of an explicit form of the model.
%
Data-driven approaches such as \cite{flaschel2021,flaschel2022} reduce this limitation by automatically choosing from a large class of parametrized analytical models, which significantly exceed the amount of models a human would examine.
So called \enquote{data-driven mechanics} approaches completely abandon the formulation of a constitutive model and purely act on discrete stress-strain data \cite{kirchdoerfer2016, carrara2020}, thus making no assumption on the constitutive model at all.
These \enquote{data-driven} methods have also been successfully applied to magnetics, where $BH$-curves are replaced by data \cite{degersem2020, galetzka2021}.
In this work, we focus on constitutive models which use artificial neural networks (ANNs) \cite{kollmannsberger2021, aggarwal2018}. 
By using them as highly flexible functions, an explicit choice of functional relationship is also bypassed.

This leads to the second characteristic of machine learning based constitutive models: their excellent \emph{flexibility}.
From the perspective of constitutive modeling, the major benefit of ANNs is their flexibility, in fact, they are universal approximators \cite{Hornik1991}. However, due to their high number of parameters and recursive definition, this goes along with a lack of traceability, which constitutive models need to some amount. 
In order to arrive at reliable model predictions it is important for a model to be physically sensible, i.e., to  fulfill certain (mathematically formulated) physical requirements. Furthermore, augmenting ANN based models with physical requirements and mathematical structure can significantly decrease the amount of data required to generalize well \cite{karniadakis2021}, whereas uninformed ANNs would require a large amount of data which is often not available.
Thus, while the specific formulations of ANN based constitutive models may be different, the importance of including constitutive conditions is generally accepted \cite{karniadakis2021, kumar2022}.

\medskip

As mentioned above, for (electro-) mechanical constitutive models these physical requirements and mathematical conditions  were developed in the last decades \cite{TruesdellNoll,haupt1977,Treloar1975}, e.g., the second law of thermodynamics \cite{Silhavy2014} and polyconvexity \cite{Ball1976,Ball1977}.
While the conditions that a constitutive model should fulfill are well known, fulfilling all of them at the same time can be seen as  \enquote{the main open problem of the theory of material behavior} (\textcite[Section 20]{TruesdellNoll}), a problem which remains for ANN based models.
To incorporate these requirements into ANN based models, here both methods known from analytical models, as well as novel methods specifically designed for ANNs are applied. 

Starting with the more conventional methods, by a special choice of \emph{input quantities}, i.e., invariants \cite{Schroeder2003, itskov2019}, multiple constitutive requirements can be fulfilled at the same time. Using invariants as input quantities for feed-forward neural networks (FFNNs) \cite{kollmannsberger2021} with scalar-valued output, highly flexible hyperelastic potentials \cite{Holzapfel2000} can be constructed \cite{Linka2020,tac2021a}. However, FFNNs are in general not convex, and thus the models proposed in \cite{Linka2020,tac2021a} do not fulfill the polyconvexity condition. For this, a special choice of \emph{network architecture} is required. Using input convex neural networks (ICNNs) \cite{Amos2016}, some of the present authors were the first to propose a model which combines the flexibility of ANNs with the polyconvexity condition, see \textcite{klein2021}, later also implemented by \cite{asad2022}, and followed by \textcite{tac2021}, who exploited special properties of neural ordinary differential equations \cite{chen2019}. Furthermore, ANNs can be \emph{embedded in a modeling framework}, by using them only as a part of the overall model and using additional analytical terms, e.g., which fulfill coercivity conditions \cite{klein2021}. 
There is a variety of ANN based models which use invariants \cite{kalina2022, Linka2020, klein2021, tac2021, thakolkaran2022}.
These methods are closely related to conventional constitutive modeling, in fact, the main difference is the use of ANNs instead of analytical formulations such as polynomials.

Additionally, there are also methods for augmenting ANN based models with constitutive requirements which are only possible in this novel machine learning framework. 
In particular, it is possible to fulfill constitutive requirements not in a \emph{hard} way, but to only \emph{approximate} them. At a first glance, this may not seem desirable, but softening constitutive requirements can be seen as a way to bypass conflicts that arise in the simultaneous fulfillment of multiple constitutive conditions. This potentially yields the additional flexibility required to represent highly challenging material behavior. 
In \cite{klein2021}, a model was proposed which only approximates the objectivity condition through data augmentation \cite{Ling2016}, which was necessary in order use the deformation gradient as model input and to get enough flexibility to represent the highly nonlinear behavior of soft lattice metamaterials.
Also, elastic models can directly be formulated to map strain to stress \cite{Fernandez2020} with the hyperelasticity condition, i.e., the existence of a potential, being approximated through a special choice of loss function \cite{weber2021}. In \cite{masi2021b}, an inelastic model was proposed which approximates the second law of thermodynamics with a physically motivated loss function.
To close, these methods for \emph{augmenting} or \emph{informing} ANNs about physical requirements are not restricted to mechanical material modeling, but also find application in other fields of physics and engineering, e.g.~\cite{moseley2020,shengze2021,liu2021e,champaney2022}.

\medskip

%
In the present work, an ANN based constitutive model for finite electro-elasticity is introduced. 
To the best of the authors' knowledge, this application of machine learning for electro-mechanically coupled constitutive models has not been studied at all yet.
Extending the work of \textcite{klein2021}, different sets of electro-mechanically coupled invariants are used as input quantities of ICNNs, which are then used as internal energy functions. The resulting model is highly flexible and \emph{physics-augmented}, i.e., it fulfills the material stability condition, as well as thermodynamic consistency, objectivity, material symmetry, and growth conditions. Depending on the considered invariants, the model can either be applied for compressible or nearly incomporessible material behavior, as well as for arbitrary material symmetry classes.
This makes it suitable for the effective constitutive modeling of electro-active composites and metamaterials subject to large elastic deformations.

\medskip

The outline of the manuscript is as follows. In \cref{sec:basics}, the basics of electro-elastic constitutive modeling relevant to this work are discussed, which are then applied to the proposed ANN based model in \cref{sec:model}. As a first proof of concept, the model is applied to data generated with an analytical transversely isotropic potential in \cref{sec:ana_pot}, where also some general properties of physics-augmented models are discussed. In \cref{sec:R1}, the model is applied for the effective material modeling of an analytically homogenized, transversely isotropic rank-one laminate composite. In \cref{sec:cub}, the model is further applied for the effective material modeling of a numerically homogenized cubic metamaterial. After the conclusion in \cref{sec:conc}, some additional information on the data generation is presented in the appendices.

\medskip

\paragraph{Notation}
Throughout this work, tensor compositions and contractions are denoted by $\left(\bA\,\bB\right)_{ij}=A_{ik}B_{kj}$, $\ba\cdot\bb=a_ib_i$, $\bA:\bB=A_{ij}B_{ij}$ and $\left(\mathbb{A}:\bA\right)_{ij}=\mathbb{A}_{ijkl}A_{kl}$, respectively, with vectors $\ba$ and $\bb$, second order tensors $\bA$ and $\bB$ and fourth order tensor $\mathbb{A}$. The tensor product is denoted by~$\otimes$, the second order identity tensor by $\bI$.
The gradient operator is denoted by $\bnabla$ and the divergence operator by $\bnabla \cdot$ . A zero as the lower index next to the nabla symbol indicates that the operation is carried out in the referential configuration.
The operator $\Cross$ is defined as $(\vect{A}\Cross\vect{B})_{iI}=\mathcal{E}_{ijk}\mathcal{E}_{IJK}A_{jJ}B_{kK}$, where $\mathcal{E}_{ijk}$ symbolises the third-order permutation tensor. 
For tensors in index notation, lower case indices $\{i, j, k\}$ will be used to represent the spatial configuration, whereas capital case
	indices $\{I, J, K\}$ will be used to represent the material description. 
The first Fr\'echet derivative of a function $f$ w.r.t.\ $\bX$ is denoted by $\nicefrac{\partial f}{\partial \bX}$.
The set of invertible second order tensors with positive determinant is denoted by $\text{GL}^+(3):=\big\{\bX \in\allowbreak \;\mathbb{R}^{3\times 3}\,\rvert\,\allowbreak \det \bX > 0\big\}$ and the special orthogonal group in $\mathbb{R}^3$ by $\SO(3):=\big\{\bX \in\allowbreak \mathbb{R}^{3\times 3}\;\rvert\allowbreak \;\bX^T\bX=\bI,\;\det \bX =1\big\}$.
The Frobenius norm is denoted by $\|*\|$.
Unimodular tensors are overlined, i.e., $\Bar{\A}$ with $\det \Bar{\A} = 1$.

\section{Nonlinear continuum electro-mechanics}
\label{sec:basics}

\begin{figure}[t!] 
	\centering 	
	\includegraphics[width=0.6\textwidth]{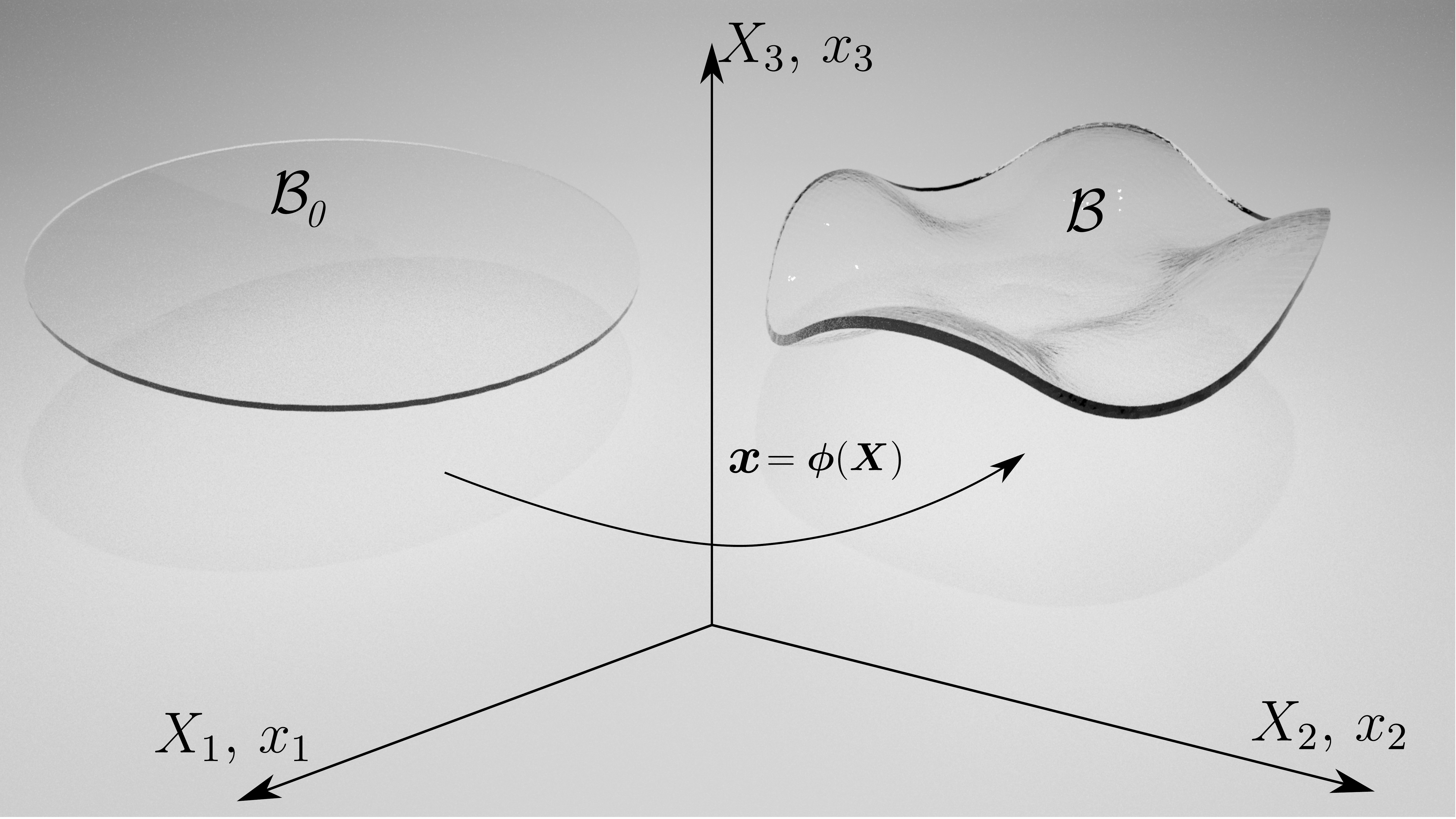}  
	\caption{Deformation of the body in the reference configuration $\mathcal{B}_0$ of an electro-mechanically coupled continuum can be induced with an electric field, leading to the deformed configuration $\mathcal{B}$.
\blue{Figure taken from \cite{Ortigosa_Multiresolution_TO}}.	}
	\label{fig:setup 1}
\end{figure}

Let us consider the deformation of an electro-mechanically coupled continuum, such as for instance a dielectric elastomer, electro-active polymer, or metamaterial \cite{dorfmann2005,Bustamante_Merodio_2011,McMeeking_Landis_2008}. 
As shown in \cref{fig:setup 1}, the body is given in its reference configuration by the open domain $\cB_0\subset\bbR^3$ with boundary
$\partial\cB_0$ and unit outward normal $\boldsymbol{n}_0$. After the deformation, it occupies a current configuration given by the domain $\cB\subset \mathbb{R}^3$ with boundary $\partial\cB$ and unit outward normal $\boldsymbol{n}$. The deformation of the electro-mechanical body 
is defined by a mapping $\boldsymbol{\phi}:\cB_0\rightarrow \mathbb{R}^3$ linking material particles $\boldsymbol{X}\in\cB_0$  to $\boldsymbol{x}=\boldsymbol{\phi}(\bX)\in\cB$.
Associated with $\boldsymbol{\phi}$, the deformation gradient tensor $\boldsymbol{F}\in\text{GL}^+(3)$ is defined as
\begin{equation}\label{eqn:F}
\boldsymbol{F}=\boldsymbol{\nabla}_0\boldsymbol{\phi}\,,\qquad F_{iI}=\frac{\partial \phi_i}{\partial X_I}\,.
\end{equation}	
Then, its determinant $J$ and its cofactor $\boldsymbol{H}$ are defined as
\begin{equation}\label{eqn:J and H}
	J=\text{det}\,\boldsymbol{F}=\blue{\frac{1}{6}\vect{F}:\left(\vect{F}\Cross \vect{F}\right)}\,,\qquad  \boldsymbol{H}\blue{=\text{Cof}\,\boldsymbol{F}}=J\boldsymbol{F}^{-T}=\frac{1}{2}\boldsymbol{F}\Cross\vect{F}\,.
\end{equation}

In the absence of inertia and magnetic effects, the system of partial differential equations
governing the behaviour of the electro-mechanical continuum comprises the conservation of linear momentum and the
compatibility equation \eqref{eqn:F}, along with the quasi-static version of the Gauss’ and Faraday’s laws. The coupled boundary value problem can be recast in a total Lagrangian formalism as
\begin{equation}\label{eqn:local form conservation of linear momentum}
	\begin{aligned}
		\left\{\begin{aligned}
			\vect{F}&=\vect{\nabla}_0\vect{\phi},&\quad &\text{in}\,\,\mathcal{B}_0\,,\\
				\vect{\nabla}_0\vect{P} + \vect{f}_0& = \vect{0}\,,&\quad &\text{in}\,\,\mathcal{B}_0\,,\\
				\vect{P}\vect{n}_0& = \vect{t}_0\,,&\quad&\text{on}\,\,\partial_{\boldsymbol{t}}\mathcal{B}_0\,,\\
				\vect{\phi} &= \bar{\vect{\phi}}\,,&\quad & \text{on}\,\,\partial_{\vect{\phi}}\mathcal{B}_0\,,\\
		\end{aligned}\right.\qquad  \,\,\,\,\,\,\,\,\,
		\left\{\begin{aligned}
				\vect{e}_0&=-\vect{\nabla}_0\varphi\,,&\quad&\text{in}\,\,\mathcal{B}_0\,,\\
				\vect{\nabla}_0\cdot\vect{d}_0 - \rho_0 &= 0\,,&\quad&\text{in}\,\,\mathcal{B}_0\,,\\
				\vect{d}_0\cdot\vect{n}_0& = - \omega_0\,,&\quad&\text{on}\,\,\partial_{\omega}\mathcal{B}_0\,,\\
				\varphi & = \bar{\varphi}\,,&\quad & \text{on} \,\,\partial_{\varphi}\mathcal{B}_0\,,
		\end{aligned}\right.
	\end{aligned}
\end{equation}
where $\vect{f}_0$ represents a body force per unit undeformed volume $\mathcal{B}_0$, $\vect{t}_0$, the traction force per unit undeformed area on $\partial_{\boldsymbol{t}}\mathcal{B}_0\subset \partial \mathcal{B}_0$,  $\bar{\vect{\phi}}$, the value of the Dirichlet boundary condition on $\partial_{\boldsymbol{\phi}}\mathcal{B}_0\subset \partial \mathcal{B}_0$, with $\partial_{\boldsymbol{t}}\mathcal{B}_0 \cup \partial_{\vect{\phi}}\mathcal{B}_0 = \partial\mathcal{B}_0$ and $\partial_{\boldsymbol{t}}\mathcal{B}_0 \cap \partial_{\vect{\phi}}\mathcal{B}_0 = \emptyset$.  Furthermore, $\rho_0$ represents an electric volume charge per unit of undeformed volume {$\mathcal{B}_0$}, and $\omega_0$ an electric surface charge  per unit of undeformed area $\partial_{\omega}\mathcal{B}_0\subset \partial\mathcal{B}_0$. 
In addition, $\vect{e}_0$ is the Lagrangian electric field vector, $\varphi:\mathcal{B}_0\rightarrow \mathbb{R}$, the scalar electric potential, and $\partial_{\varphi}\mathcal{B}_0$, the part of the boundary {$\partial\mathcal{B}_0$} where essential electric potential boundary conditions are applied so that $\partial_{\omega}\mathcal{B}_0 \cup\partial_{\varphi}\mathcal{B}_0 = \partial\mathcal{B}_0$ and $\partial_{\omega}\mathcal{B}_0\cap\partial_{\varphi}\mathcal{B}_0 = \emptyset$.
Finally,  $\vect{P}$ and $\vect{d}_0$ represent the first Piola-Kirchhoff stress tensor and the Lagrangian electric displacement field, respectively.

\medskip

\subsection{Constitutive equations of electro-elasticity}

In reversible nonlinear electro-mechanics, i.e., finite deformation electro-elasticity, the constitutive relationships required for the closure of \cref{eqn:local form conservation of linear momentum} can be defined in terms of the internal energy $e$ per unit undeformed volume $\cB_0$ \cite{gil2016,ortigosa2016}. This energy potential depends solely on the deformation gradient tensor $\vect{F}$ and on the material electric displacement field $\vect{d}_0$, namely 
\begin{equation}
	e\colon\text{GL}^+(3)\times\bbR^3\rightarrow\bbR\,,\qquad\left(\bF,\,\bd_0\right)\mapsto e\left(\bF,\,\bd_0\right)\,.
\end{equation}
For simplicity, we write $e=e(\vect{\mathcal{U}})$ with $\vect{\mathcal{U}}=\{\vect{F},\vect{d}_0\}$. Differentiation of the internal energy 
with respect to its arguments yields the first Piola-Kirchhoff stress tensor and the material electric field  as
\begin{equation} \label{eq:P_e}
\vect{P}(\vect{\mathcal{U}})=\frac{\partial e(\vect{\mathcal{U}})}{\partial\vect{F}}\,,\qquad 
\vect{e}_0(\vect{\mathcal{U}})=\frac{\partial e(\vect{\mathcal{U}})}{\partial\vect{d}_0}\,. 
\end{equation}

Following the work of \cite{gil2016,ortigosa2016}, where an extension of the concept of polyconvexity \cite{Ball1976,Ball1977} was proposed into the field of nonlinear electro-mechanics, we consider internal energies which are polyconvex.
Thus, $e$ can be written as
\begin{equation}\label{eq:mvc}
	\begin{aligned}
    e\left(\vect{\mathcal{U}}\right)&=\cP\left(\vect{\mathcal{V}}\right)\,,\qquad \vect{\mathcal{V}}=\left(\bF,\,\bH,\,J,\,\bd_0,\,\bd\right)\in\bbR^{25}
\,,%
	\end{aligned}
\end{equation}
where the \blue{spatial vector field} $\vect{d}$ is defined as $\vect{d}=\vect{F}\vect{d}_0$ and 
$\cP$ is a convex function with respect to its $25$ arguments.

\begin{remark}
\blue{Notice that the fifth argument in the definition of the polyconvex internal energy in \cref{eq:mvc}, namely the vector field $\boldsymbol{d}$, must not be confused with the spatial or Eulerian counterpart of the material electric displacement field $\boldsymbol{d}_0$, here denoted as $\vect{d}^{\ast}$. The relationship between $\vect{d}_0$ and $\vect{d}^{\ast}$ can be obtained through the standard Piola transformation $\vect{d}_0=\vect{H}^T\vect{d}^{\ast}$, i.e., $\vect{d}^{\ast}=J^{-1}\vect{d}$.
Furthermore, the Eulerian material electric field $\vect{e}_0$ is related to its spatial counterpart $\vect{e}^{\ast}$ through the  push-forward transformation $\vect{e}_0=\vect{F}^T\vect{e}^{\ast}$.}

\blue{In addition to the electric displacement field and the electric field, it is customary to introduce the polarization vector, which is defined in the material setting according to}
\begin{equation}\blue{
    \vect{p}_0=\vect{d}_0 - \varepsilon_0 J \vect{C}^{-1}\vect{e}_0,\qquad \vect{C}=\vect{F}^T\vect{F}, 
}\end{equation}
\blue{with $\varepsilon_0$ being the electric permittivity of vacuum. The push-forward relationship relating the material polarization vector field $\vect{p}_0$ to its spatial counterpart $\vect{p}^{\ast}$ is $\vect{p}_0=\vect{H}^T\vect{p}^{\ast}$.}

\end{remark}

For sufficiently differentiable functions $\cP$, the polyconvexity condition can be formulated in terms of the Hessian of $\cP$, namely
\begin{equation}\label{eqn:polyconvexity in second derivatives}
	\delta\vect{\mathcal{V}}:\frac{\partial^2\cP}{\partial\vect{\mathcal{V}}\partial\vect{\mathcal{V}}}:\delta\vect{\mathcal{V}}\geq 0\quad \forall\, \delta\vect{\mathcal{V}}\in\mathbb{R}^{25}\,,
\end{equation}
where all the components of the (symmetric) Hessian operator $\frac{\partial^2\cP}{\partial\vect{\mathcal{V}}\partial\vect{\mathcal{V}}}$ are given as
\begin{equation}
\frac{\partial^2\cP}{\partial\vect{\mathcal{V}}\partial\vect{\mathcal{V}}}=\begin{bmatrix}
	\frac{\partial^2\cP}{\partial \vect{F}\partial \vect{F}}  &  \frac{\partial^2\cP}{\partial \vect{F}\partial \vect{H}}  &   {\frac{\partial^2\cP}{\partial \vect{F}\partial J}}  &  {\frac{\partial^2\cP}{\partial \vect{F}\partial \vect{d}_0}}  &  {\frac{\partial^2\cP}{\partial \vect{F}\partial \vect{d}}} \\
	  &  
\frac{\partial^2\cP}{\partial \vect{H}\partial \vect{H}}  &  {\frac{\partial^2\cP}{\partial \vect{H}\partial J}}  &   {\frac{\partial^2\cP}{\partial \vect{H}\partial \vect{d}_0}}  &  {\frac{\partial^2\cP}{\partial \vect{H}\partial \vect{d}}} \\
&
&
{\frac{\partial^2\cP}{\partial J\partial J}}  &  {\frac{\partial^2\cP}{\partial J\partial \vect{d}_0}}  &  {\frac{\partial^2\cP}{\partial J\partial \vect{d}}} \\
&\text{symm.}
&
&
{\frac{\partial^2\cP}{\partial \vect{d}_0\partial \vect{d}_0}}  &  {\frac{\partial^2\cP}{\partial \vect{d}_0\partial \vect{d}}} \\
&&&&{\frac{\partial^2\cP}{\partial \vect{d}\partial \vect{d}}} 	  
\end{bmatrix}\,.
\end{equation}

The polyconvexity condition is sufficient to entail the rank-one convexity or ellipticity condition, 
which for energies with second order differentiability can be written as
\begin{equation}\label{eqn:ellipticity condition}
\delta\vect{\mathcal{U}}:\frac{\partial^2 e(\vect{\mathcal{U}})}{\partial\vect{\mathcal{U}}\partial\vect{\mathcal{U}}}:\delta\vect{\mathcal{U}}\geq 0\,,\qquad \forall\, \delta\vect{\mathcal{U}}=\{\vect{u}\otimes\vect{V},\vect{V}_{\perp}\}\,,\qquad \vect{V}_{\perp}\cdot\vect{V}=0\,,
\end{equation}
with
\begin{equation}\label{eqn:ellipticity}
\frac{\partial^2{e}}{\partial\vect{\mathcal{U}}\partial\vect{\mathcal{U}}}=\begin{bmatrix}
\frac{\partial^2\cP}{\partial \vect{F}\partial \vect{F}}  &  \frac{\partial^2\cP}{\partial \vect{F}\partial \vect{d}_0}  \\
\frac{\partial^2\cP}{\partial \vect{d}_0\partial \vect{F}}  &  {\frac{\partial^2\cP}{\partial \vect{d}_0\partial \vect{d}_0}}  
\end{bmatrix}\,.
\end{equation}
When written as in \cref{eqn:ellipticity}, the rank-one condition can be referred to as the ellipticity or Legendre-Hadamard condition, which guarantees the propagation of real travelling plane waves in the material, and the well-posedness of the boundary value problem defined in \cref{eqn:local form conservation of linear momentum} \cite{Marsden_book,Schroder_2005}.

From the perspective of constitutive modeling, ellipticity is desired rather than polyconvexity, as ellipticity ensures a favorable numerical behavior of the model. Polyconvexity, on the other side, stems from a more theoretical context, e.g., its associated coercivity conditions make assumptions on the model behavior far outside a practically relevant deformation range \cite{kruzik2019}. However, polyconvexity is more straightforward to include in the model formulation than the ellipticity condition, and is therefore used in the present work, c.f.~\cite{klein2021}.

\begin{remark}\label{remark legendre}

Polyconvexity in the sense of \cref{eq:mvc} permits to conclude that $e(\vect{\mathcal{U}})=\cP(\vect{\mathcal{V}})$ is in fact convex with respect to $\vect{d}_0$, namely
\begin{equation}
\delta\vect{d}_0\cdot\frac{\partial^2 e(\vect{\mathcal{U}})}{\partial\vect{d}_0\partial\vect{d}_0}\cdot\delta\vect{d}_0\geq 0,\qquad \forall \,\delta\vect{d}_0\in\mathbb{R}^3\,.
\end{equation}
This in turn guarantees the existence of \blue{the total or amended free energy density}
\begin{equation}
\Psi:\text{GL}^+(3)\times \mathbb{R}^3\rightarrow \mathbb{R}\,,\qquad (\vect{F},\vect{e}_0)\mapsto \Psi\left(\bF,\,\be_0\right)\,.
\end{equation}
For simplicity, we write $\Psi=\Psi(\vect{\mathcal{W}})$, with $\vect{\mathcal{W}}=\{\vect{F},\vect{e}_0\}$, where $\Psi$ is related to the polyconvex internal energy density $e(\vect{\mathcal{U}})$ by means of the Legendre transformation
\begin{equation}
\Psi(\vect{\mathcal{W}})=-\sup_{\vect{d}_0}\Big\{\vect{e}_0\cdot\vect{d}_0 - e(\vect{\mathcal{U}})\Big\}\,.
\end{equation}

\end{remark}

\begin{remark}
\blue{The internal energy density $e(\vect{F},\vect{d}_0)$ can be additively decomposed into a polarization term and a Maxwell term \cite{Bustamante_Dorfmann_Ogden_2009_variational,Dormann_Ogden_2006,Bustamante_Hossain_2021,Vu_Steinmann_Possart_2007,Vu_Steinmann_2012} according to}
\begin{equation*}\blue{
    e(\vect{F},\vect{d}_0)=e_P(\vect{F},\vect{d}_0) + e_M(\vect{F},\vect{d}_0)\,,
}\end{equation*}
\blue{where the Maxwell term, when the material is immersed in vacuum, is given by} 
\begin{equation*}\blue{
    e_M(\vect{F},\vect{d}_0)=\frac{1}{2J\varepsilon_0}\|\vect{d}\|^2\,.
}\end{equation*}
\blue{However, according to \cref{eq:P_e}, the constitutive equations relating $\{\vect{F},\vect{e}_0\}$ and $\{\vect{P},\vect{d}_0\}$ need to be given in terms of the total or amended energy representation $e(\vect{F},\vect{d}_0)$. Furthermore, the requirement of polyconvexity is imposed on the total or amended internal energy density $e(\vect{F},\vect{d}_0)$.}

\end{remark}

\subsection{Invariant representation of the internal energy density}\label{sec:objectivity and material symmetry}

The internal energy density $e(\vect{\mathcal{U}})$ must obey to the principle of objectivity or material frame indifference \cite{TruesdellNoll}, i.e., invariance with respect to rotations $\vect{Q}\in\SO(3)$ in the spatial configuration, which can be formulated as
\begin{equation}\label{eqn:objectivity}
e\left(\bQ\,\bF,\,\bd_0\right)=e\left(\bF,\bd_0\right) \quad \forall\,\bF\in \GL^+(3),\,\bd_0\in\bbR^3, \,\bQ\in\SO(3)\,.
\end{equation}
Notice in above \cref{eqn:objectivity} that only the deformation gradient tensor $\vect{F}$ is multiplied by arbitrary rotations $\vect{Q}$, as this is a two-point tensor, and hence, lying in both spatial and material configurations. This is not the case for the material electric field $\vect{d}_0$, which is solely related to the reference or material configuration, and hence, already implying objectivity. 

In addition, the strain energy density must take into account the underlying features of the material symmetry group $\mathcal{G}\subseteq \blue{O(3)}$ characterizing the isotropic or anisotropic behaviour of the material. The material symmetry condition \cite{Haupt2002} can be represented as
\begin{equation}\label{eqn:anisotropic_restriction}
        e(\bF\,\bQ,\,\bQ\,\bd_0)=e(\bF,\bd_0) \quad \forall\,\bF\in \GL^+(3),\,\bd_0\in\bbR^3, \,\bQ\in\cG\subseteq \blue{O(3)} \,.
\end{equation}
Making use of the concept of structural tensors, and representation theorems of tensor functions, the authors in \cite{Ebbing_2009,Schroeder2008,Schroder_2003} considered second order structural tensors $\vect{G}^{\mathcal{G}}$ and fourth order structural tensors $\vect{\mathbb{G}}^{\mathcal{G}}$ to characterize an underlying material symmetry group $\mathcal{G}\subseteq \blue{O(3)}$, hence, satisfying the following invariance relations with respect to transformations $\vect{Q}\in\mathcal{G}\subseteq \blue{O(3)}$, namely
\begin{equation}
\left.
\begin{aligned}
\vect{Q}\vect{G}^{\mathcal{G}}\vect{Q}^T&=\vect{G}_{\mathcal{G}}\\
\vect{Q}\boxtimes \vect{Q}:\vect{\mathbb{G}}^{\mathcal{G}}:\vect{Q}^T\boxtimes \vect{Q}^T&=\vect{\mathbb{G}}^{\mathcal{G}}
\end{aligned}\right\}\,\,\forall\, \vect{Q}\in \mathcal{G}\subseteq \blue{O(3)}\,,
\end{equation}
with 
\begin{equation}
\left(\vect{Q}\boxtimes \vect{Q}:\vect{\mathbb{G}}^{\mathcal{G}}:\vect{Q}^T\boxtimes \vect{Q}^T\right)_{IJKL}=Q_{IR}Q_{JS}Q_{KT}Q_{LU}\left(\mathbb{G}^{\mathcal{G}}\right)_{R S T U}\,.
\end{equation}
By making use of the isotropicisation theorem \cite{zheng1994}, strain energy density functions complying with the material symmetry condition can be defined \cite{Ebbing_2009} by re-expressing the anisotropic restriction \eqref{eqn:anisotropic_restriction} for every $\boldsymbol{Q}\in \blue{O(3)}$ as
\begin{equation}\label{eqn:material symmetry final}
e^{\mathcal{G}}(\boldsymbol{F}\boldsymbol{Q},\boldsymbol{Q}\boldsymbol{d}_0,\boldsymbol{Q}\boldsymbol{G}^{\mathcal{G}}\boldsymbol{Q}^T,\,\vect{Q}\boxtimes \vect{Q}:\vect{\mathbb{G}}^{\mathcal{G}}:\vect{Q}^T\boxtimes \vect{Q}^T)=e^{\mathcal{G}}(\vect{F},\vect{d}_0,\vect{Q}^{\mathcal{G}},\vect{\mathbb{G}}^{\mathcal{G}})\,.
\end{equation}

In addition to the objectivity and material symmetry conditions in  \cref{eqn:objectivity,eqn:anisotropic_restriction}, respectively, the internal energy must comply with the following conditions
\begin{equation}
	\left.e\left(\vect{\mathcal{U}}\right)\right\vert_{\vect{F}=\bI,\vect{d}_0=\vect{0}}=0\,,\qquad
		\left.\frac{\partial e\left(\vect{\mathcal{U}}\right)}{\partial\vect{F}}\right\vert_{\vect{F}=\bI,\vect{d}_0=\vect{0}}=\vect{0}\,,\qquad 
		\left.\frac{\partial e\left(\vect{\mathcal{U}}\right)}{\partial\vect{d}_0}\right\vert_{\vect{F}=\bI,\vect{d}_0=\vect{0}}=\vect{0}\,,		
\end{equation}
which essentially entail that the internal energy, the first Piola-Kirchhoff stress tensor, and the material electric field must vanish in the origin, i.e., for $\vect{F}=\bI$ and $\vect{d}_0=\vect{0}$.

\medskip

By formulating constitutive models in terms of invariants \cite{itskov2019}, they can fulfill both the material symmetry and objectivity condition by construction. To ensure polyconvexity, see~\cref{eqn:polyconvexity in second derivatives}, it is important to use invariants which are convex in $\vect{\mathcal{V}}$, see~\cref{eq:mvc}. 
Furthermore, the compressibility properties of a material must be considered, i.e., if it is compressible or nearly incompressible. In the compressible case, the growth condition $e(\vect{\mathcal{U}}) \rightarrow \infty \,\text{as}\, J \rightarrow 0^+$ must be fulfilled, while for nearly incompressible materials $J \approx 1$ must hold. In order to fulfill these conditions, an additional volumetric term $\cP_{\text{vol}}(J)$ is added to the internal energy. Overall, the internal energy, c.f.~\cref{eq:mvc}, can be expressed as
\begin{equation}\label{eq:e_full}
e(\vect{\mathcal{U}})=\cP(\vect{\mathcal{V}})=\widetilde{\cP}(\vect{\mathcal{I}})+\cP_{\text{vol}}(J)\,.
\end{equation}
where $\vect{\mathcal{I}}:\vect{\mathcal{V}}\rightarrow \mathbb{R}^n$ are invariants of the symmetry group under consideration and $\cP_{\text{vol}}(J)$ is chosen according to the compressible or nearly incompressible nature of the material under consideration.
The growth condition for compressible materials can be fulfilled by the term 
\begin{equation}\label{eqn:Wvol1}
\cP_{\text{vol}}(J)=\alpha\left(J+\frac{1}{J}-2\right)^2\,,
\end{equation}
while the condition $J\approx 1$ for nearly incompressible materials can be fulfilled by the term 
\begin{equation}\label{eqn:Wvol2}
\cP_{\text{vol}}(J)=\alpha\left(J-1\right)^2\,.
\end{equation}
In both cases, $\cP_{\text{vol}}(J)$ is convex w.r.t.\ $J$ and thus polyconvex, and $\alpha>0$ is a (possibly large) penalty parameter with units of stress. We now introduce invariants for all symmetry groups considered in the present work. For a detailled description of the symmetry groups see \textcite{zheng1994}.

\subsubsection{Isotropic electro-mechanics}

For nearly incompressible materials, invariants can either be formulated in the full deformation gradient $\bF$ or in its isochoric part $\bar{\bF}=J^{-1/3}\bF$. The latter is motivated from the purely mechanical, nearly incompressible, isotropic case. 
For this special case, which can describe a wide range of elastomeric materials, the two isochoric invariants 
\begin{equation}\label{inv_ii}
\begin{aligned}
      \bar{I}_1&= J^{-2/3}\|\bF\|^2\,,& \qquad 
      \bar{I}_2&=J^{-2}\|\bH\|^{3}\,,
    \end{aligned}
\end{equation}
can be constructed. 
Note that while $\bar{I}_1$ is polyconvex, the typically employed isochoric invariant $\bar{I}_2^*=J^{-4/3}\|\bH\|^{2}$ is not. Thus, we here use the slightly adapted invariant $\bar{I}_2=(\bar{I}_2^*)^{3/2}$, which is again polyconvex \cite{Hartmann2003}.
%
Setting $\vect{\mathcal{I}}=\{\bar{I}_1,\bar{I}_2 \}$ in \cref{eq:e_full}, the multiplicative decomposition of the deformation gradient and the additive split of the internal energy have two implications: first of all, isochoric deformations will only produce a deviatoric Cauchy stress, which represents a physically sensible behavior for isotropic bodies, c.f.~\textcite{sansour2008}. Furthermore, this correspondence between isochoric deformation and deviatoric stress is convenient for numerical applications, c.f.~\textcite[Section 10.1]{wriggers2008}. 

However, when anisotropy or electrical effects are considered, this correspondence is not a physically sensible assumption anymore \cite{sansour2008}. Beside that, even when constructing the invariants for this case with the isochoric part of the deformation gradient, this stress correspondence would not hold anymore.
Therefore, both the anisotropic and mixed invariants are constructed with the full deformation gradient also for the nearly incompressible case, while the isotropic, isochoric invariants are constructed with the isochoric part of the deformation gradient for the nearly incompressible case. Hence, the following two possibilities are considered for the internal energy in the additive decomposition in \cref{eq:e_full} 

\begin{equation}\label{pg_iso}
\widetilde{\cP}=\widetilde{\cP}(\vect{\mathcal{I}}):\bbR^5\rightarrow \mathbb{R}
\quad\text{with}\quad
\vect{\mathcal{I}}=\left\{\begin{aligned}
\vect{\mathcal{I}}^{\text{iso}}=&\{{I}_1,{I}_2,I_4,I_5,J\}\,,& \quad& (\text{Compressible case})\\
\bar{\vect{\mathcal{I}}}^{\text{iso}}=&\{\bar{I}_1,\bar{I}_2,I_4,I_5,J\}\,,& \quad& (\text{Nearly incompressible case})\\
\end{aligned}\right.,
\end{equation} 
with the further invariants
\begin{equation}\label{inv_iso}
\begin{aligned}
   {I_1}&= \|\vect{F}\|^2\,,&\qquad 
   {I_2}&= \|\vect{H}\|^{2}\,,&\qquad
    I_4 &= \|\bd_0\|^2\,,&\qquad 
    I_5 &= \|\bd\|^2\,.
    \end{aligned}
\end{equation}
Proof of convexity of all the mechanical invariants featuring in \eqref{inv_iso} can be found in \cite{Hartmann2003, Schroeder2008}. 
Note that, even for $J\approx 1$, it is important to include $J$, which becomes clear when considering standard representation theorems \cite{itskov2019}.
%
%
%

\begin{remark}\label{non_pc_iso}
	An additional invariant, which completes the integrity basis of isotropic electro-elasticity, is the following%
	\begin{equation}
	I_{6}=	\|\vect{Hd}_0\|^2\,.
	\end{equation}
	However, above invariant is not polyconvex, as the condition specified in \cref{eqn:polyconvexity in second derivatives} is not satisfied, namely
	\begin{equation}
	\begin{aligned}
	\delta\vect{\mathcal{V}}:
	\frac{\partial ^2 I_{6}}{\partial \vect{\mathcal{V}}\partial \vect{\mathcal{V}}}:\delta\vect{\mathcal{V}} = 2\|\delta\vect{H}\,\vect{d}_0+\vect{H}\,\delta\vect{d}_0\| ^2 + 4(\delta\vect{H}\,\delta\vect{d}_0)\cdot(\vect{H}\,\vect{d}_0)\,,
	\end{aligned}
	\end{equation} 	
	Taking 	$\delta\vect{H}=\vect{H}$ and $\delta\vect{d}_0=-\vect{d}_0$ yields
	\begin{equation}
	\delta\vect{\mathcal{V}}:
	\frac{\partial ^2 I_6}{\partial \vect{\mathcal{V}}\partial \vect{\mathcal{V}}}:\delta\vect{\mathcal{V}}= -4\|\bH\,\bd_0\|^2\,,
	\end{equation}
	which is clearly negative and hence, non-convex with respect to the elements of the  set $\vect{\mathcal{V}}$. Therefore,  $I_{6}$ is not included in the list of polyconvex invariants as specified in \cref{pg_iso}. 
	
	\blue{Notice however, that, as shown in \cite{gil2016}, it would be possible to include a polyconvex version of the invariant $I_6$. This can be achieved through the following ``polyconvexification'' technique, which entails the definition of the overall polyconvex invariant $I_6^{\ast}$, as }

\begin{equation}\blue{
	    I_6^{\ast}=\alpha^2 I_2^2 + \beta^2 I_4^2 + 2\alpha\beta I_6\,,\qquad \text{with}\quad\alpha,\beta >0\,.
}\end{equation}
\blue{$I_6^{\ast}$ satisfies that the following Hessian operator}
\begin{equation}\blue{
    \begin{bmatrix}
    \frac{\partial^2 I_6^{\ast}}{\partial \vect{H}\partial \vect{H}} & \frac{\partial^2 I_6^{\ast}}{\partial \vect{H}\partial \vect{d}_0}\\
\frac{\partial^2 I_6^{\ast}}{\partial \vect{d}_0\partial \vect{H}} & \frac{\partial^2 I_6^{\ast}}{\partial \vect{d}_0\partial \vect{d}_0}    
    \end{bmatrix}    
}\end{equation}
\blue{is positive definite $\forall \vect{H},\vect{d}_0$.}

\blue{As it can be observed in \cref{sec:ana_pot,sec:R1,sec:cub}, in this work we are able to obtain reasonably accurate approximations of complex constitutive behaviors using an incomplete set of polyconvex invariants for the formulation of the internal energy density $e(\vect{F},\vect{d}_0)$. Thus, our preference is to advocate for simplicity whenever permitted, restricting to the current incomplete but polyconvex set of invariants, rather than including the additional, sophisticated polyconvexified invariant.}

\blue{Only when the invariant-based formulation presented in \cref{sec:model} might not be sufficient to accurately reproduce data, alternative approaches might be needed. For instance, in \cite{klein2021}, an approach where the inputs of the neural network are the components of the deformation gradient tensor itself instead of invariants, is also presented. In this approach, however, the satisfaction of material frame indifference needs to be ``learned'' by the neural network (i.e., it cannot be satisfied a priori),  in contrast to the current invariant-based formulation. Comparing both invariant-based and deformation gradient-based approaches in electro-mechanics will be a subject of future work.}

\end{remark}

\subsubsection{Transversely isotropic material symmetry}\label{sec:TI}

A single second order structural tensor $\bG^{\text{ti}}$ suffices to characterise the material symmetry group $\mathcal{D}_{\infty h}$ of transverse isotropy \cite{Zheng1993, Zheng1993a, Ebbing2010}. Assuming a preferred (unitary) direction $\vect{n}$ \blue{in the reference configuration}, it is defined as
\begin{equation}\label{eq:ti_struct}
    \bG^{\text{ti}}=\boldsymbol{n}\otimes\boldsymbol{n}.
\end{equation}
In addition to the isotropic invariants $\vect{\mathcal{I}}^{\text{iso}}$ from \cref{inv_iso}, using $\bG^{\text{ti}}$, for this material symmetry group the  polyconvex invariants 
\begin{equation}
\begin{aligned}\label{eq:ti_invar}
\vect{\mathcal{J}}^{\text{ti}}&=\left\{J^{\text{ti}}_1,J^{\text{ti}}_2,J^{\text{ti}}_3\right\} \\
\text{with}\qquad
    J^{\text{ti}}_1 &= \big\|\bF\,\bG^{\text{ti}}\big\|^2\,,\qquad & J^{\text{ti}}_2 &= \big\|\bH\,\bG^{\text{ti}}\big\|^{2}\,,\qquad
    J^{\text{ti}}_3 &= \text{tr}\Big(\left(\vect{d}_0\otimes\vect{d}_0\right)\,\vect{G}^{\text{ti}}\Big)\,,     
    \end{aligned}
\end{equation}
can be defined, which comply with the objectivity and material symmetry conditions, c.f.~\cref{eqn:objectivity,eqn:material symmetry final}, respectively.
Therefore, the internal energy in the additive decomposition in \cref{eq:e_full} is defined for this particular material symmetry group as 
\begin{equation}\label{pg_ti}
\widetilde{\cP}=\widetilde{\cP}(\vect{\mathcal{I}}): \mathbb{R}^9\rightarrow \mathbb{R} \quad \text{with}\quad
\vect{\mathcal{I}} = 
\left\{\begin{aligned}
\big\{\vect{\mathcal{I}}^{\text{iso}},\vect{\mathcal{J}}^{\text{ti}}\big\}\,, \quad& (\text{Compressible case})\\
\big\{\bar{\vect{\mathcal{I}}}^{\text{iso}},\vect{\mathcal{J}}^{\text{ti}}\big\}\,, \quad& (\text{Nearly incompressible case})\\
\end{aligned}\right. \,.
\end{equation}

\subsubsection{Cubic material symmetry}

For the cubic material symmetry group $\mathcal{O}_{h}$, a single fourth order structural tensor $\bbG^{\text{cub}}$ suffices to characterise the underlying material symmetry attributes. It is defined as \cite{Schroeder2008}
\begin{equation}\label{eq:cub_struct}
    \bbG^{\text{cub}}=\sum_{i=1}^3 \ba_i\otimes\ba_i\otimes\ba_i\otimes\ba_i\,.
\end{equation}
Therefore, in addition to the isotropic invariants $\vect{\mathcal{I}}^{\text{iso}}$ from \cref{inv_iso}, using $\bbG^{\text{cub}}$, for the cubic material symmetry group the polyconvex invariants 
\begin{equation}\label{eq:cub_invar}
	\begin{aligned}
	\vect{\mathcal{J}}^{\text{cub}} &= 
	\left\{J^{\text{cub}}_1,J^{\text{cub}}_2,J^{\text{cub}}_3\right\} \\
	\text{with}\qquad
    J^{\text{cub}}_1 &= \bF^T\bF\colon\bbG^{\text{cub}}\colon\bF^T\bF\,, \qquad& J^{\text{cub}}_2 &= \bH^T\bH\colon\bbG^{\text{cub}}\colon\bH^T\bH\,, \\
    J^{\text{cub}}_3 &=\bd_0\otimes\bd_0\colon\bbG^{\text{cub}}\colon\bI\,,
	\end{aligned}
\end{equation}
can be defined \cite[Section 3.7]{Ebbing2010}, which comply with the objectivity and material symmetry conditions, c.f.~\cref{eqn:objectivity,eqn:material symmetry final}, respectively.
Therefore, the internal energy in the additive decomposition in \cref{eq:e_full} is defined for this particular material symmetry group as 
\begin{equation}\label{pg_cub}
\widetilde{\cP}=\widetilde{\cP}(\vect{\mathcal{I}}): \mathbb{R}^8\rightarrow \mathbb{R} \quad \text{with}\quad
\vect{\mathcal{I}} = 
\left\{\begin{aligned}
\big\{\vect{\mathcal{I}}^{\text{iso}},\vect{\mathcal{J}}^{\text{cub}}\big\}\,, \quad& (\text{Compressible case})\\
\big\{\bar{\vect{\mathcal{I}}}^{\text{iso}},\vect{\mathcal{J}}^{\text{cub}}\big\}\,, \quad& (\text{Nearly incompressible case})\\
\end{aligned}\right. .
\end{equation}

\begin{remark}
Using the fourth order cubic structural tensor from \cref{eq:cub_struct}, an additional invariant can be constructed as
\begin{equation}
J_{4}^{\text{cub}}=	\vect{d}_0\otimes \vect{d}_0:\vect{\mathbb{G}}^{\text{cub}}:\vect{F}^T\vect{F}\,.
\end{equation}
However, this invariant is not polyconvex, as \cref{eqn:polyconvexity in second derivatives} is not satisfied, namely
\begin{equation}
\begin{aligned}
\delta\vect{\mathcal{V}}:
\frac{\partial ^2 J_{4}^{\text{cub}}}{\partial \vect{\mathcal{V}}\partial \vect{\mathcal{V}}}:\delta\vect{\mathcal{V}} =&2\|\vect{u}+\vect{v}\|^2 + 4\vect{u}\cdot\vect{v}
\end{aligned}
\end{equation} 	
with $\vect{u}=\sum_{i=1}^3\left(\delta\vect{d}_0\cdot \vect{a}_i\right)\vect{Fa}_i$ and $\vect{v}=\sum_{i=1}^3\left(\vect{d}_0\cdot \vect{a}_i\right)\delta\vect{Fa}_i$. 
Taking $\vect{v}=-\vect{u}$, i.e., e.g.\ $\delta\bd_0=-\bd_0$ and $\delta\bF=\bF$, yields
\begin{equation}
\delta\vect{\mathcal{U}}:
\frac{\partial ^2 J_{4}^{\text{cub}}}{\partial \vect{\mathcal{U}}\partial \vect{\mathcal{U}}}:\delta\vect{\mathcal{U}}=- 4\|\vect{u}\|^2\,,
\end{equation}
which is clearly negative and hence, non convex with respect to the elements of the  set $\vect{\mathcal{V}}$. Therefore, $J_{4}^{\text{cub}}$ is not included in the list of polyconvex invariants as specified in \cref{eq:cub_invar}.

\end{remark}

\section{Physics-augmented neural network constitutive model} 
\label{sec:model}

In the previous section, we discussed the form of the internal energy for compressible and nearly incompressible material behavior, c.f.~\cref{eq:e_full}, both formulated in terms of electro-mechanically coupled invariants. 
In both cases, the internal energy $e(\boldsymbol{\cU})$ comprises the polyconvex term $\widetilde{\cP}(\boldsymbol{\cV})$, which models the behavior of the specific material, and an additional analytical term $\cP_{\text{vol}}(J)$, which ensures the desired volumetric behavior. 
The volumetric term can easily be analytically defined as in \cref{eqn:Wvol1,eqn:Wvol2} for both compressible and nearly incompressible materials, and thus requires no further treatment. 
In its most general form, the term $\widetilde{\cP}$ is a convex function of the quantities $\vect{\mathcal{V}}$ defined in the polyconvexity condition, c.f.~\cref{eq:mvc}. Instead of formulating $\widetilde{\cP}$ directly in $\vect{\mathcal{V}}$, it is commonly formulated in invariants of $\vect{\mathcal{V}}$, c.f.~\cref{sec:objectivity and material symmetry}. In this way, the objectivity and material symmetry conditions are fulfilled by construction. However, it also leads to a loss of information, as it is not possible to formulate a complete integrity basis in terms of polyconvex invariants for electro-elasticity, c.f.~\cref{non_pc_iso}.

\medskip

Instead of choosing an analytical formulation for the term $\widetilde{\cP}$, we aim to exploit the excellent approximation properties of neural networks \cite{Hornik1991} and use input convex feed-forward neural networks (ICNNs) \cite{Amos2016}.
In a nutshell, feed-forward neural networks (FFNNs) can be seen as a composition of several vector-valued functions, where the components of the vectors are referred to as nodes or neurons, and the function in each neuron is referred to as activation function. 
More explicitly, the FFNN with the vector-valued input $\bX=:\bA_{0}\in\mathbb{R}^{n^{[0]}}$, output $\bY=:\bA_{H+1}\in\mathbb{R}^{n^{[H+1]}}$, and $H$ hidden layers is given by
\begin{equation}
\begin{aligned}\label{eq:FFNN}
\bA_h=\sigma^{[h]}\left({{\bW}}^{[h]}\bA_{h-1}+\bb^{[h]}\right)\in&\;\mathbb{R}^{n^{[h]}},\quad h=1,\dotsc,H+1\,.
\end{aligned}
\end{equation}
Here, ${{\bW}}^{[h]}\in\mathbb{R}^{n^{[h]}\times n^{[h-1]}}$ are the weights and $\bb^{[h]}\in\mathbb{R}^{n^{[h]}}$ the biases in each layer. Together, they form the set of parameters $\bp$ of the neural network, which is optimized in the calibration process to fit a given dataset. In the 
layers $\bA_h$, the scalar activation functions $\sigma^{[h]}$ are applied in a component-wise manner.
By using a scalar-valued output, i.e.\ $n^{[H+1]}=1$, and  the special choice of activation functions as 
\begin{itemize}
    \item neuron-wise convex in the first layer \\
    (here: the \emph{Softplus} function $\sigma^{[1]}(x)=s(x)=\log (1+ \exp (x))$),
    \item neuron-wise convex and non-decreasing in every subsequent layer\\ 
    (here: \emph{Softplus} $\sigma^{[h]}(x)=s(x)$ with non-negative weights ${{\bW}}^{[h]}_{ij}\geq 0$ for $h=2,\ldots,H$),
    \item and scalar-valued, convex and non-decreasing in the output layer\\
    (here: $n^{[H+1]}=1$ and a linear activation function $\sigma^{[H+1]}(x)=x$ with  ${{\bW}}^{[H+1]}_{ij}\geq 0$),
\end{itemize}
the FFNN is convex by construction and considered as an ICNN, see \textcite[Appendix A]{klein2021} for a proof. 
In the following, we use the short notation $\cS\cP(\bX;\,n^{[1]},\ldots,n^{[H]})$ to describe FFNN architectures and $\cS\cP^+(\bX;\,n^{[1]},\ldots,n^{[H]})$ to describe ICNN architectures, with input $\bX$ and $H$ hidden layers.
For both FFNNs and ICNNs we apply (restricted) \emph{Softplus} activation functions in all hidden layers, which are infinitely continuously differentiable, i.e.\ $\cS\cP,\,\cS\cP^+\in\cC^\infty(\mathbb{R}^{n^{[0]}},\mathbb{R}^{n^{[H+1]}})$. 
A more extensive introduction to neural networks can be found in e.g. \cite{kollmannsberger2021, aggarwal2018}. 

\medskip

Now, following \textcite{klein2021}, we use a set of invariants as described in \cref{sec:basics} as input quantities of an ICNN, which represents the term $\widetilde{\cP}:=\cS\cP^+$ as part of the internal energy $e$.   The resulting potential
\begin{equation}\label{ML_model}
    e\left(\bF,\,\bd_0\right)
    =\cP(\vect{\mathcal{V}})
    =\cS\cP^+(\boldsymbol{\cI};\,n^{[1]},\ldots,n^{[H]})+\cP_{\text{vol}}(J)\,
\end{equation}
fulfills all constitutive requirements specified in \cref{sec:basics}. Here, $\boldsymbol{\cI}$ denotes a vector of invariants, which corresponds to the arguments of $\widetilde{\cP}$ as specified in \cref{sec:objectivity and material symmetry}.  
Note that for this special application also the first layer must be neuron-wise non-decreasing, as the invariants are non-linear functions of the quantities we want to be convex in, c.f.~\textcite[Remark A.10]{klein2021}.
Then, in particular the polyconvexity condition is fulfilled by using sets of convex invariants, and preserving the convexity by using input convex neural networks.

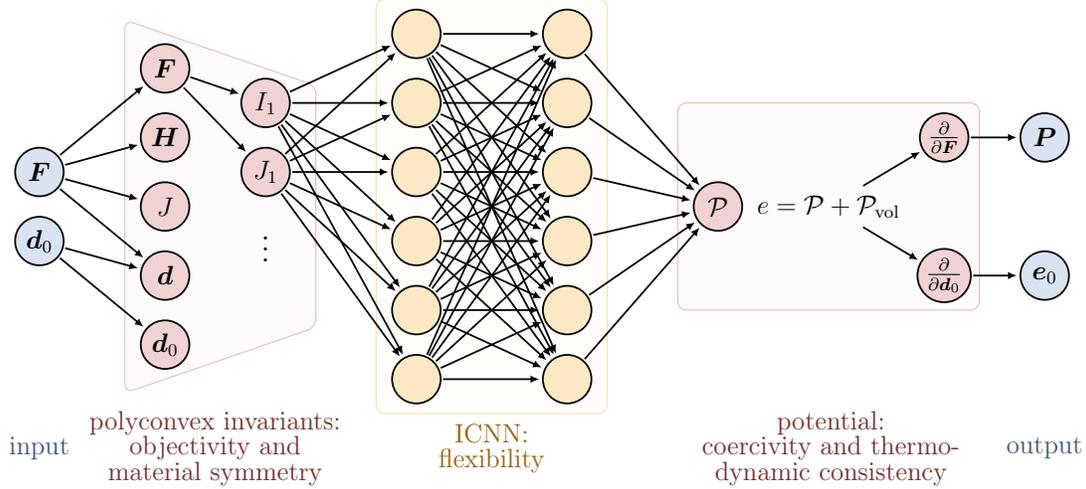
\begin{figure}[t!]
\centering
\resizebox{0.9\textwidth}{!}{
\tikzsetnextfilename{ANN_model}

\begin{tikzpicture}[x=1.6cm,y=1.1cm]
  \large
  \def\NC{6} 
  \def\nstyle{int(\lay<\Nnodlen?(\lay<\NC?min(2,\lay):3):4)} 
  \tikzset{ 
    node 1/.style={node in_out},
    node 2/.style={node inv_pot},
    node 3/.style={node icnn},
  }
  
    \draw[color33!40,fill=color33,fill opacity=0.02,rounded corners=4]
    (1.6,-2.7) --++ (0,5.4) --++ (1.9,-0.95) --++ (0,-3.5) -- cycle;
          \draw[color22!40,fill=color22,fill opacity=0.02,rounded corners=4]
    (4.1,-3) rectangle++ (2.3,6.0);
              \draw[color33!40,fill=color33,fill opacity=0.02,rounded corners=4]
    (7.1,-1.5) rectangle++ (3,3);
  
 \node[node 1, outer sep=0.6] (1-1) at (0.75,0.5) {$\boldsymbol{F}$};
 \node[node 1, outer sep=0.6] (1-2) at (0.75,-0.5) {$\boldsymbol{d}_0$};
        
 \node[node 2, outer sep=0.6] (2-1) at (2,2) {$\boldsymbol{F}$};
 \node[node 2, outer sep=0.6] (2-2) at (2,1) {$\boldsymbol{H}$};
 \node[node 2, outer sep=0.6] (2-3) at (2,0) {$J$};
 \node[node 2, outer sep=0.6] (2-4) at (2,-1) {$\boldsymbol{d}$};
 \node[node 2, outer sep=0.6] (2-5) at (2,-2) {$\boldsymbol{d}_0$};
        
    \draw[connect arrow] (1-1) -- (2-1);
  \draw[connect arrow] (1-1) -- (2-2);
  \draw[connect arrow] (1-1) -- (2-3);
  \draw[connect arrow] (1-1) -- (2-4);
  \draw[connect arrow] (1-2) -- (2-4);
 \draw[connect arrow] (1-2) -- (2-5);

 \node[node 2, outer sep=0.6] (3-1) at (3,1.5) {$I_1$};
 \node[node 2, outer sep=0.6] (3-2) at (3,0.5) {$J_1$};
 \node[scale=1.2] (3-5) at (3,-0.5) {$\vdots$};

 \draw[connect arrow] (2-1) -- (3-1);
 \draw[connect arrow] (2-1) -- (3-2);

\def\N{6}
    \foreach \i [evaluate={\y=\N/2-\i+0.5;}] in {1,...,\N}{ 
     \node[node 3,outer sep=0.6] (4-\i) at (4.5,\y) {};
    
    \foreach \j in {1,2}{
    \draw[connect arrow]  (3-\j) -- (4-\i);
    }
    
      }
     \foreach \i [evaluate={\y=\N/2-\i+0.5;}] in {1,...,\N}{ 
     \node[node 3,outer sep=0.6] (5-\i) at (6,\y) {};
    \foreach \j in {1,...,\N}{
    \draw[connect arrow]  (4-\j) -- (5-\i);
    }
      }
      

    \node[node 2, outer sep=0.6] (6-1) at (7.5,0) {$\mathcal{P}$};
       \foreach \j in {1,...,\N}{
    \draw[connect arrow]  (5-\j) -- (6-1);
}

  \node[] (7-1) at (8.6, 0)
  {$e=\mathcal{P}+\mathcal{P}_{\text{vol}}$};
  
   \node[node 2, outer sep=0.6] (8-1) at (9.75,1) {$\frac{\partial}{\partial \boldsymbol{F}}$};
    \draw[connect arrow]  (7-1) -- (8-1);
    
       \node[node 2, outer sep=0.6] (8-2) at (9.75,-1) {$\frac{\partial}{\partial \boldsymbol{d}_0}$};
    \draw[connect arrow]  (7-1) -- (8-2);
    
    \node[node 1, outer sep=0.6] (9-1) at (10.75,1) {${ \boldsymbol{P}}$};
    \draw[connect arrow]  (8-1) -- (9-1);
    
       \node[node 1, outer sep=0.6] (9-2) at (10.75,-1) {${ \boldsymbol{e}_0}$};
    \draw[connect arrow]  (8-2) -- (9-2);

      \node[align=center,color21!60!black] at (0.75,-3.5) {input};
      \node[align=center,color21!60!black] at (10.75,-3.5) {output};
      \node[align=center,color33!60!black] at (2.5,-3.5) {polyconvex invariants:\\[-0.2em]objectivity and\\[-0.2em]material symmetry};
      \node[align=center,color22!60!black] at (5.25, -3.5) {ICNN:\\[-0.2em]flexibility};
      \node[align=center,color33!60!black] at (8.625,-3.5) {potential:\\[-0.2em]coercivity and thermo-\\[-0.2em]dynamic consistency};
 
  \end{tikzpicture}
}
\caption{Illustration of the physics-augmented, ANN based constitutive model.  Note that the polyconvexity of $\cP$ must be preserved in every layer of the model. Figure generated with \url{https://tikz.net/neural_networks/}.}
\label{fig:ANN_model}
\end{figure}

Then, as gradients of the energy potential $e$, see \cref{eq:P_e}, the stress and electrical field can be computed via automatic differentiation, which is widely available in machine learning libraries. By construction, the conditions $\bP(\uuI,\,\bnull)=\bnull$ and $\be_0(\uuI,\,\bnull)=\bnull$ are not fulfilled, and including these conditions in the model formulation would not be straightforward. However, as the data with which the model is calibrated fulfills these conditions, the model will approximate them after the calibration process.
In \cref{fig:ANN_model}, the overall structure and flow of this physics-augmented ANN model is illustrated.

\medskip

In \cref{ML_model}, the term $\cP_{\text{vol}}$ depends on the volumetric behavior of the material and is prescribed as an analytical term, while $\cS\cP^+$ is calibrated to data of a specific material. Throughout this work, we use calibration datasets of the form
\begin{equation}\label{eq:dataset}
    D_i=\left\{\left(\bF^{i,\,1},\,\hat{\bd}_0^{i,\,1};\,\hat{\bP}^{i,\,1},\,\hat{\be}_{0}^{i,\,1}\right),\,\left(\bF^{i,\,2},\,\hat{\bd}_0^{i,\,2};\,\hat{\bP}^{i,\,2},\,\hat{\be}_{0}^{i,\,2}\right),\,\dotsc\right\}\,,\qquad D=\bigcup_i D_i
\end{equation}
where $D_i$ are different loading paths and $D$ is the overall calibration dataset. Note that throughout this work all quantities $\ast$ are made dimensionless, which is indicated as $\hat{\ast}$, see \cref{sec:scaling} for further details.
To calibrate the model $\square$ on the dataset $D$ and find its parameters $\bp$, we choose to minimize the loss function defined in terms of the mean squared error
\begin{equation}\label{MSE}
\begin{aligned}
\text{MSE}^\square(\bp)=\sum_i \frac{1}{\#(D_i)}\frac{1}{w_i} \sum_{j} \bigg[
\left\|\hat{\bP}^{i,j}-\hat{\bP}^\square\left(\bF^{i,j},\,\hat{\bd}_0^{i,j};\,\bp\right)\right\|^2 
+  \left\|\hat{\be}_0^{i,j}-\hat{\be}_0^\square\left(\bF^{i,j},\,\hat{\bd}_0^{i,j};\,\bp\right)\right\|^2 \bigg]\,,
\end{aligned}
\end{equation}
where $\#(D_i)$ denotes the number of data points in $D_i$ and $w_i$ is the norm
\begin{equation}
    w_i=\frac{1}{\#(D_i)}\sum_j  \left\|\hat{\bP}^{i,\,j}\right\|\,.
\end{equation}
Note that the weights $w_i$ are introduced in order to balance the importance of different load paths within the MSE, see \cite{fernandez2021}.
In contrast to the dataset introduced in \cref{eq:dataset}, the homogenization of composites and  metamaterials not only allows to compute $\bP$ and $\be_0$, but also the internal energy $e$. Basically, this information could also be included in the loss function. However, previous examinations showed that this hardly increases the prediction quality of the model, c.f.~\cite{Fernandez2020,klein2021}. Also, this would lead to a loss of generality, as experimental investigations only provide the data introduced in \cref{eq:dataset}.
Therefore, the internal energy is calibrated only through its gradients, which can be considered as a variant of the Sobolev training strategy \cite{vlassis2020a,vlassis_sobolev_2021}. 

\section{Application to a transversely isotropic potential} 
\label{sec:ana_pot}

As a first proof of concept, we apply the ANN based constitutive model proposed in \cref{sec:model} to fitting an analytical electro-elastic potential with transversely isotropic material symmetry.
Furthermore, we investigate how the consideration of constitutive requirements in the model formulation influences its properties, especially generalization and flexibility.

\subsection{Data generation}

The transversely isotropic potential considered in this section includes the invariants described in \cref{sec:TI} for this particular material symmetry group. Specifically, the following invariant-based additive decomposition is considered
\begin{equation}\label{p_ti}
\begin{aligned}
e^{\text{ti}}(\vect{\mathcal{U}})=\mathcal{P}^{\text{ti}}(\vect{\mathcal{V}})=\widetilde{\mathcal{P}}(\vect{\mathcal{I}}^{\text{iso}}) + \widetilde{\mathcal{P}}(\vect{\mathcal{J}}^{\text{ti}})+\cP_{\text{vol}}(J)\,,
\end{aligned}
\end{equation}
with 
\begin{equation}\label{eqn:ti model}
\begin{aligned}
\widetilde{\mathcal{P}}(\vect{\mathcal{I}}^{\text{iso}})&=\frac{\mu_1}{2}\bar{I}_1 + \frac{\mu_2}{2}\bar{I}_2^* - \mu_3\log J + \frac{1}{2\varepsilon_1}\frac{I_5}{J}\,,\\   
\widetilde{\mathcal{P}}(\vect{\mathcal{J}}^{\text{ti}})&=
\frac{\mu_3}{2}\left(\frac{\left(J^{\text{ti}}_1\right)^{a_1}}{a_1}+\frac{\left(J^{\text{ti}}_2\right)^{a_2}}{a_2}\right)   +\frac{1}{2\varepsilon_2}J_3^{\text{ti}}\,,\\
\cP_{\text{vol}}(J)&=\frac{\lambda}{2}\left(J-1\right)^2\,,
\end{aligned}
\end{equation}
where $\mu_1,\mu_2,\mu_3,\lambda$ represent mechanical material parameters with units of stress (Pa) and $\varepsilon_1,\varepsilon_2$ electric permittivity-type material parameters with units of $\text{C}\text{V}^{-1}\text{m}^{-1}$.
In particular, here the following specific values are used
\begin{equation}
\begin{aligned}
\mu_2&=\mu_1\,,&\qquad \mu_3&=3\mu_1\,,&\qquad a_1&=2\,&   \quad a_2&=2\,,\\ \lambda&=10^3\mu_1\,,&\qquad
 \varepsilon_2&=2\varepsilon_1\,,&\qquad \vect{n}&=\begin{pmatrix}
0&0&1
\end{pmatrix}^T\,.
\end{aligned}
\end{equation}

Note that usually the difference between the values of the mechanical parameters $\mu_1,\mu_2,\mu_3,\lambda$ and the electro-mechanical parameters $\varepsilon_1,\varepsilon_2$ can be of several orders of magnitude. This scaling difference can entail difficulties for the construction of a suitable neural network capturing both underlying physics, namely mechanical and electro-mechanical (in fact, it is challenging for any numerical optimization approach for parameter fitting). With the aim of circumventing this potential drawback, we make use of the energy scaling procedure described in~\cref{sec:scaling}.

While real-world investigations of material behavior require (cumbersome) experiments, this analytical potential can either be directly evaluated for a given deformation gradient $\bF\in\text{GL}^+(3)$ and electric displacement field $\bd_0\in\bbR^3$, or be (numerically) equilibrated for stress-strain states such as uniaxial tension, biaxial tension, pure shear, etc. Here, we choose the former approach to generate data for the calibration of machine learning based constitutive models.
In order to examine some special model properties, the global behavior of \cref{p_ti} is sampled for a wide range of physically admissible deformations, which are described in \cref{sec:F_sampling}. Altogether, this results in a fairly large dataset with 1.5 million points. As we will show, such a large dataset may not be necessary to calibrate (physics-augmented) machine learning models, however, it will be useful to examine some specific model properties.
Finally, for this specific example, based on the analytical potential in \cref{p_ti,eqn:ti model}, for each deformation gradient tensor $\vect{F}$ and electric displacement field $\vect{d}_0$, we simply evaluate the derivatives of the analytical potential $e^{\text{ti}}(\vect{\mathcal{U}})$  according to \cref{eq:P_e} in order to obtain their respective work conjugates $\vect{P}$ and $\vect{e}_0$, which are required for evaluating the mean squared error as defined in \cref{MSE}.

\subsection{Model calibration}
\label{sec:calib}

\paragraph{Model A.}
This is the \emph{physics-augmented}, invariant-based model as proposed in \cref{sec:model}, which fulfills all of the constitutive requirements as described in \cref{sec:basics}. As we assume transversely isotropic, nearly incompressible material behavior, we use the invariant vector
\begin{equation} \label{eqn:TI_model_A_invar}
    \boldsymbol{\cI}=\left(\Bar{I}_1,\,\Bar{I}_2,\,J,\,-J,\,\hat{I}_5,\,J_1^*,\,J_2^*\right)\in\bbR^7
\end{equation}
with the invariants as described in \cref{sec:objectivity and material symmetry}, and the scaling as described in \cref{sec:scaling}.
The transversely isotropic invariants are slightly adapted according to $J_1^*=J_1+0.35\,I_1$ and $J_2^*=J_2+0.35\,I_2$, which can also be obtained by replacing the structural tensor from \cref{eq:ti_struct} with $\bG^{\text{ti}}=\text{diag}\left(0.35,\,0.35,\,1\right)$. 
Furthermore, here all invariants are polyconvex, whereas in \cref{eqn:ti model} the non-polyconvex isochoric invariant $\bar{I}^*_2$ is used, c.f.~\cref{inv_ii}.
As the invariants are non-linear in the quantities of the polyconvexity condition, also the first layer of the ICNN must have non-decreasing activation functions and thus non-negative weights for the \emph{Softplus} function, c.f.~\textcite[Remark A.10]{klein2021}. The only exception is $J$, which is the only invariant quantity included in the polyconvexity condition, c.f.~\cref{eq:mvc}. Thus, the weights in the first layer which act on the invariant $J$ may be positive or negative. This is pragmatically implemented by adding the additional invariant $-J$.

With the choice of invariants as specified in \cref{eqn:TI_model_A_invar}, the approximation of the analytical potential from  \cref{p_ti} does \emph{not} reduce to approximating a (mostly) linear function, as the input quantities of the ICNN differ from the ones used in \cref{p_ti}.  In particular, the invariant $J_3$ is not used and all other invariants differ slightly in their concrete form. When choosing all the invariants as they appear in~\cref{p_ti}, the ICNN would model a linear function in most of the invariants, which would be trivial and thus not be a good benchmark case. In fact, for this case, even significantly smaller calibration datasets than the ones used throughout this section would yield a perfect approximation of~\cref{p_ti}.

For this model, we choose two network architectures: model ${\text{A}^1}$ consisting of only one hidden layer with 8 nodes ($H=1, n^{[1]}=8$) and model ${\text{A}^2}$ consisting of two hidden layers with 16 nodes in each layer ($H=2, n^{[h]}=16$), for which we use the short notations ${\text{A}^1}=\cS\cP^+( \boldsymbol{\cI};\,8)$ and ${\text{A}^2}=\cS\cP^+( \boldsymbol{\cI};\,16,16)$, respectively. The former has a total amount of 73 free parameters, the latter has 417 free parameters.

\paragraph{Model B.}
In addition to the physics-augmented model from \cref{sec:model}, we introduce a second, \emph{uninformed} machine learning model, which includes considerably less mathematical structure, i.e., it incorporates fewer physical requirements. For this purpose, we use a FFNN without any restrictions on the activation functions and weights, which is thus not convex. Here, we also use the \emph{Softplus} activation functions, but with unrestricted weights.
Now, we directly map $\bC=\bF^T\bF$ and $\bd_0$ to $\bP$ and $\be_0$, i.e.
\begin{equation}
    \left(\bP,\,\be_0\right)=\cS\cP(\bC,\,\bd_0;\,n^{[1]},\ldots,n^{[H]})\,.
\end{equation}
Note that in contrast to the model proposed in \cref{sec:model}, the neural network in the above equation is \emph{not} used to model an internal energy, c.f.~\cref{ML_model}, but directly outputs the mechanical stress field and the electrical field.
Since formulated in $\bC$, this model is objective, but beside that it does not fulfill any of the other constitutive requirements introduced in \cref{sec:basics}. In fact, it is not even hyperelastic, i.e., it is not necessarily thermodynamically consistent. 

For this model, we also choose two network architectures: model ${\text{B}^1}$ consisting of two hidden layers with 16 nodes in each layer ($H=2, n^{[h]}=16$) and model ${\text{B}^2}$ consisting of four layers with 64 nodes in each layer ($H=4, n^{[h]}=64$), for which we use the short notations ${\text{B}^1}=\cS\cP(\bC,\bd_0;16,16)$ and $\text{B}^2=\cS\cP(\bC,\bd_0;64,64,64,64)$, respectively. 
The former has a total amount of 684 free parameters, the latter has $14,092$ free parameters. In both cases, the output layer consists of a 12-dimensional layer with linear activation functions.

\paragraph{Calibration details.}
As described above, the overall dataset $D$ comprises of 1.5 million data points. 
It is divided into subsets $D_i$, where each one consists of a fixed deviatoric direction, c.f.~\cref{sec:F_sampling}. 
Overall, this results in $300,000$ sub-datasets with 50 data points each.
In this way, we are able to weight the MSE of the different loading paths according to \cref{MSE}, which is important as the magnitude of the paths greatly differs.
The calibration datasets are composed of different amounts of sub-datasets $n_{\text{data}}= 2^n,\,n= 4,5,\ldots,11$, resulting in calibration dataset sizes between 400 and $102,400$ points.
For each calibration dataset size $n_{\text{data}}$, each model calibration was initialized 10 times. For each initialization, different random sub-datasets were used for the calibration dataset, by ordering the sub-datasets from $1$ to $300,000$ and generating random sets of integers to choose out of them. Also, for each initialization, the model parameters, i.e., the weights and biases, were initialized randomly with the default settings of TensorFlow. For the following investigations the MSE after the calibration process is evaluated  on the overall dataset for all 1.5 million data points, meaning that the test dataset equals the overall dataset.

\medskip

The models are implemented in TensorFlow 2.5.0, using Python 3.9.9. The optimization is carried out using the ADAM algorithm with default settings, except for the learning rate which is set to $0.005$. The calibration is carried out for $5,000$ iterations (epochs), which showed to be sufficient for converged results.
The full batch of training data is used with a batch size of 400.
For the reproducibility of the results, in addition to the data, we provide the calibrated parameters of the models on the GitHub repository \url{https://github.com/CPShub/sim-data}. For general details on the reproducibility of the models, see also \textcite[Sect.~4.2]{klein2021}.

\subsection{Model evaluation}

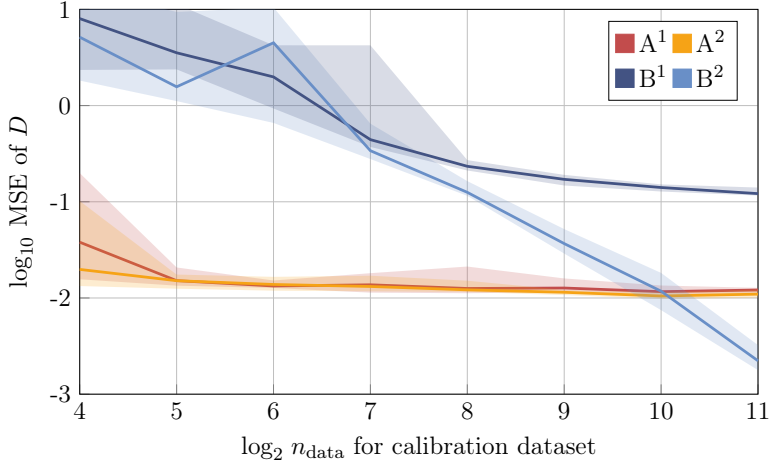
\begin{figure}[t!] 
\centering
\resizebox{0.65\textwidth}{!}{
    \tikzsetnextfilename{TI_MSE}

\begin{tikzpicture}
\pgfplotsset{
  set layers,
  mark layer=axis tick labels
}
\begin{groupplot}[
	group style = {group size = 1 by 1, vertical sep = 0.02*\textwidth,
								horizontal sep = 0.02*\textwidth},
	cycle list name=mycolorlist_MSE,
	xtick pos = left,
	ytick pos = left,
	grid = major
	]
  
  \nextgroupplot[align=center,legend pos = north east,legend columns=2,ylabel = {$\text{log}_{10}$ MSE of $D$},
width=0.72*\textwidth, height=0.45*\textwidth, 
xtick={4,5,6,7,8,9,10,11},xticklabels={4,5,6,7,8,9,10,11},
ytick={-3,-2,-1,0,1},
yticklabels={-3,-2,-1,0,1},
ymin = -3, ymax = 1.0,
xmin = 4, xmax = 11,
xlabel={$\text{log}_{2} \;n_{\text{data}}$ for calibration dataset},
legend cell align = {left}
] 

\foreach \num in {1, 2,3,4}{
 \addplot coordinates { (-2,-2) (-3,-3) };
 };

\addlegendentry{${\text{A}^1}$}
\addlegendentry{${\text{A}^2}$}
\addlegendentry{${\text{B}^1}$}
\addlegendentry{${\text{B}^2}$}

\pgfplotstableread{plot_data/TI/MSE_M4_8_1.txt}\data
\addplot [mark=none,fill=color33, draw=none, opacity=0.2] table {\data};
\pgfplotstableread{plot_data/TI/MSE_M4_8_1_median.txt}\data
\addplot [color33, fill=none,mark=none, very thick] table {\data};

\pgfplotstableread{plot_data/TI/MSE_M4_16_2.txt}\data
\addplot [mark=none,fill=color22, draw=none, opacity=0.2] table {\data};
\pgfplotstableread{plot_data/TI/MSE_M4_16_2_median.txt}\data
\addplot [color22, fill=none,mark=none, very thick] table {\data};

\pgfplotstableread{plot_data/TI/MSE_M1_64_4.txt}\data
\addplot [mark=none,fill=color21, draw=none, opacity=0.2] table {\data};

\pgfplotstableread{plot_data/TI/MSE_M1_16_2.txt}\data
\addplot [mark=none,fill=color12, draw=none, opacity=0.2] table {\data};

\pgfplotstableread{plot_data/TI/MSE_M1_16_2_median.txt}\data
\addplot [color12, fill=none,mark=none, very thick] table {\data};

\pgfplotstableread{plot_data/TI/MSE_M1_64_4_median.txt}\data
\addplot [color21, fill=none,mark=none, very thick] table {\data};

\end{groupplot}
\end{tikzpicture}
}
\caption{Transversely isotropic potential, evaluation of the models. ${\text{A}^1}=\cS\cP^+( \boldsymbol{\cI};\,8)$ and ${\text{A}^2}=\cS\cP^+( \boldsymbol{\cI};\,16,16)$ denote the physics-augmented models, while ${\text{B}^1}=\cS\cP(\bC,\bd_0;16,16)$ and ${\text{B}^2}=\cS\cP(\bC,\bd_0;64,64,64,64)$ denote the uninformed models with less mathematical structure.
For the different models, the MSE of the overall dataset $D$ is evaluated depending on the number $n_{\text{data}}$ of sub-datasets  $D_i$ used for calibration. The spreads of all 10  initializations are shown by shaded areas, whereas the continuous lines indicate the medians. 
}
\label{fig:MSE_TI}
\end{figure}

In \cref{fig:MSE_TI}, the results of all different model calibrations are shown, with the logarithm of the number of sub-datasets $n_{\text{data}}$ used for calibration on the horizontal axis and the logarithmic MSE of the overall dataset on the vertical axis. The two  invariant-based models A are visualized in red colors, while the uninformed  models B are displayed in blue colors.
As the calibration was initialized multiple times for each $n_{\text{data}}$, the corresponding MSEs form an area containing all results, with the median visualized with a continuous line.

For the model B, the small network architecture $\text{B}^1$  
does not provide enough flexibility, consequently the MSE stagnates at a relatively high value around $10^{-1}$. For the bigger network structure $\text{B}^2$
, however, the MSE decreases consistently with increasing calibration dataset size, and could probably decrease even further with more data.
In contrast to that, the invariant-based approaches can cope with far smaller network architectures and calibration datasets. In fact, increasing the network size from model $\text{A}^1$ to $\text{A}^2$ 
hardly improves the model quality. However, the MSE converges quite fast to around $10^{-2}$ and does not further decrease for an increasing calibration dataset or model architecture. 
For both models, the spread of the MSE for multiple initializations is decreasing for increasing calibration datasets. This was also observed in \textcite{klein2021}, where multiple initializations of the same machine learning model architecture resulted in quite similar MSEs of the test dataset.

\medskip

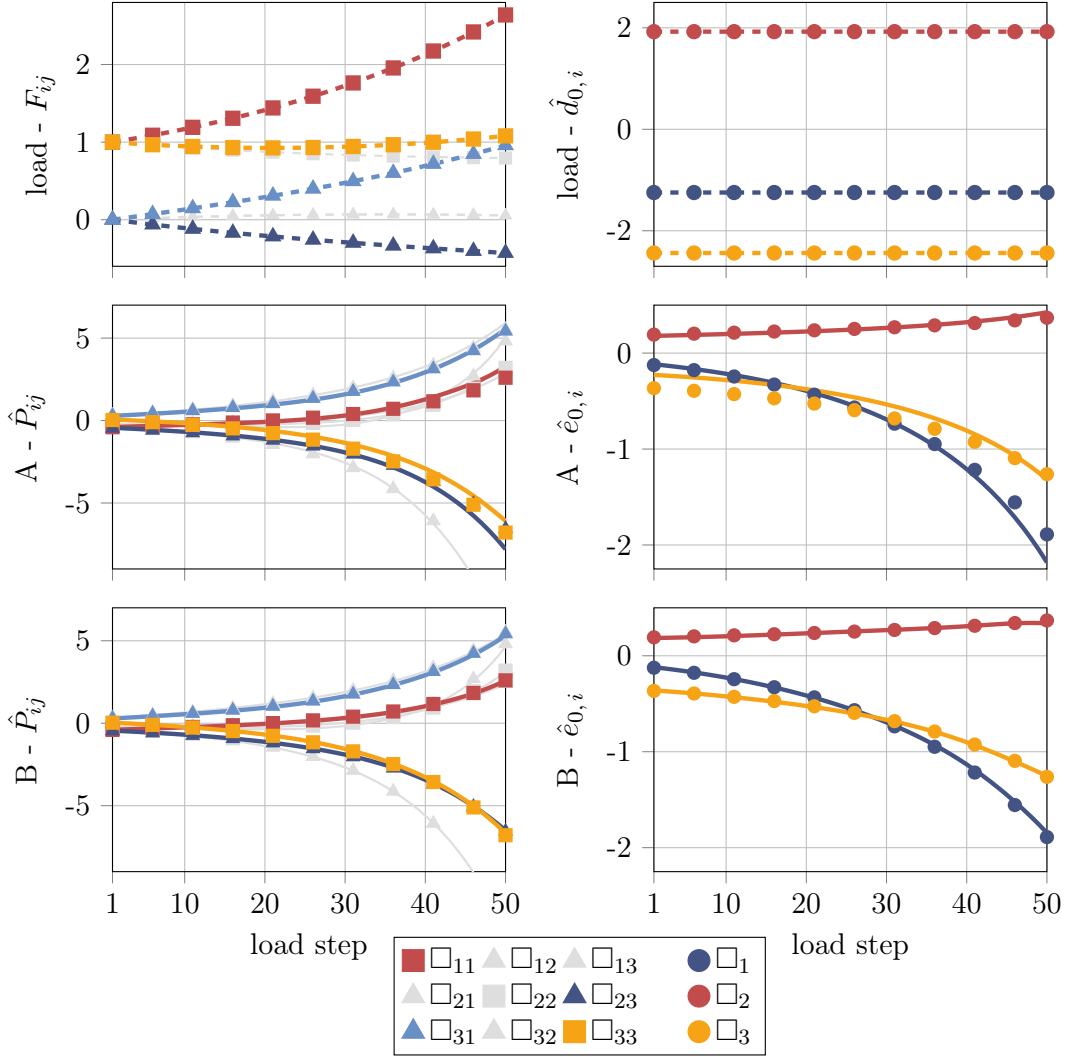
\begin{figure}[t!]
\centering
\resizebox{0.9\textwidth}{!}{
\tikzsetnextfilename{TI_eval}

\pgfplotstableset{
create on use/time_steps/.style={create col/copy column from table={plot_data/TI/model_eval/time_steps.txt}{0}}
}

\begin{tikzpicture}
\pgfplotsset{set layers}
\begin{groupplot}[
	group style = {group size = 2 by 3, vertical sep = 0.5 cm,
								horizontal sep = 0.115*\textwidth},
	cycle list name=mycolorlist_TI,
	xtick align = outside,
	ytick align = outside,
	xtick pos = left,
	]
	
		    \nextgroupplot[ylabel = {load - $F_{ij}$}, ytick={0,1,2},yticklabels={0,1,2},
width=0.4*\textwidth, height=0.3*\textwidth,
xtick={-5},xticklabels={},
ymin = -0.6, ymax = 2.80,
xmin = 1, xmax = 50,
ytick pos = left,
grid = major,
xtick={1,10,20,30,40,50},xticklabels={,,,,,},
xtick pos = left,
] 
\foreach \num in {1,2,3,4,5,6,7,8,9,10,11,12}{
 \addplot coordinates { (-2,-2) (-3,-3) };
 };

        \foreach \num in {1,2,3,4,7,0,5,6,8}{
    \addplot table [x =time_steps,y index=\num] 
    {plot_data/TI/model_eval/F.txt};
    };



    	    \nextgroupplot[legend to name = grouplegend, legend columns = 8, ylabel = {load - $\hat{d}_{0,\,i}$}, ytick={-2,0,2},yticklabels={-2,0,2},
width=0.4*\textwidth, height=0.3*\textwidth,
xtick={1,10,20,30,40,50},xticklabels={,,,,,},
ymin = -2.7, ymax = 2.50,
xmin = 1, xmax = 50,
ytick pos = left,
grid = major]

\foreach \num in {11, 12, 13}{
 \addplot coordinates { (-2,-2) (-3,-3) };
  \addlegendentryexpanded{$\square_{\num}$}
 };
 \addlegendimage{empty legend}\addlegendentry{}
 \addlegendimage{empty legend}\addlegendentry{}
 \addlegendimage{empty legend}\addlegendentry{}
 \addlegendimage{empty legend}\addlegendentry{}
   \addplot coordinates { (-2,-2) (-3,-3) }; \addlegendentryexpanded{$\square_{1}$}
\foreach \num in {21, 22, 23}{
 \addplot coordinates { (-2,-2) (-3,-3) };
 \addlegendentryexpanded{$\square_{\num}$}
 };
 \addlegendimage{empty legend}\addlegendentry{}
 \addlegendimage{empty legend}\addlegendentry{}
 \addlegendimage{empty legend}\addlegendentry{}
  \addlegendimage{empty legend}\addlegendentry{}
    \addplot coordinates { (-2,-2) (-3,-3) }; \addlegendentryexpanded{$\square_{2}$}
    \foreach \num in {31, 32, 33}{
 \addplot coordinates { (-2,-2) (-3,-3) };
 \addlegendentryexpanded{$\square_{\num}$}
 };
 \addlegendimage{empty legend}\addlegendentry{}
 \addlegendimage{empty legend}\addlegendentry{}
 \addlegendimage{empty legend}\addlegendentry{}
  \addlegendimage{empty legend}\addlegendentry{}
    \addplot coordinates { (-2,-2) (-3,-3) }; \addlegendentryexpanded{$\square_{3}$}
 
 \foreach \num in {1,2,3,4,5,6,7,8,9}{
 \addplot coordinates { (-2,-2) (-3,-3) };
 };

        \foreach \num in {0,1,2}{
    \addplot table [x =time_steps,y index=\num] 
    {plot_data/TI/model_eval/D.txt};
    };
    
        \nextgroupplot[ylabel = {A - $\hat{P}_{ij}$}, ytick={-5,0,5},yticklabels={-5,0,5},
width=0.4*\textwidth, height=0.3*\textwidth,
ymin = -9, ymax = 7,
xmin = 1, xmax = 50,
ytick pos = left,
grid = major,
xtick={1,10,20,30,40,50},xticklabels={,,,,,}] 
 
\foreach \num in {1,2,3,4,5,6,7,8,9,10,11,12,13,14,15,16,17,18,19,20,21,22,23,24}{
 \addplot coordinates { (-2,-2) (-3,-3) };
 };

        \foreach \num in {1,2,3,4,7,0,5,6,8}{
    \addplot table [x =time_steps,y index=\num] 
    {plot_data/TI/model_eval/P.txt};
    };
        \foreach \num in {1,2,3,4,7,0,5,6,8}{
    \addplot table [x =time_steps, y index=\num] {plot_data/TI/model_eval/W_I/P_m.txt};
    };

    \nextgroupplot[legend pos = north west,legend columns=1, ylabel = {A - $\hat{e}_{0,\,i}$}, ytick={-2,-1,0},yticklabels={-2,-1,0},
width=0.4*\textwidth, height=0.3*\textwidth,
ymin = -2.250, ymax = 0.50,
xmin = 1, xmax = 50,
ytick pos = left,
legend pos = north west,
grid = major,
xtick={1,10,20,30,40,50},xticklabels={,,,,,}]

\foreach \num in {1,2,3,4,5,6,7,8,9,10,11,12,13,14,15,16,17,18,19,20,21,22,23,24,25,26,27,28,29,30,31,32,33,34,35,36,37,38,39,40,41,42}{
 \addplot coordinates { (-2,-2) (-3,-3) };
 };

    \foreach \num in  {0, 1, 2}{
    \addplot table [x =time_steps,y index=\num, only marks] 
    {plot_data/TI/model_eval/E.txt};
    };
        \foreach \num in {0,1,2}{
    \addplot table [dashed, x =time_steps,y index=\num,  mark=none] {plot_data/TI/model_eval/W_I/E_m.txt};
    };
    
\nextgroupplot[legend to name = grouplegend, legend columns = 8,ylabel = {B - $\hat{P}_{ij}$}, ytick={-5,0,5},yticklabels={-5,0,5},
width=0.4*\textwidth, height=0.3*\textwidth,
ymin = -9, ymax = 7,
xmin = 1, xmax = 50,
ytick pos = left,
grid = major,
xtick={1,10,20,30,40,50},xticklabels={1,10,20,30,40,50},
xlabel={load step}] 

\foreach \num in {11, 12, 13}{
 \addplot coordinates { (-2,-2) (-3,-3) };
  \addlegendentryexpanded{$\square_{\num}$}
 };
 \addlegendimage{empty legend}\addlegendentry{}
 \addlegendimage{empty legend}\addlegendentry{}
 \addlegendimage{empty legend}\addlegendentry{}
 \addlegendimage{empty legend}\addlegendentry{}
   \addplot coordinates { (-2,-2) (-3,-3) }; \addlegendentryexpanded{$\square_{1}$}
\foreach \num in {21, 22, 23}{
 \addplot coordinates { (-2,-2) (-3,-3) };
 \addlegendentryexpanded{$\square_{\num}$}
 };
 \addlegendimage{empty legend}\addlegendentry{}
 \addlegendimage{empty legend}\addlegendentry{}
 \addlegendimage{empty legend}\addlegendentry{}
  \addlegendimage{empty legend}\addlegendentry{}
    \addplot coordinates { (-2,-2) (-3,-3) }; \addlegendentryexpanded{$\square_{2}$}
    \foreach \num in {31, 32, 33}{
 \addplot coordinates { (-2,-2) (-3,-3) };
 \addlegendentryexpanded{$\square_{\num}$}
 };
 \addlegendimage{empty legend}\addlegendentry{}
 \addlegendimage{empty legend}\addlegendentry{}
 \addlegendimage{empty legend}\addlegendentry{}
  \addlegendimage{empty legend}\addlegendentry{}
    \addplot coordinates { (-2,-2) (-3,-3) }; \addlegendentryexpanded{$\square_{3}$}

\foreach \num in {1,2,3,4,5,6,7,8,9,10,11,12}{
 \addplot coordinates { (-2,-2) (-3,-3) };
 };

        \foreach \num in {1,2,3,4,7,0,5,6,8}{
    \addplot table [x =time_steps,y index=\num, only marks] 
    {plot_data/TI/model_eval/P.txt};
    };
        \foreach \num in {1,2,3,4,7,0,5,6,8}{
    \addplot table [x =time_steps, y index=\num] {plot_data/TI/model_eval/P_model/P_m.txt};
    };

    \nextgroupplot[legend pos = north west,legend columns=1, ylabel = {B - $\hat{e}_{0,\,i}$}, ytick={-2,-1,0},yticklabels={-2,-1,0},
width=0.4*\textwidth, height=0.3*\textwidth,
ymin = -2.25, ymax = 0.5,
xmin = 1, xmax = 50,
ytick pos = left,
grid = major,
xtick={1,10,20,30,40,50},xticklabels={1,10,20,30,40,50},
xlabel={load step}]

\foreach \num in {1,2,3,4,5,6,7,8,9,10,11,12,13,14,15,16,17,18,19,20,21,22,23,24,25,26,27,28,29,30,31,32,33,34,35,36,37,38,39,40,41,42}{
 \addplot coordinates { (-2,-2) (-3,-3) };
 };

    \foreach \num in  {0, 1, 2}{
    \addplot table [x =time_steps,y index=\num, only marks] 
    {plot_data/TI/model_eval/E.txt};
    };
        \foreach \num in {0,1,2}{
    \addplot table [dashed, x =time_steps,y index=\num,  mark=none] {plot_data/TI/model_eval/P_model/E_m.txt};
    };
    
\end{groupplot}
\node at ($(group c1r3.east)!0.5!(group c2r3.west) + (0 cm, -3.3 cm) $) {\ref{grouplegend}}; 

\end{tikzpicture}
}
\caption{Transversely isotropic potential, load case and evaluation of the models. Points depict data from the analytical potential, while continuous lines depict the evaluation of the models. A denotes the physics-augmented model $\text{A}^1$, while B denotes the uninformed model $\text{B}^2$. All quantities are dimensionless.}
\label{fig:TI_eval}
\end{figure}

Out of the $300,000$ sub-datasets $D_i$, one load case is now used to evaluate the mechanical stress and electrical field for both calibrated models. As shown in \cref{fig:TI_eval} in the top row, in this specific load case, all three components of the electrical displacement field $\bd_0$ are non-zero but constant, while the deformation gradient consists of an increasing value of $F_{11}$ and strong shear components.

The invariant-based model $\text{A}^1$ is evaluated for $n_{\text{data}}=32=2^5$, the uninformed model $\text{B}^2$ is evaluated for $n_{\text{data}}=2,048=2^{11}$.
Each model was calibrated one time, which resulted in a $\log_{10}$ MSE of the overall dataset of $-1.77$ for $\text{A}^1$ and $-2.76$ for $\text{B}^2$.
The $\log_{10}$ MSE for the single load case visualized below is $-1.63$ for $\text{A}^1$ and $-3.26$ for $\text{B}^2$. Therefore, this case is representative for the global $\log_{10}$ MSE of the model $\text{A}^1$, while it overestimates the prediction quality of $\text{B}^2$.

In the middle row of \cref{fig:TI_eval}, the resulting stress and electrical field for model $\text{A}^1$ are shown. For both quantities, all components are activated and highly nonlinear. Beside slight deviations, e.g., in the $P_{33}$ component, this very small, physics-augmented model shows excellent results for all components.
In the bottom row of \cref{fig:TI_eval}, the results for the model $\text{B}^2$ are shown. This fairly large, uninformed model also shows excellent results for both the mechanical stress and the electrical field, even outperforming the results of $\text{A}^1$.
Overall, both models perform excellent on this really challenging evaluation case. 

\subsection{Influence of the physical augmentation on the model}

Overall, the invariant-based, physics-augmented model A can cope with much smaller network architectures and calibration datasets than the uninformed model B. However, it is less flexible and thus has a slightly worse prediction quality for very big calibration datasets. This shows two effects that including constitutive requirements has on the model: 
First of all, it provides a pronounced mathematical structure which helps to generalize the model very fast. By including fundamental physical principles in the model, such as the second law of thermodynamics, the model behavior is already specified to some amount, thus simplifying the calibration process for a specific material. In that sense, constitutive conditions can be seen as an \emph{inductive bias} \cite{haussler1988}.

While the use of invariants provides the model with a lot of structure and fulfills several constitutive requirements by construction, this is also a potential bottleneck: in a purely mechanical framework it is possible to formulate a complete functional basis in terms of (polyconvex) invariants  for the transverse isotropy used in this example \cite{zheng1994}.
However, this is not possible for all symmetry groups, especially for electro-mechanically coupled material behavior, c.f.~\cref{non_pc_iso}. Overall, the invariant layer may lead to a loss of information: even when it is possible to construct a complete functional basis in invariants for a symmetry group, some of its elements may not be polyconvex and can thus not be used. Also, for some anisotropy classes, no complete functional basis in invariants is known at all \cite{zheng1994}. This loss of information can only be partially compensated by the flexibility of the neural network.
This leads to the second point: the loss of flexibility. Constitutive requirements often restrict the possible function space of the model, e.g., when loosing information in the invariant layer or when assuming an internal energy as \emph{convex}.

However, the restrictions made in this work are either physically well based, e.g., the use of an internal energy which ensures the second law of thermodynamics, or they have a strong mathematical motivation, such as the polyconvexity condition which ensures material stability. Therefore, they should not be seen as strong restrictions of the model behavior, rather, they lead to reliable, i.e., physically sensible, model predictions, while also improving the generalization properties of the model.
Thus, in the trade-off between structure and flexibility, structure should be prioritized as far as possible.
Only if it is inevitable, only when more flexibility is required, should the structure of the model be weakened, e.g., by fulfilling some constitutive requirements only in an approximate fashion, c.f.~\cite{klein2021}. 
In terms of constitutive requirements, the invariant based model proposed in \cref{sec:model} has the same amount of structure as most analytical models found in literature, c.f.~\cite{steinmann2012}. Only for some analytical models, it is possible to arrive at even more interpretable formulations in the sense that the model parameters have a physical meaning, such as shear and compression modulus for the hyperelastic Hencky model \cite{Hencky1928,Hencky1929,neff2016, agn_neff2020axiomatic}.

\medskip

That the model proposed in \cref{sec:model} fulfills fundamental physical principles is the first reason to trust its predictions -- the second one is its excellent prediction of test cases. 
Generally, it is not sufficient to calibrate the model to a given dataset, it is just as important to evaluate it for load paths which are not included in the calibration dataset. Only in this way the \emph{generalization} of the model can be assessed. In this academic example it is straightforward to evaluate the global behavior of the potential \cref{p_ti},  and thus to assess the generalization of the proposed machine learning model on a global scale, c.f.~\cref{fig:MSE_TI}. However, this is in general not possible due to strong practical limitations of experimental investigations \cite{baaser2011} or the computational effort of numerical investigations of complex microstructures \cite{Jamshidian2020}. In practice, usually only small test datasets can be created. The deformations used for this should be as general as possible, e.g., by combining both normal and shear components in the deformation gradient, c.f.~\cref{fig:TI_eval}, and within the deformation range in which the model will be used. Thereby, it can be assumed that the test case is representative for the global behavior of the material.
When the model is then able to predict the material behavior for this test case it can be assumed that it can generally predict its global behavior.
Note that this strategy is well known and usually implemented in machine learning, as datasets are typically separated into calibration and test sets \cite{aggarwal2018}.

Due to the reasons mentioned above, uniform sampling of the input space will not be applied in the following sections. It should be noted that some machine learning approaches directly act on global force and displacement data generated with finite element simulations \cite{kalina2022, flaschel2021}. However, these methods are restricted to data generated with FE simulations and are thus less general.
Instead, in the following sections, deformation scenarios which are commonly applied in experimental investigations will be used for the calibration dataset, such as uniaxial tension, and additional test cases with general deformation modes, c.f.~\cite{Fernandez2020}.

\section{Application to an analytically homogenized rank-one laminate}
\label{sec:R1}

After the more academic example in the previous section, the proposed model is now applied to data of an analytically homogenized rank-one laminate. In doing so, we demonstrate the applicability of the model to the challenging effective behaviors of electro-mechanical composite materials.

\subsection{Data generation}

\begin{figure}[t!]
	\centering		\includegraphics[width=0.8\textwidth]{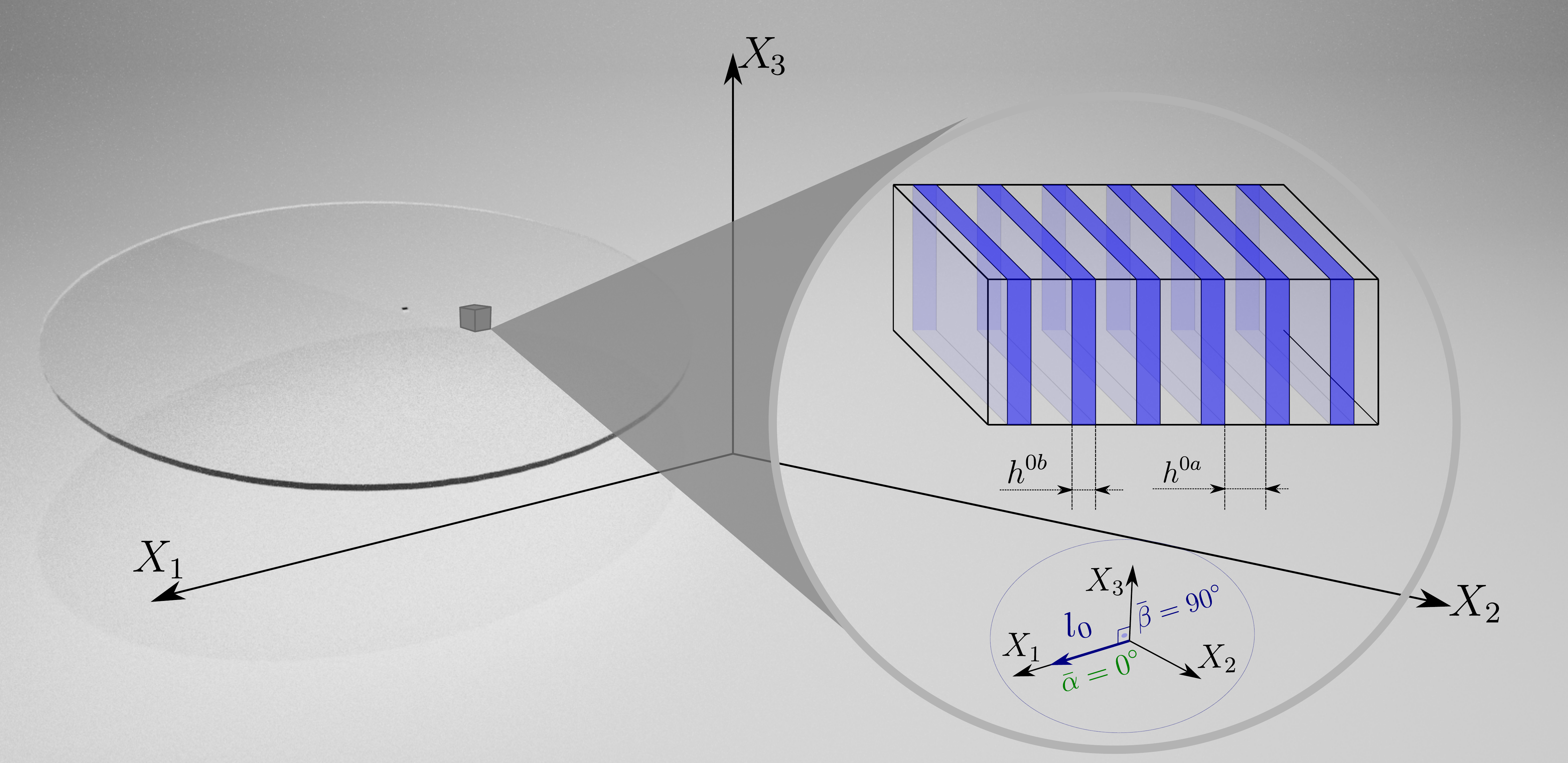} 
	\caption{Description of the microstructure of the rank-one laminate: the purple constituent corresponds to phase $b$, inserted within phase $a$ according to a lamination orientation perpendicular to the vector $\vect{l}_0$.}
	\label{fig: normal vector}
\end{figure}

The structure of a biphasic rank-one laminated DE composite is comprised of constituents or phases $a$ and $b$, as illustrated in \cref{fig: normal vector}, c.f.~\cite{marin2021}. 
The two phases $a$ and $b$ occupy volume fractions  $c^a=h_0^a/\left(h_0^a+h_0^b\right)$ and $c^b=1-c^a$, respectively, where $h_0^a$ and $h_0^b$ denote the thicknesses of the phases in the initial configuration $\mathcal{B}_0$ (with small $h_0^{a},h_0^{b}$ to comply with homogenization theory). 
The interface between both phases is characterised by the normal vector $\vect{l}_0$, which is spherically parametrized as $\vect{l}_0=\left(\sin\bar{\beta}\cos\bar{\alpha},\, \sin\bar{\beta}\sin\bar{\alpha},\, \cos\bar{\beta}\right)^{T}$.

For the homogenization, we assume the existence of the microscopic fields $\vect{F}^a,\vect{d}_0^a$ and $\vect{F}^b,\vect{d}_0^b$ in each phase $a$ and $b$. 
The internal energy density of both phases $a$ and $b$, namely $e^{a,b}(\vect{F}^{a,b},\vect{d}^{a,b}_{0})$, admits a decomposition similar to that in \cref{eq:e_full}, but in terms of invariants of either $\{\vect{F}^a,\vect{d}_0^a\}$ or $\{\vect{F}^b,\vect{d}_0^b\}$, namely
\begin{equation}\label{eqn:additive decomposition ROL}
e^{a,b}(\vect{F}^{a,b},\vect{d}^{a,b}_{0})
=\mathcal{P}^{a,b}(\vect{\mathcal{V}}^{a,b})
=\widetilde{\mathcal{P}}^{a,b}(\vect{\mathcal{I}}^{a,b})+ \mathcal{P}_{\text{vol}}^{a,b}(J^{a,b})\,.
\end{equation}
Here, the specific form of both contributions in \cref{eqn:additive decomposition ROL} is assumed as
%
%
\begin{equation}\label{eqn:the model for ROL}
\begin{aligned}
\widetilde{\mathcal{P}}^{a}(\vect{\mathcal{I}}^{a}) &=
\frac{\mu_1}{2}{{I}}^{a}_{1} + \frac{\mu_2}{2}{{I}}_2^{a} + \frac{{I}^a_5}{2\varepsilon J^a}\,,
\qquad & 
\mathcal{P}_{\text{vol}}^{a}(J^{a}) &=
\frac{\lambda}{2}(J^a-1)^2\,,
\\
\widetilde{\mathcal{P}}^{b}(\vect{\mathcal{I}}^{b}) &=		f_m\Big(\frac{\mu_1}{2}{{I}}^b_{1} + \frac{\mu_2}{2}{{I}}^b_{2}\Big) + \frac{{I}^b_{5}}{2 f_e\varepsilon J^b}\,,
\qquad & 
{\mathcal{P}_{\text{vol}}^{b}}(J^{b}) &=
f_m\frac{\lambda}{2}(J^b-1)^2\,,
\end{aligned}	
\end{equation}
where $f_m,f_e$ represent the mechanical and electro-mechanical contrast parameters relating the material properties of phases $a$ and $b$. 
Clearly, both matrix and inclusion constituents are described by means of isotropic internal energy densities. 
Specifically, we use of the following material parameters
\begin{equation}
\mu_2=0.1\mu_1\,,\qquad \lambda=50\mu_1\,,\qquad f_m=20\,,\qquad f_e=2\,,\qquad c_a=0.5\,,\qquad \vect{l}_0=\begin{pmatrix}
0&0&1
\end{pmatrix}^T.
\end{equation}

The detailed derivation of the homogenized relationship between the microscopic fields $\bF^{a,b},\bd_0^{a,b}$, as well as $e^{a,b}$ and $\bP^{a,b},\be_0^{a,b}$, and their macroscopic counterparts $\vect{F},\vect{d}_0,e,\bP,\be_0$ can be found in \cref{sec:ROL_appendix}. 
Note that the scaling procedure described in \cref{sec:scaling} with respect to the material parameters $\mu_1,\varepsilon$ is carried out  in order to facilitate the calibration of the  ANN based constitutive model by reducing the significant dissimilarity between mechanical and electro-mechanical parameters.

\medskip

For the data generation, the following 10 load cases for the macroscopic fields $\vect{F}$ and $\vect{d}_0$ are considered: (1) purely mechanical uniaxial tension, (2) purely mechanical pure shear load, (3--4) the minimum mechanical value for uniaxial tension, and then two datasets with an electric displacement field applied in $X_1$ and $X_3$ direction, respectively, (5--6) the maximum mechanical value for uniaxial tension, and then two datasets with an electric displacement field applied in $X_1$ and $X_3$ direction, respectively, (7--8) the minimum mechanical value for shear load, and then two datasets with an electric displacement field applied in $X_1$ and $X_3$ direction, respectively, and (9--10) the maximum mechanical value for shear load, and then two datasets with an electric displacement field applied in $X_1$ and $X_3$ direction, respectively.
With each single load case consisting of 100 data points, this results in an overall calibration dataset size of $1,000$ datapoints.
For the explicit definition of  the deformation gradients of these load cases, see \cite{Fernandez2020}.
In addition, a biaxial test case and a mixed shear-tension test case were generated, both with variable electric displacement fields.

\subsection{Model calibration}

For the invariant-based ANN model proposed in \cref{sec:model}, we use the input
\begin{equation}
    \boldsymbol{\cI}=\left(I_1,\,I_2,\,J,\,-J,\,\hat{I}_4,\,\hat{I}_5,\,J_1^{\text{ti}},\,J_2^{\text{ti}},\,\hat{J}_3^{\text{ti}}\right)\in\bbR^{9}\,,
\end{equation}
with the invariants as described in \cref{sec:objectivity and material symmetry}, and the scaling as described in \cref{sec:scaling}. The input consists of both isotropic and transversely isotropic invariants.
For the ICNNs, we choose two network architectures: one consisting of only one hidden layer with 8 nodes ($H=1, n^{[1]}=8$) and one consisting of two hidden layers with 16 nodes in each layer ($H=2, n^{[h]}=16$), for which we use the short notations $\cS\cP^+(  \boldsymbol{\cI};\,8)$ and  $\cS\cP^+(  \boldsymbol{\cI};\,16,16)$, respectively. The former has a total amount of 89 free parameters, the latter has 449 free parameters.
Each model architecture was initialized three times, using different randomly initialized model parameters each time. The calibration is carried out for $2,500$ epochs.
The full batch of training data is used with TensorFlow's default learning rate and default batch size.
For the remaining calibration details, see \cref{sec:calib}.

\subsection{Model evaluation}

\begin{table}[t!]
\centering
\begin{tabular}{llcll}
\toprule
Architecture & \multicolumn{2}{l}{$\log_{10}$ MSE}  \\
&calibration&test\\ \midrule
$\cS\cP^+(  \boldsymbol{\cI};\,8)$  & -2.85 & -1.78 \\
---\textquotedbl---&  -2.87 & -1.77  \\
---\textquotedbl---& -2.77& -1.76  \\

$\cS\cP^+(  \boldsymbol{\cI};\,16, 16)$   & -2.81 & -1.61 \\
---\textquotedbl---&  -2.76& -1.60 \\
---\textquotedbl---& -2.75& -1.59 \\
\bottomrule
\end{tabular}
\caption{MSEs of the calibrated ANN based models for the rank-one laminate.}
\label{tb:loss_rank_one}
\end{table}

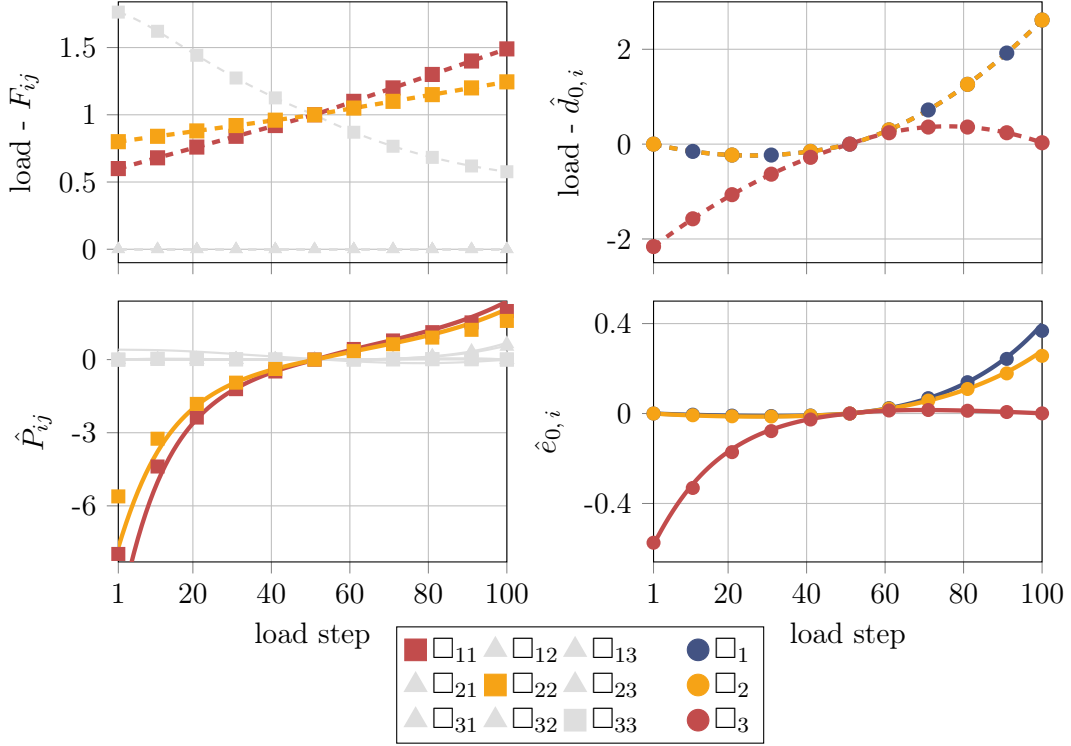
\begin{figure}[tp!]
\centering

\resizebox{0.9\textwidth}{!}{
\tikzsetnextfilename{rank_one_biax}

\pgfplotstableset{
create on use/time_steps/.style={create col/copy column from table={plot_data/rank_one/load_step.txt}{0}}
}

\begin{tikzpicture}
\pgfplotsset{set layers}
\begin{groupplot}[
	group style = {group size = 2 by 2, vertical sep = 0.5 cm,
								horizontal sep = 0.115*\textwidth},
	cycle list name=mycolorlist_r1_biax,
	xtick align = outside,
	ytick align = outside,
	xtick pos = left,
	]
	
		    \nextgroupplot[legend to name = grouplegend, legend columns=8, ylabel = {load - $F_{ij}$}, ytick={0,0.5,1,1.5},yticklabels={0,0.5,1,1.5},
width=0.4*\textwidth, height=0.3*\textwidth,
ymin = -0.1, ymax = 1.840,
xmin = 1, xmax = 100,
ytick pos = left,
grid = major,
xtick={1,20,40,60,80,100},xticklabels={,,,,,},
xtick pos = left,
] 

\foreach \num in {11, 12, 13}{
 \addplot coordinates { (-2,-2) (-3,-3) };
  \addlegendentryexpanded{$\square_{\num}$}
 };
 \addlegendimage{empty legend}\addlegendentry{}
 \addlegendimage{empty legend}\addlegendentry{}
 \addlegendimage{empty legend}\addlegendentry{}
 \addlegendimage{empty legend}\addlegendentry{}
   \addplot coordinates { (-2,-2) (-3,-3) }; \addlegendentryexpanded{$\square_{1}$}
\foreach \num in {21, 22, 23}{
 \addplot coordinates { (-2,-2) (-3,-3) };
 \addlegendentryexpanded{$\square_{\num}$}
 };
 \addlegendimage{empty legend}\addlegendentry{}
 \addlegendimage{empty legend}\addlegendentry{}
 \addlegendimage{empty legend}\addlegendentry{}
  \addlegendimage{empty legend}\addlegendentry{}
    \addplot coordinates { (-2,-2) (-3,-3) }; \addlegendentryexpanded{$\square_{2}$}
    \foreach \num in {31, 32, 33}{
 \addplot coordinates { (-2,-2) (-3,-3) };
 \addlegendentryexpanded{$\square_{\num}$}
 };
 \addlegendimage{empty legend}\addlegendentry{}
 \addlegendimage{empty legend}\addlegendentry{}
 \addlegendimage{empty legend}\addlegendentry{}
  \addlegendimage{empty legend}\addlegendentry{}
    \addplot coordinates { (-2,-2) (-3,-3) }; \addlegendentryexpanded{$\square_{3}$}

        \foreach \num in {1,2,3,5,6,7,8,0,4}{
    \addplot table [x =time_steps,y index=\num] 
    {plot_data/rank_one/F_test1_do_var.txt};
    };

	    \nextgroupplot[legend pos = outer north east, legend columns = 8, ylabel = {load - $\hat{d}_{0,\,i}$}, ytick={-2,0,2},yticklabels={-2,0,2},
width=0.4*\textwidth, height=0.3*\textwidth,
ymin = -2.5, ymax = 3,
xmin = 1, xmax = 100,
ytick pos = left,
grid = major,
xtick={1,20,40,60,80,100},xticklabels={,,,,,}]

\foreach \num in {1,2,3,4,5,6,7,8,9,10,11,12}{
 \addplot coordinates { (-2,-2) (-3,-3) };
 }
 
 \foreach \num in {1,2,3,4,5,6,7,8,9}{
 \addplot coordinates { (-2,-2) (-3,-3) };
 };

        \foreach \num in {0,1,2}{
    \addplot table [x =time_steps,y index=\num] 
    {plot_data/rank_one/D_test1_do_var.txt};
    };
    
    \nextgroupplot[legend style={at={(axis cs: 1,2.5)},anchor=north west}, legend columns=8, ylabel = {$\hat{P}_{ij}$}, ytick={-6,-3,0},yticklabels={-6,-3,0},
width=0.4*\textwidth, height=0.3*\textwidth,
ymin = -8.3, ymax = 2.40,
xmin = 1, xmax = 100,
ytick pos = left,
grid = major,
xlabel={load step},
xtick={1,20,40,60,80,100},xticklabels={1,20,40,60,80,100}]

\foreach \num in {1,2,3,4,5,6,7,8,9,10,11,12,13,14,15,16,17,18,19,20,21,22,23,24}{
 \addplot coordinates { (-2,-2) (-3,-3) };
 };

        \foreach \num in {1,2,3,5,6,7,8,0,4}{
    \addplot table [x =time_steps,y index=\num] 
    {plot_data/rank_one/P_test1_do_var.txt};
    };
        \foreach \num in {1,2,3,5,6,7,8,0,4}{
    \addplot table [x =time_steps, y index=\num] {plot_data/rank_one/P_m_test1_do_var.txt};
    };

    \nextgroupplot[legend pos = north west,legend columns=1, ylabel = {$\hat{e}_{0,\,i}$}, ytick={-0.4,0,0.4},yticklabels={-0.4,0,0.4},
width=0.4*\textwidth, height=0.3*\textwidth,
ymin = -0.66, ymax = 0.50,
xlabel={load step},
xmin = 1, xmax = 100,
ytick pos = left,
grid = major,
xtick={1,20,40,60,80,100},xticklabels={1,20,40,60,80,100}]

\foreach \num in {1,2,3,4,5,6,7,8,9,10,11,12,13,14,15,16,17,18,19,20,21,22,23,24,25,26,27,28,29,30,31,32,33,34,35,36,37,38,39,40,41,42}{
 \addplot coordinates { (-2,-2) (-3,-3) };
 };

    \foreach \num in  {0, 1, 2}{
    \addplot table [x =time_steps,y index=\num, only marks] 
    {plot_data/rank_one/E_test1_do_var.txt};
    };
        \foreach \num in {0,1,2}{
    \addplot table [dashed, x =time_steps,y index=\num,  mark=none] {plot_data/rank_one/E_m_test1_do_var.txt};
    };

\end{groupplot}
\node at ($(group c1r2.east)!0.5!(group c2r2.west) + (0 cm, -3.3 cm) $) {\ref{grouplegend}}; 

\end{tikzpicture}
}
\caption{Evaluation of the biaxial test case for the rank-one laminate. Points depict homogenization data, while lines depict the evaluation of the machine learning model. 
All quantities are dimensionless.}
\label{fig:r1_biax}
\end{figure}

In \cref{tb:loss_rank_one}, the MSEs of the different model initializations are provided. Both  architectures show excellent results for the calibration dataset with slightly worse MSEs for the test dataset. Overall, the model with the smaller network architecture is already sufficiently flexible, in fact, it performs even better on the test dataset than the model with the bigger architecture. 

\begin{figure}[tp!]
\centering

\resizebox{0.9\textwidth}{!}{
\tikzsetnextfilename{rank_one_mixed}

\pgfplotstableset{
create on use/time_steps/.style={create col/copy column from table={plot_data/rank_one/load_step.txt}{0}}
}

\begin{tikzpicture}
\pgfplotsset{set layers}
\begin{groupplot}[
	group style = {group size = 2 by 2, vertical sep = 0.5 cm,
								horizontal sep = 0.115*\textwidth},
	cycle list name=mycolorlist_r1_mixed,
	xtick align = outside,
	ytick align = outside,
	xtick pos = left,
	]
	
		    \nextgroupplot[legend to name = grouplegend, legend columns=8, ylabel = {load - $F_{ij}$}, ytick={0,0.5,1,1.5},yticklabels={0,0.5,1,1.5},
width=0.4*\textwidth, height=0.3*\textwidth,
ymin = -0.4, ymax = 1.6,
xmin = 1, xmax = 100,
ytick pos = left,
grid = major,
xtick={1,20,40,60,80,100},xticklabels={,,,,,},
xtick pos = left,
] 

\foreach \num in {11, 12, 13}{
 \addplot coordinates { (-2,-2) (-3,-3) };
  \addlegendentryexpanded{$\square_{\num}$}
 };
 \addlegendimage{empty legend}\addlegendentry{}
 \addlegendimage{empty legend}\addlegendentry{}
 \addlegendimage{empty legend}\addlegendentry{}
 \addlegendimage{empty legend}\addlegendentry{}
   \addplot coordinates { (-2,-2) (-3,-3) }; \addlegendentryexpanded{$\square_{1}$}
\foreach \num in {21, 22, 23}{
 \addplot coordinates { (-2,-2) (-3,-3) };
 \addlegendentryexpanded{$\square_{\num}$}
 };
 \addlegendimage{empty legend}\addlegendentry{}
 \addlegendimage{empty legend}\addlegendentry{}
 \addlegendimage{empty legend}\addlegendentry{}
  \addlegendimage{empty legend}\addlegendentry{}
    \addplot coordinates { (-2,-2) (-3,-3) }; \addlegendentryexpanded{$\square_{2}$}
    \foreach \num in {31, 32, 33}{
 \addplot coordinates { (-2,-2) (-3,-3) };
 \addlegendentryexpanded{$\square_{\num}$}
 };
 \addlegendimage{empty legend}\addlegendentry{}
 \addlegendimage{empty legend}\addlegendentry{}
 \addlegendimage{empty legend}\addlegendentry{}
  \addlegendimage{empty legend}\addlegendentry{}
    \addplot coordinates { (-2,-2) (-3,-3) }; \addlegendentryexpanded{$\square_{3}$}

        \foreach \num in {2,5,6,7,8,0,1,3,4}{
    \addplot table [x =time_steps,y index=\num] 
    {plot_data/rank_one/F_test3_do_var.txt};
    };

	    \nextgroupplot[legend pos = outer north east, legend columns = 8, ylabel = {load - $\hat{d}_{0,\,i}$}, ytick={-2,0,2},yticklabels={-2,0,2},
width=0.4*\textwidth, height=0.3*\textwidth,
ymin = -3.2, ymax = 3,
xmin = 1, xmax = 100,
ytick pos = left,
grid = major,
xtick={1,20,40,60,80,100},xticklabels={,,,,,}]

\foreach \num in {1,2,3,4,5,6,7,8,9,10,11,12}{
 \addplot coordinates { (-2,-2) (-3,-3) };
 }
 
 \foreach \num in {1,2,3,4,5,6,7,8,9}{
 \addplot coordinates { (-2,-2) (-3,-3) };
 };

        \foreach \num in {0,1,2}{
    \addplot table [x =time_steps,y index=\num] 
    {plot_data/rank_one/D_test3_do_var.txt};
    };
    
    \nextgroupplot[legend style={at={(axis cs: 1,2.5)},anchor=north west}, legend columns=8, ylabel = {$\hat{P}_{ij}$}, ytick={-4,-2,0,2},yticklabels={-4,-2,0,2},
width=0.4*\textwidth, height=0.3*\textwidth,
ymin = -5, ymax = 2.2,
xmin = 1, xmax = 100,
ytick pos = left,
grid = major,
xlabel={load step},
xtick={1,20,40,60,80,100},xticklabels={1,20,40,60,80,100}]

\foreach \num in {1,2,3,4,5,6,7,8,9,10,11,12,13,14,15,16,17,18,19,20,21,22,23,24}{
 \addplot coordinates { (-2,-2) (-3,-3) };
 };

        \foreach \num in {2,5,6,7,8,0,1,3,4}{
    \addplot table [x =time_steps,y index=\num] 
    {plot_data/rank_one/P_test3_do_var.txt};
    };
        \foreach \num in {2,5,6,7,8,0,1,3,4}{
    \addplot table [x =time_steps, y index=\num] {plot_data/rank_one/P_m_test3_do_var.txt};
    };
    
    \nextgroupplot[legend pos = north west,legend columns=1, ylabel = {$\hat{e}_{0,\,i}$}, ytick={-0.3,0,0.3},yticklabels={-0.3,0,0.3},
width=0.4*\textwidth, height=0.3*\textwidth,
ymin = -0.4, ymax = 0.55,
xlabel={load step},
xmin = 1, xmax = 100,
ytick pos = left,
grid = major,
xtick={1,20,40,60,80,100},xticklabels={1,20,40,60,80,100}]

\foreach \num in {1,2,3,4,5,6,7,8,9,10,11,12,13,14,15,16,17,18,19,20,21,22,23,24,25,26,27,28,29,30,31,32,33,34,35,36,37,38,39,40,41,42}{
 \addplot coordinates { (-2,-2) (-3,-3) };
 };

    \foreach \num in  {0,1, 2}{
    \addplot table [x =time_steps,y index=\num, only marks] 
    {plot_data/rank_one/E_test3_do_var.txt};
    };
        \foreach \num in {0,1,2}{
    \addplot table [dashed, x =time_steps,y index=\num,  mark=none] {plot_data/rank_one/E_m_test3_do_var.txt};
    };

\end{groupplot}
\node at ($(group c1r2.east)!0.5!(group c2r2.west) + (0 cm, -3.3 cm) $) {\ref{grouplegend}}; 

\end{tikzpicture}
}
\caption{Evaluation of the shear-tension test case for the rank-one laminate. Points depict homogenization data, while lines depict the evaluation of the machine learning model. 
All quantities are dimensionless.}
\label{fig:r1_mixed}
\end{figure}
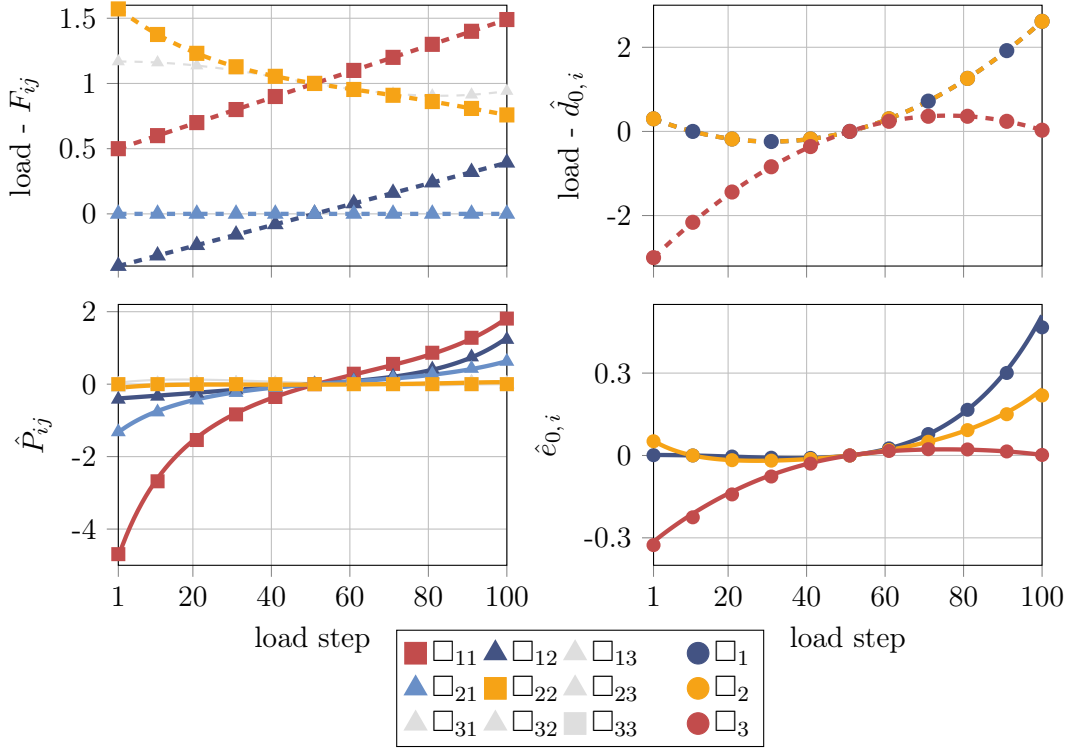

In \cref{fig:r1_biax,fig:r1_mixed}, the evaluations of the two test cases are shown for the model with the best test MSE. The first test case in \cref{fig:r1_biax} comprises of a biaxial deformation gradient and a variable electric displacement field. The second test case in \cref{fig:r1_mixed} comprises of a mixed shear-tension deformation gradient and a variable electric displacement field. Both cases are highly challenging and especially the mixed case represents a general deformation case, extrapolating the model to data not seen in the calibration dataset. For both cases the model shows excellent results, with only slight deviations for high biaxial compression values in \cref{fig:r1_biax}. 

Thus, this example shows how the physics-augmented neural network constitutive model is able to generalize with small amounts of data, and that even very small network architectures can provide sufficient flexibility.

\section{Application to a numerically homogenized cubic metamaterial}
\label{sec:cub}

Finally, a more complex microstructure than the one in \cref{sec:R1} is considered. Thus, the homogenization of the effective material behavior  of the cubic metamaterial, which serves as data for the model calibration, is carried out numerically.
This shows the application of the proposed machine learning model to cases where no analytical homogenization is possible, and the straightforward applicability towards other symmetry groups.

\subsection{Data generation}
\label{sec:cub:data}

\begin{figure}[t!]
	\centering
	\includegraphics[width=0.7\textwidth]{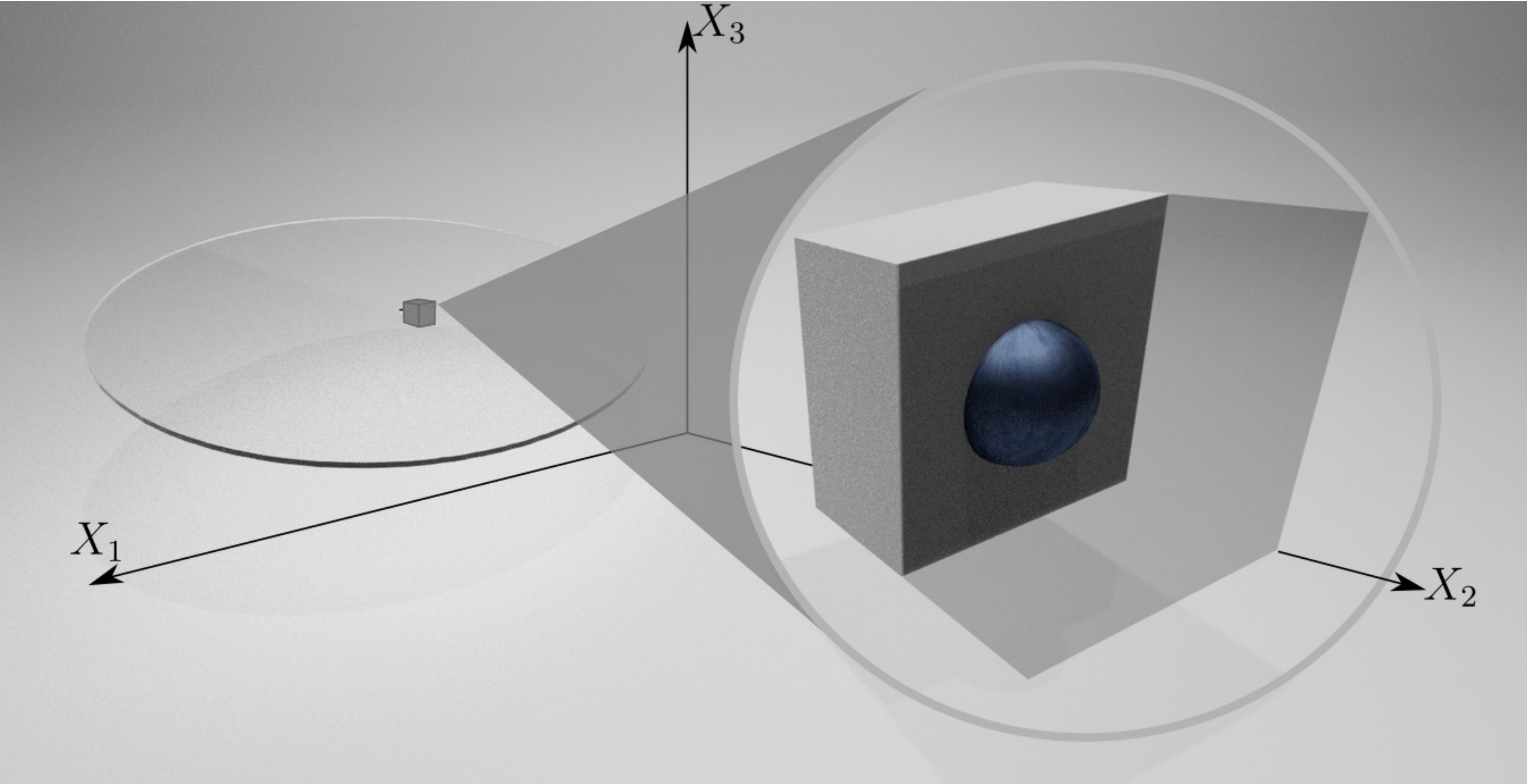}  
	\caption{RVE for the cubic metamaterial, comprising a cubic matrix with spherical inclusion.}
	\label{fig:cubic cell}
\end{figure}

We consider a metamaterial with cubic symmetry, which is characterized by a representative volume element (RVE) with a cubic matrix and a spherical inclusion, see \cref{fig:cubic cell}. The  radius of the inclusion is a quarter of the cell length. 

In this case, the spatial and material coordinates in the RVE are denoted as $\vect{x}_{\mu}:\mathcal{B}_{\mu}\rightarrow \mathbb{R}^3$ and $\vect{X}_{\mu}:\mathcal{B}_{0_{\mu}}\rightarrow \mathbb{R}^3$, respectively, where $\mathcal{B}_{\mu}$ represents the deformed configuration of the undeformed RVE configuration $\mathcal{B}_{0_{\mu}}$. In addition, we assume the existence of a microscopic electric potential $\varphi_{\mu}:\mathcal{B}_{0_{\mu}}\rightarrow \mathbb{R}$. With this, it is possible to define the deformation gradient tensor and the electric field in the RVE, denoted as $\vect{F}_{\mu}$ and $\vect{e}_{0_{\mu}}$, as
\begin{equation}\label{eqn_:micro fields}
\vect{F}_{\mu}=\frac{\partial \vect{x}_{\mu}}{\partial\vect{X}_{\mu}}\,,\qquad 
\vect{e}_{0_{\mu}}=-\frac{\partial\varphi_{\mu}}{\partial\vect{X}_{\mu}}\,.
\end{equation}

The macroscopic deformation gradient $\vect{F}$ and electric field $\vect{e}_0$ are defined in terms of volume averages of their microscopic counterparts $\vect{F}_{\mu}$ and $\vect{e}_{0_{\mu}}$  according to, c.f.~\cite{miehe2016},
\begin{equation}\label{eqn:homogenisation of F}
\begin{aligned}	
\vect{F} &= \frac{1}{V_{\mu}}\int_{\mathcal{B}_{0_{\mu}}}\vect{F}_{\mu}\left({\vect{X}_{{\mu}}}\right)\,dV_{\mu}
=\frac{1}{V_{\mu}}\int_{\partial\mathcal{B}_{0_{\mu}}}\vect{x}_{\mu}\left({\vect{X}_{{\mu}}}\right)\otimes \vect{n}_{0_{\mu}}\,dV_{\mu}\,,
\\
\vect{e}_0& = \frac{1}{V_{\mu}}\int_{\mathcal{B}_{0_{\mu}}}\vect{e}_{0_{\mu}}\left({\vect{X}_{{\mu}}}\right)\,dV_{\mu}
=\frac{1}{V_{\mu}}\int_{\partial\mathcal{B}_{0_{\mu}}}\varphi_{\mu}\left({\vect{X}_{{\mu}}}\right)\vect{n}_{0_{\mu}}\,dV_{\mu}\,.
\end{aligned}
\end{equation}
Likewise, the macroscopic stress and electric displacement fields $\bP$ and $\bd_0$, respectively, can be defined as averages of their microscopic counterparts, namely $\vect{P}_{\mu}$ and  $\vect{d}_{0_{\mu}}$, as
\begin{equation}
\vect{P}=\frac{1}{V_{\mu}}\int_{\mathcal{B}_{0_{\mu}}}\vect{P}_{\mu}\,dV_{\mu}\,,\qquad
\vect{d}_0=\frac{1}{V_{\mu}}\int_{\mathcal{B}_{0_{\mu}}}\vect{d}_{0_{\mu}}\,dV_{\mu}\,.
\end{equation}

For a detailed derivation of the computational homogenization procedure for this type of microstructure, the reader is referred to  \cref{sec:RVE homogenization}. 
Here, we just want to further introduce the microscopic material models required for the homogenization.
The microscopic internal energy density $e_{\mu}(\vect{F}_{\mu},\vect{d}_{0_{\mu}})$ admits a decomposition similar to that in \cref{eq:e_full}, but in terms of invariants of microscopic fields, namely
\begin{equation}\label{eqn:additive decomposition micro}
e_{\mu}(\vect{F}_{\mu},\vect{d}_{0_{\mu}})=\mathcal{P}_{\mu}(\vect{\mathcal{V}}_{\mu})=\widetilde{\mathcal{P}}(\vect{\mathcal{I}}_{\mu})+ \mathcal{P}_{\text{vol}_{{\mu}}}(J_{\mu})\,.
\end{equation}
The specific form of both contributions in \cref{eqn:additive decomposition micro} depends on the material coordinate $\vect{X}_{\mu}\in\mathcal{B}_{0_{\mu}}$ according to
\begin{equation}\label{eqn:the model for cubic cell}
\begin{aligned}
\widetilde{\mathcal{P}}_{\mu}(\vect{X}_{\mu},\vect{\mathcal{I}}_{\mu})&=\left\{\begin{aligned}
&	\frac{\mu_1}{2}{I_1}_{\mu} + \frac{\mu_2}{2}{I_2}_{\mu} + \frac{1}{2\varepsilon J_{\mu}}{I_5}_{\mu}\,,&\quad &\vect{X}_{\mu}\in \mathcal{B}^{\text{mat.}}_{0_{\mu}}\\
&		f_m\Big(\frac{\mu_1}{2}{I_1}_{\mu} + \frac{\mu_2}{2}{I_2}_{\mu}\Big) + \frac{1}{2 f_e\varepsilon J_{\mu}}{I_5}_{\mu}\,,&\quad &\vect{X}_{\mu}\in \mathcal{B}^{\text{incl.}}_{0_{\mu}}
\end{aligned}\right.,
\\  {\mathcal{P}_{\text{vol}}}_{\mu}(\vect{X}_{\mu},J_{\mu})&=\left\{\begin{aligned}&-\frac{\mu_1+2\mu_2}{2}\log(J_{\mu}) + \frac{\lambda}{2}(J_{\mu}-1)^2,&\quad &\vect{X}_{\mu}\in \mathcal{B}^{\text{mat.}}_{0_{\mu}}\\
&f_m\Big(-\frac{\mu_1+2\mu_2}{2}\log(J_{\mu}) + \frac{\lambda}{2}(J_{\mu}-1)^2\Big),&\quad &\vect{X}_{\mu}\in \mathcal{B}^{\text{incl.}}_{0_{\mu}}
\end{aligned}\right. .	
\end{aligned}	
\end{equation}
Here, $\mathcal{B}_{0_{\mu}}^{\text{mat.}}$ and $\mathcal{B}_{0_{\mu}}^{\text{incl.}}$ denote the subdomains of $\cB_{0_\mu}$ corresponding to the matrix and the spherical inclusion, respectively, see \cref{fig:cubic cell}.  
Clearly, both matrix and inclusion constituents are described by means of isotropic internal energy densities. 
Furthermore, $f_m,f_e$ represent the mechanical and electro-mechanical contrast parameters relating the material properties of both matrix and inclusion. Specifically, we have made use of the following definition for the material parameters
\begin{equation}
\mu_2=\mu_1\,,\qquad \lambda=30\mu_1\,,\qquad f_m=2\,,\qquad f_e=5\,.
\end{equation}
Note that the scaling procedure described in \cref{sec:scaling} with respect to the material parameters $\mu_1,\varepsilon$ is carried out  in order to facilitate the calibration of the ANN based constitutive model, reducing the significant dissimilarity between mechanical and electro-mechanical parameters.

\medskip

\begin{figure}[t!]
	\centering
	\subfloat[Finite element mesh of the RVE
        \label{fig:cubic cell:fe}]{%
		\includegraphics[width=0.33\textwidth]{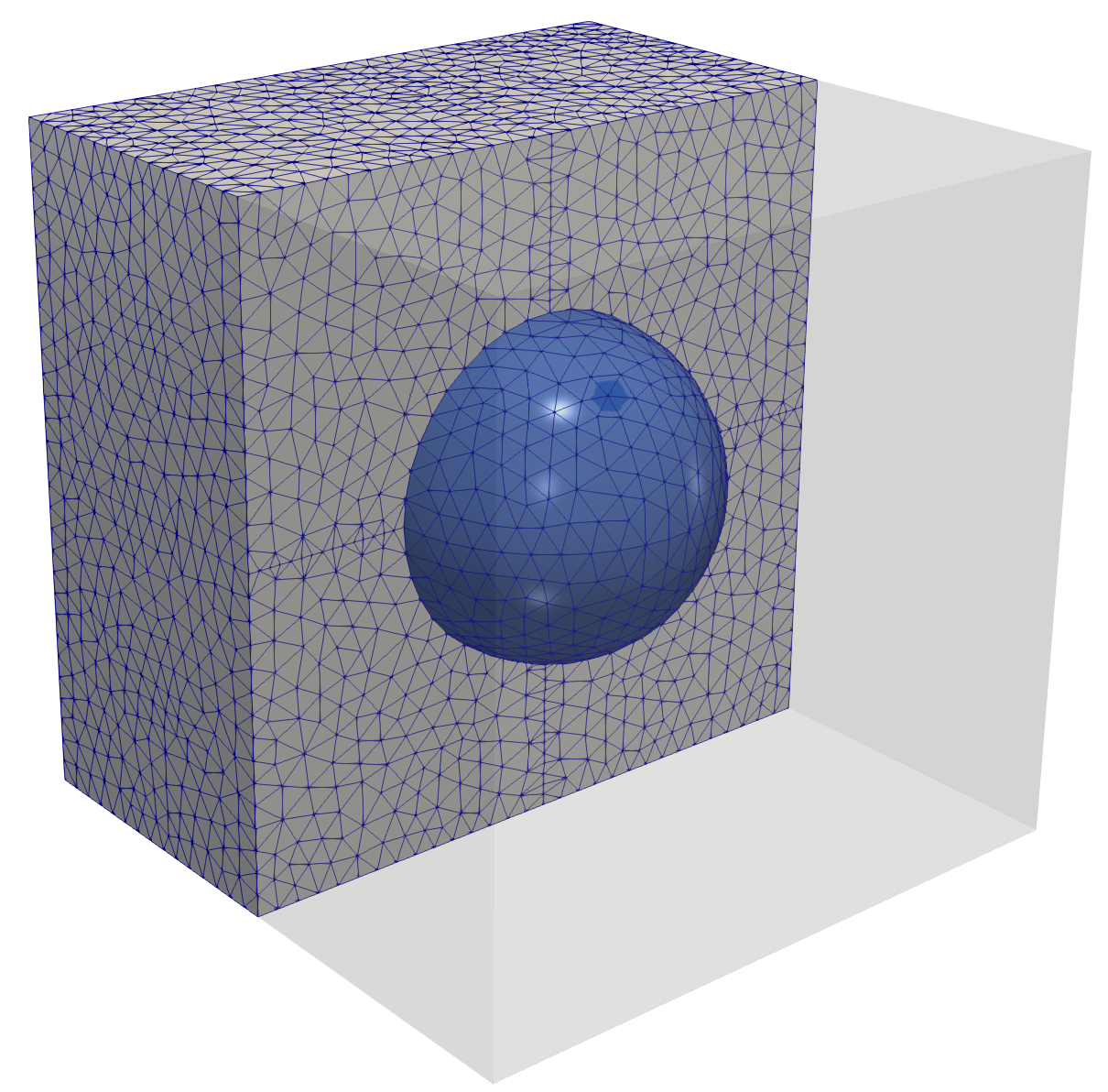}%
		\includegraphics[width=0.08\textwidth]{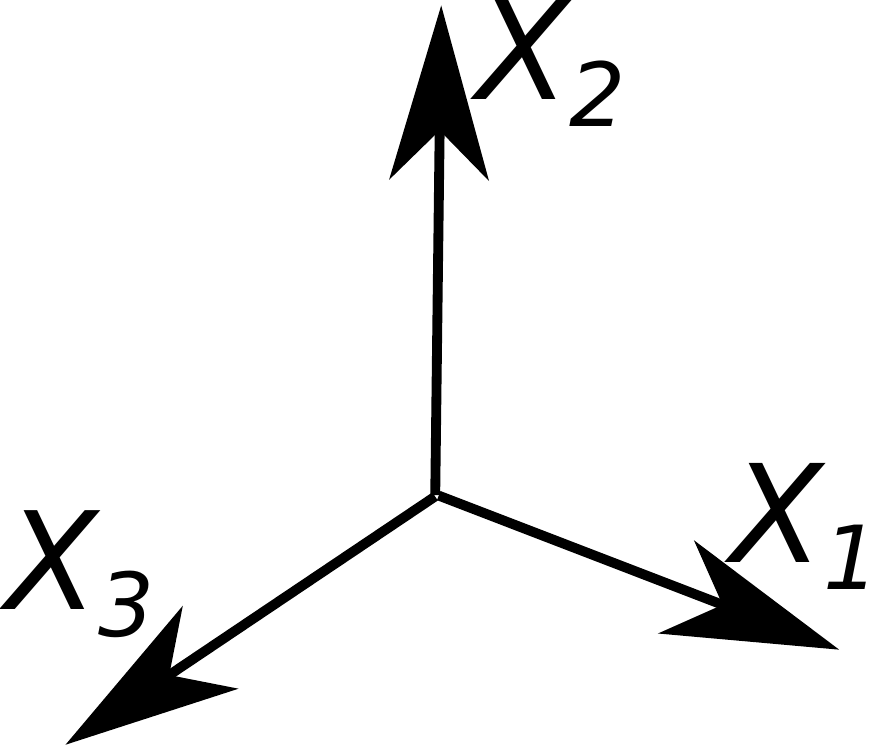}%
	}%
	\subfloat[$(\hat{\vect{P}}_{\mu})_{13}$ for the test case
        \label{fig:cubic cell:testP}]{%
		\includegraphics[width=0.282\textwidth]{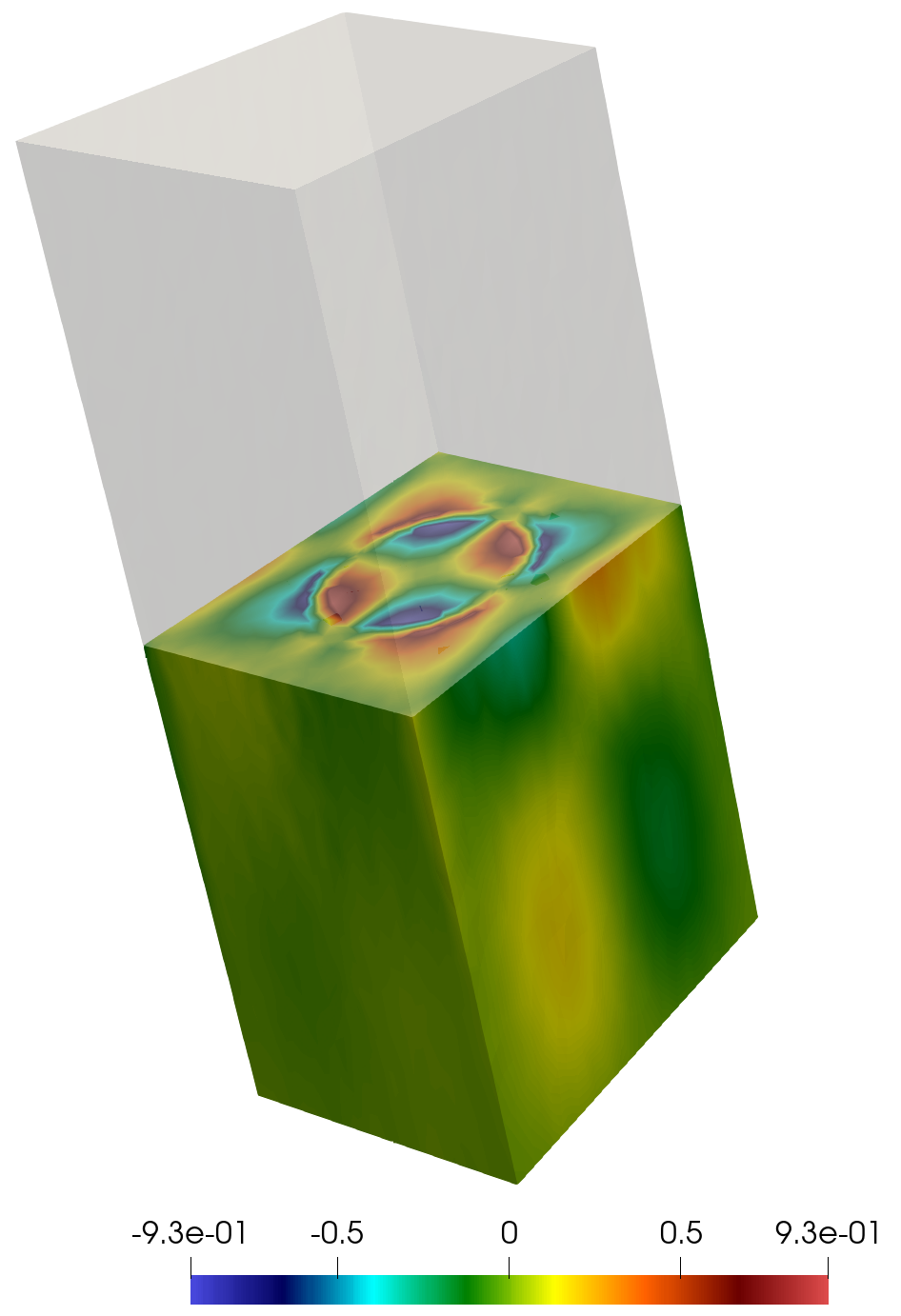}%
	}%
	\subfloat[$(\hat{\vect{e}}_{0_{\mu}})_3$ for the test case
        \label{fig:cubic cell:testE}]{%
		\includegraphics[width=0.30\textwidth]{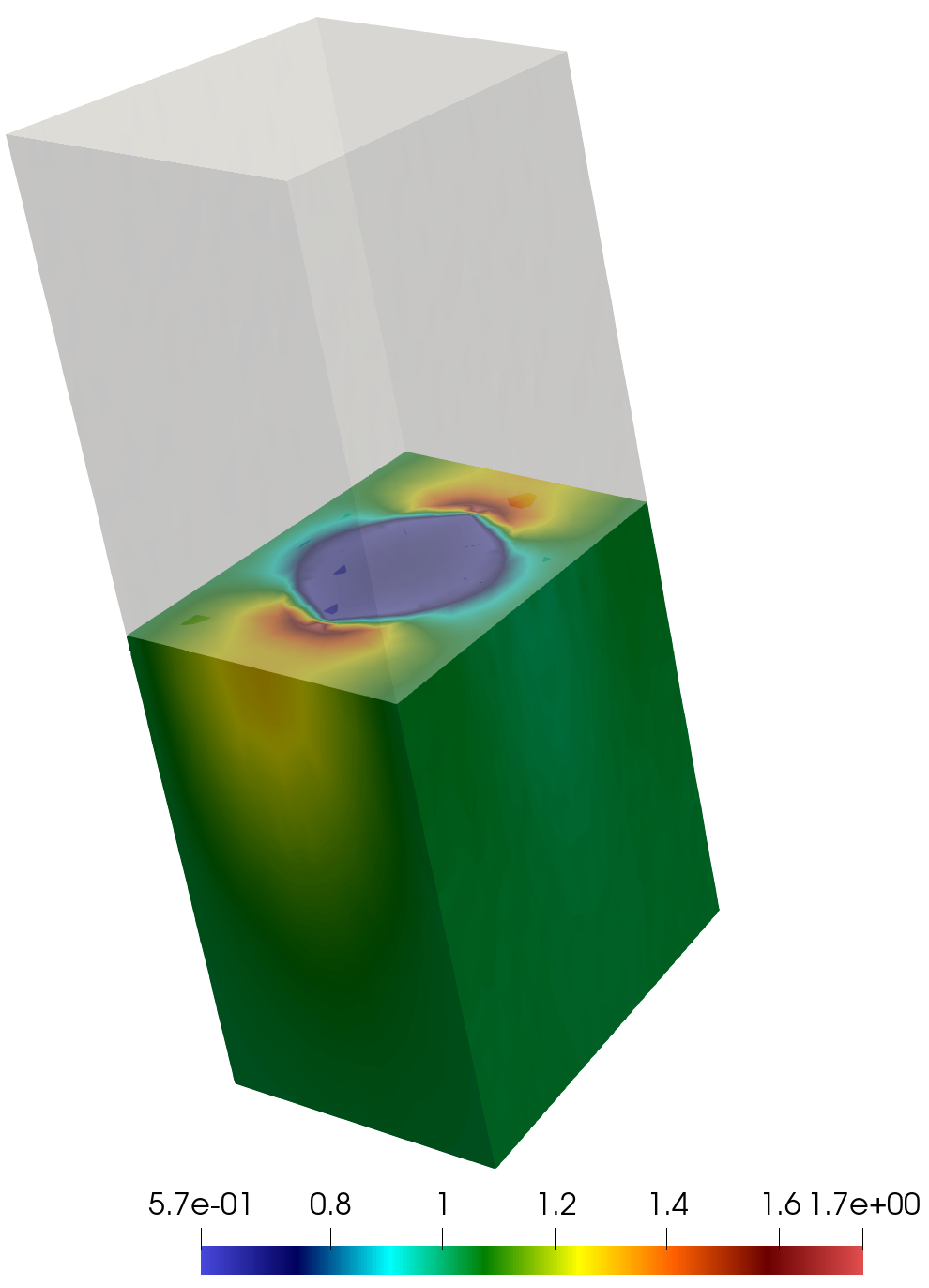}%
	}%
	\caption{Microstructure simulation of the cubic metamaterial. (a) For the the numerical homogenization, the RVE is discretized with quadratic tetrahedral finite elements. (b) and (c) show the highly deformed RVE for the test case at load step 1, i.e., for $\bF_{11}=0.5, {(\hat{d}_0)}_3=2$.}
%
	\label{fig:cubic cell Results}
\end{figure}

For the data generation, the following 6 load cases for the macroscopic fields $\vect{F}$ and $\bd_0$ are considered: (1) purely mechanical uniaxial tension, (2) purely mechanical simple shear, (3) the minimum mechanical value for uniaxial tension, with an electric displacement field applied in the $X_3$-direction, (4) the maximum mechanical value for uniaxial tension, with an electric displacement field applied in the $X_3$-direction, (5) the minimum mechanical value for shear load, with an electric displacement field applied in the $X_1$-direction, and (6) and the maximum mechanical value for shear load, with an electric displacement field applied in the $X_1$-direction.
With each single load case consisting of 100 data points, this results in an overall calibration dataset size of $600$ data points.
%
In addition, a complex test case is generated, which is illustrated in \cref{fig:cubic cell Results,fig:cub_eval}.

The finite element discretization of the RVE geometry in \cref{fig:cubic cell}, used in order obtain solution of the micro-fluctuation fields (refer to \cref{sec:RVE homogenization}) is shown in \cref{fig:cubic cell Results}(a). 
While \cref{fig:cub_eval} shows the load paths and evaluation of the test case, which will be discussed in detail below in \cref{sec:cubic cell:eval}, \cref{fig:cubic cell:testP,fig:cubic cell:testE} visualize the deformed state at load step 1, i.e., for $\bF_{11}=0.5, {(\hat{\bd}_0)}_3=2$ (see \cref{sec:scaling} for the definition of the scaled fields). 
Specifically, the microscopic (scaled) first Piola-Kirchhoff stress tensor $(\hat{\vect{P}}_{\mu})_{13}$ and the microscopic (scaled) electric field $(\hat{\vect{e}}_{0_{\mu}})_3$ are visualized over half of the RVE in its deformed configuration. 
For both stress and electric field, highly varying patterns can be observed, which illustrates the complexity of the behavior of the microstructure.

\subsection{Model calibration}

For the invariant-based ANN model proposed in \cref{sec:model}, we use the input
\begin{equation}
    \boldsymbol{\cI}=\left(I_1,\,I_2,\,J,\,-J,\,\hat{I}_4,\,\hat{I}_5,\,J_1^{\text{cub}},\,J_2^{\text{cub}},\,\hat{J}_3^{\text{cub}}\right)\in\bbR^{9}\,,
\end{equation}
with the invariants as described in \cref{sec:objectivity and material symmetry} and the scaling as described in \cref{sec:scaling}. 
Again, we choose two ICNN network architectures: one consisting of only one hidden layer with 8 nodes ($H=1, n^{[1]}=8$) and one consisting of two hidden layers with 16 nodes in each layer ($H=2, n^{[h]}=16$), for which we use the short notations $\cS\cP^+(  \boldsymbol{\cI};\,8)$ and $\cS\cP^+(  \boldsymbol{\cI};\,16,16)$, respectively. The former has a total amount of 89 free parameters, the latter has 449 free parameters.
Each model architecture was initialized three times, using different randomly initialized model parameters each time. The calibration is carried out for $2,500$ epochs.
The full batch of training data is used with TensorFlow's default learning rate and default batch size.
For the remaining calibration details, see \cref{sec:calib}.

\subsection{Model evaluation}
\label{sec:cubic cell:eval}

\begin{table}[t!]
\centering
\begin{tabular}{llcll}
\toprule
Architecture & \multicolumn{2}{l}{$\log_{10}$ MSE}  \\
&calibration&test\\ \midrule
$\cS\cP^+(  \boldsymbol{\cI};\,8)$  & -3.98 & -2.82  \\
---\textquotedbl---& -4.02 & -2.74  \\
---\textquotedbl---&  -3.97 & -2.72 \\
$\cS\cP^+(  \boldsymbol{\cI};\,16,16)$   & -4.10 & -2.76 \\
---\textquotedbl---& -4.12& -2.76 \\
---\textquotedbl---&  -4.07& -2.70 \\
\bottomrule
\end{tabular}
\caption{MSEs of the calibrated ANN based models for the cubic metamaterial}
\label{tb:loss_cub}
\end{table}

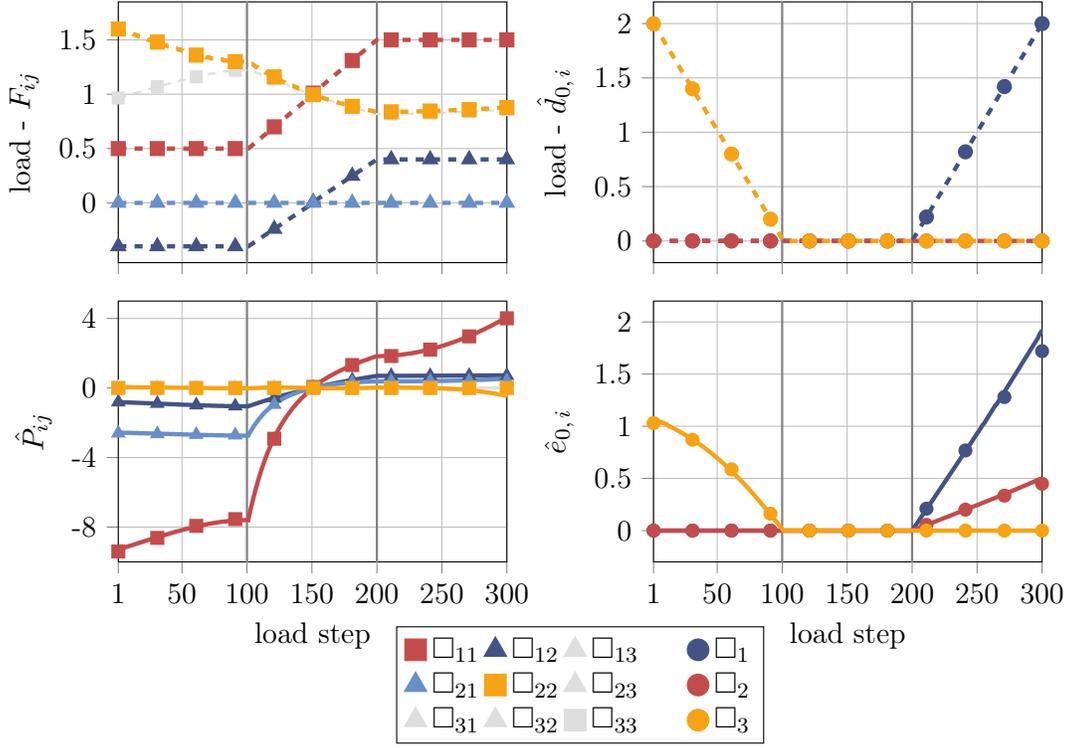
\begin{figure}[t!]
\centering

\resizebox{0.9\textwidth}{!}{
\tikzsetnextfilename{cub_eval}

\pgfplotstableset{
create on use/time_steps/.style={create col/copy column from table={plot_data/cub/load_step.txt}{0}}
}

\begin{tikzpicture}
\pgfplotsset{set layers}
\begin{groupplot}[
	group style = {group size = 2 by 2, vertical sep = 0.5 cm,
								horizontal sep = 0.115*\textwidth},
	cycle list name=mycolorlist_cub,
	xtick align = outside,
	ytick align = outside,
	xtick pos = left,
	]
	
		    \nextgroupplot[legend to name = grouplegend, legend columns=8, ylabel = {load - $F_{ij}$}, ytick={0,0.5,1,1.5},yticklabels={0,0.5,1,1.5},
width=0.4*\textwidth, height=0.3*\textwidth,
ymin = -0.55, ymax = 1.85,
xmin = 1, xmax = 300,
ytick pos = left,
grid = major,
xtick={1,50,100,150,200,250,300},xticklabels={,,,,,},
xtick pos = left,
] 

\draw[draw = gray, thick] (100, -1) -- (100, 2);
\draw[draw = gray, thick] (200, -1) -- (200, 2);

\foreach \num in {11, 12, 13}{
 \addplot coordinates { (-2,-2) (-3,-3) };
  \addlegendentryexpanded{$\square_{\num}$}
 };
 \addlegendimage{empty legend}\addlegendentry{}
 \addlegendimage{empty legend}\addlegendentry{}
 \addlegendimage{empty legend}\addlegendentry{}
 \addlegendimage{empty legend}\addlegendentry{}
   \addplot coordinates { (-2,-2) (-3,-3) }; \addlegendentryexpanded{$\square_{1}$}
\foreach \num in {21, 22, 23}{
 \addplot coordinates { (-2,-2) (-3,-3) };
 \addlegendentryexpanded{$\square_{\num}$}
 };
 \addlegendimage{empty legend}\addlegendentry{}
 \addlegendimage{empty legend}\addlegendentry{}
 \addlegendimage{empty legend}\addlegendentry{}
  \addlegendimage{empty legend}\addlegendentry{}
    \addplot coordinates { (-2,-2) (-3,-3) }; \addlegendentryexpanded{$\square_{2}$}
    \foreach \num in {31, 32, 33}{
 \addplot coordinates { (-2,-2) (-3,-3) };
 \addlegendentryexpanded{$\square_{\num}$}
 };
 \addlegendimage{empty legend}\addlegendentry{}
 \addlegendimage{empty legend}\addlegendentry{}
 \addlegendimage{empty legend}\addlegendentry{}
  \addlegendimage{empty legend}\addlegendentry{}
    \addplot coordinates { (-2,-2) (-3,-3) }; \addlegendentryexpanded{$\square_{3}$}

        \foreach \num in {2,5,6,7,8,0,3,1,4}{
    \addplot table [x =time_steps,y index=\num] 
    {plot_data/cub/F.txt};
    };


	    \nextgroupplot[legend pos = outer north east, legend columns = 8, ylabel = {load - $\hat{d}_{0,\,i}$}, ytick={0,0.5,1,1.5,2},yticklabels={0,0.5,1,1.5,2},
width=0.4*\textwidth, height=0.3*\textwidth,
ymin = -0.2, ymax = 2.2,
xmin = 1, xmax = 300,
ytick pos = left,
grid = major,
xtick={1,50,100,150,200,250,300},xticklabels={,,,,,}]

\draw[draw = gray, thick] (100, -1) -- (100, 5);
\draw[draw = gray, thick] (200, -1) -- (200, 5);

\foreach \num in {1,2,3,4,5,6,7,8,9,10,11,12}{
 \addplot coordinates { (-2,-2) (-3,-3) };
 }
 
 \foreach \num in {1,2,3,4,5,6,7,8,9}{
 \addplot coordinates { (-2,-2) (-3,-3) };
 };

        \foreach \num in {0,1,2}{
    \addplot table [x =time_steps,y index=\num] 
    {plot_data/cub/D.txt};
    };
    
    \nextgroupplot[legend style={at={(axis cs: 1,2.5)},anchor=north west}, legend columns=8, ylabel = {$\hat{P}_{ij}$}, ytick={-8,-4,0,4},yticklabels={-8,-4,0,4},
width=0.4*\textwidth, height=0.3*\textwidth,
ymin = -10, ymax = 5,
xmin = 1, xmax = 300,
ytick pos = left,
grid = major,
xlabel={load step},
xtick={1,50,100,150,200,250,300},xticklabels={1,50,100,150,200,250,300}]

\draw[draw = gray, thick] (100, -20) -- (100, 20);
\draw[draw = gray, thick] (200, -20) -- (200, 20);
 
\foreach \num in {1,2,3,4,5,6,7,8,9,10,11,12,13,14,15,16,17,18,19,20,21,22,23,24}{
 \addplot coordinates { (-2,-2) (-3,-3) };
 };

        \foreach \num in {2,5,6,7,8,0,3,1,4}{
    \addplot table [x =time_steps,y index=\num] 
    {plot_data/cub/P.txt};
    };
        \foreach \num in {2,5,6,7,8,0,1,3,4}{
    \addplot table [x =time_steps, y index=\num] {plot_data/cub/P_m.txt};
    };
    
    \nextgroupplot[legend pos = north west,legend columns=1, ylabel = {$\hat{e}_{0,\,i}$}, ytick={0,0.5,1,1.5,2},yticklabels={0,0.5,1,1.5,2},
width=0.4*\textwidth, height=0.3*\textwidth,
ymin = -0.3, ymax = 2.2,
xlabel={load step},
xmin = 1, xmax = 300,
ytick pos = left,
grid = major,
xtick={1,50,100,150,200,250,300},xticklabels={1,50,100,150,200,250,300}]

\draw[draw = gray, thick] (100, -1) -- (100, 5);
\draw[draw = gray, thick] (200, -1) -- (200, 5);
 
\foreach \num in {1,2,3,4,5,6,7,8,9,10,11,12,13,14,15,16,17,18,19,20,21,22,23,24,25,26,27,28,29,30,31,32,33,34,35,36,37,38,39,40,41,42}{
 \addplot coordinates { (-2,-2) (-3,-3) };
 };

    \foreach \num in  {0, 1, 2}{
    \addplot table [x =time_steps,y index=\num, only marks] 
    {plot_data/cub/E.txt};
    };
        \foreach \num in {0,1,2}{
    \addplot table [dashed, x =time_steps,y index=\num,  mark=none] {plot_data/cub/E_m.txt};
    };

\end{groupplot}
\node at ($(group c1r2.east)!0.5!(group c2r2.west) + (0 cm, -3.3 cm) $) {\ref{grouplegend}}; 

\end{tikzpicture}
}
\caption{Evaluation of the test case for the cubic metamaterial. Points depict homogenization data, while lines depict the evaluation of the ANN based models. All quantities are dimensionless.}
\label{fig:cub_eval}
\end{figure}

In \cref{tb:loss_cub}, the MSEs of the different model initializations are provided. Both architectures show excellent results for the calibration dataset with slightly worse MSEs for the test dataset. Overall, the model with the smaller network architecture is sufficiently flexible and using a bigger architecture does not improve the results.

In \cref{fig:cub_eval},  the evaluation of the test case is shown. For the load steps 101 to 200, the load is purely mechanical, with a deformation gradient consisting of both normal and shear components, thus representing a general deformation state. For both the minimum and the  maximum of the applied deformation gradient, an additional electric displacement field is applied in different directions in load steps 1 to 100 and 201 to 300, respectively.
The prediction of the model for the stress tensor is excellent, both for purely mechanical and electro-mechanical load cases. Also, the prediction of the electrical field is excellent, with only slight visible deviations.

Once again, this example demonstrates how the physics-augmented neural network constitutive model is able to represent the coupling between electrical and mechanical effects, and furthermore shows the straightforward applicability to arbitrary symmetry groups.

\section{Conclusion}
\label{sec:conc}

In the present work, a machine learning based constitutive model for reversible electro-mechanically coupled material behavior at finite deformations is proposed. Using different sets of invariants as inputs, an electro-elastic internal energy density is formulated as a convex neural network. In this way, the model fulfills the polyconvexity condition which ensures material stability, as well as thermodynamic consistency, objectivity, material symmetry, and growth conditions. Depending on the considered invariants, this physics-augmented machine learning model can either be applied for compressible or nearly incompressible material behavior, as well as for arbitrary material symmetry classes.

\medskip

In \cref{sec:ana_pot}, the model is calibrated to data generated with an analytical potential, where the input space $\bF\in\text{GL}^+(3)$, $\bd_0\in\bbR^3$ is sampled using an algorithm proposed by \textcite{kunc2019}. 
Using this example, the trade-off between structure and flexibility is discussed, which is inherent to constitutive modeling. 
First of all, including constitutive requirements in the model formulation leads to reliable, i.e., physically sensible model predictions. This can be achieved, for example, by using invariants.
But more than that, including constitutive requirements provides the model a pronounced mathematical structure, which helps to generalize with moderate amounts of data. This is especially important for neural networks, which would otherwise require a large amount of data for the calibration.
In that sense, constitutive conditions can be seen as an \emph{inductive bias} \cite{haussler1988}.
On the other side, including constitutive requirements possibly leads to a loss of flexibility, e.g., when for a given anisotropy no complete basis in (polyconvex) invariants can be constructed.
However, the restrictions made in this work are either physically well based, e.g., the use of an internal energy which ensures the second law of thermodynamics, or they have a strong mathematical motivation, such as the polyconvexity condition which ensures material stability. Therefore, they should not be seen as strong assumptions on the model behavior, rather they lead to reliable model predictions, while also improving the generalization properties of the model.

\medskip

The applicability and versatility of this physics-augmented neural network constitutive model is demonstrated by calibrating it to two more datasets.
In \cref{sec:R1}, the model is applied for the effective constitutive modeling of an analytically homogenized, transversely isotropic rank-one laminate.
In \cref{sec:cub}, a cubic metamaterial is numerically homogenized, and the generated data is then used to calibrate the effective material model.
In both cases, the calibration datasets have a fairly small size, and the model predictions are excellent. This shows both the excellent flexibility of the model, which is required to capture the complex behavior of electro-mechanically coupled microstructures, as well as the straightforward applicability towards arbitrary symmetry groups.

\medskip

Considering the infinitely continuously differentiable neural network cores and the capability of automatic differentiation, an implementation of the proposed model into a finite element framework will be straightforward, c.f.~\cite{kalina2022}. With the material stability condition fulfilled by construction, it can be expected to have a favorable numerical behavior, c.f.~\cite{asad2022}.
Thus, in future work we aim to apply the physics-augmented machine learning models for macroscopic finite element simulations of nonlinear EAP composites and metamaterials. 
Furthermore, due to the similar mathematical structure, the extension of the model towards electro-magneto-mechanically coupled material behavior \cite{bustamate2021} should be relatively straightforward.

\vspace*{3ex}

\noindent
\textbf{Conflict of interest.} The authors declare that they have no conflict of interest.
\vspace*{1ex}

\noindent
\textbf{Acknowledgments.} 
Dominik K.\ Klein and Oliver Weeger acknowledge funding from the Deutsche Forschungsgemeinschaft (DFG – German Research Foundation) -- grant no.~460684687, and support by the Graduate School CE within the Centre for Computational Engineering at the Technical University of Darmstadt. 
Rogelio Ortigosa acknowledges the financial support through the contract 21132/SF/19, Fundación S\'eneca,
Regi\'on de Murcia (Spain), through the program Saavedra Fajardo. Rogelio Ortigosa and Jes\'us Mart\'inez-Frutos are funded by
Fundaci\'on S\'eneca (Murcia, Spain) through grant 20911/PI/18. They  also acknowledge the
financial support through grant PID2021-125687OA-I00.

\vspace*{1ex}

\noindent
\textbf{Data availability.}
The authors provide access to the simulation data required to reproduce the results through the public GitHub repository \url{https://github.com/CPShub/sim-data}.

\appendix
\section*{Appendix}
\numberwithin{equation}{section}

\section{Scaling of internal energy density functions}\label{sec:scaling}

This section describes the scaling procedure used in this work for the definition of the various internal energy functions used throughout Sections \ref{sec:ana_pot}, \ref{sec:R1} and \ref{sec:cub}. This procedure will be initially described and particularized for the analytical transversely isotropic potential shown in Section \ref{sec:ana_pot}, and generalized afterwards to the other two cases addressed in Sections \ref{sec:R1} and \ref{sec:cub}.

\paragraph{Transversely isotropic model.}
Typically, in an electro-mechanical constitutive model such as the one defined in \cref{eqn:ti model}, the difference between the mechanical parameters $\mu_1,\mu_2,\mu_3,\lambda$ and the electrical parameters $\varepsilon_1,\varepsilon_2$ is of several orders of magnitude. 
This scaling difference can entail numerical difficulties for the construction and calibration of a suitable neural network capturing both underlying physics, namely mechanical and electro-mechanical. 
With the aim of circumventing this potential drawback, for the particular model in \cref{eqn:ti model} we define the scaled internal energy density $\hat{e}^{\text{ti}}(\vect{F},\hat{\vect{d}}_0)$, in terms of the scaled electric displacement field $\hat{\vect{d}}_0$, as
\begin{equation}\label{eqn:scaled_ti}
	\hat{e}^{\text{ti}}(\hat{\vect{\cU}}) 
	= \hat{e}^{\text{ti}}(\vect{F},\hat{\vect{d}}_0)
	= \frac{1}{\mu_1}e(\vect{F},\vect{d}_0)\qquad \text{with}\qquad
		\hat{\vect{d}}_0=\frac{\vect{d}_0}{\sqrt{\mu_1\varepsilon_1}}, \quad \hat{\vect{\cU}}=\big\{\bF,{\hat\bd}_0\big\}\,.
\end{equation}
Note that the components of $\bF$ are dimensionless and $\cO(1)$, thus $\bF$ does not need to be scaled.

The scaled internal energy in \cref{eqn:scaled_ti} admits the same additive decomposition as in \cref{p_ti}, namely
%
%
\begin{equation}\label{p_ti_scaled}
	\begin{aligned}
		\hat{e}^{\text{ti}}(\hat{\vect{\mathcal{U}}})
		=\hat{\mathcal{P}}(\hat{\vect{\mathcal{V}}})
		=\hat{\widetilde{\mathcal{P}}}(\hat{\bar{\vect{\mathcal{I}}}}^{\text{iso}}) + \hat{\widetilde{\mathcal{P}}}(\hat{\vect{\mathcal{J}}}^{\text{ti}})+\hat{e}_{\text{vol}}(J)\,,
	\end{aligned}
\end{equation}
where $\hat{\vect{\cV}}=\big\{\bF,\bH,J,{\hat\bd}_0,\hat\bd\big\}$ with $\hat\bd=\bF{\hat\bd}_0$. Each scaled contribution is defined as
\begin{equation}
\begin{aligned}
\hat{\widetilde{\mathcal{P}}}(\hat{\bar{\vect{\mathcal{I}}}}^{\text{iso}})	&=\frac{\bar{I}_1}{2} + \frac{\mu_2}{2\mu_1}\bar{I}_2 - \frac{\mu_3}{\mu_1}\log J + \frac{\hat{I}_5}{2J}\,,\\   
\hat{\widetilde{\mathcal{P}}}(\hat{{\vect{\mathcal{J}}}}^{\text{ti}})&=
\frac{\mu_3}{2\mu_1}\left(\frac{\left(J^{\text{ti}}_1\right)^{a_1}}{a_1}+\frac{\left(J^{\text{ti}}_2\right)^{a_2}}{a_2}\right)   +\frac{\varepsilon_1}{2\varepsilon_2}\hat{J}_3^{\text{ti}}\,,\\
\hat{e}_{\text{vol}}(J)&=\frac{\lambda}{2\mu_1}\left(J-1\right)^2\,,
\end{aligned}
\end{equation}
where $\hat{\bar{\vect{\mathcal{I}}}}^{\text{iso}}$ and $\hat{\vect{\mathcal{J}}}^{\text{ti}}$ are comprised of scaled (dimensionless) invariants,
\begin{equation}
		\hat{\bar{\vect{\mathcal{I}}}}^{\text{iso}}=\{\bar{I}_1,\bar{I}_2,\hat{I}_4,\hat{I}_5,J\},\qquad
		\hat{\vect{\mathcal{J}}}^{\text{ti}}=\{J_1^{\text{ti}},J_2^{\text{ti}},\hat{J}_3^{\text{ti}}\},
\end{equation} 
with the new scaled invariants $\{\hat{I}_4,\hat{I}_5,\hat{J}_3^{\text{ti}},\hat{J}_4^{\text{ti}}\}$ defined as
\begin{equation}
\hat{I}_4=\|\hat{\vect{d}}_0\|^2\,,\qquad \hat{I}_5=\|\hat{\vect{d}}\|^2\,,\qquad 
\hat{J}^{\text{ti}}_3=\text{tr}\Big(\left(\hat{\vect{d}}_0\otimes\hat{\vect{d}}_0\right)\,\vect{G}^{\text{ti}}\Big)\,, \qquad
\hat{J}^{\text{ti}}_4 = \text{tr}\Big(\left(\hat{\bd}\otimes\hat{\bd}\right)\,\vect{G}^{\text{ti}}\Big)\,.
\end{equation}

The scaled (dimensionless) material electric field $\hat{\vect{e}}_0$ and the true material electric field $\vect{e}_0$ can be obtained from their respective associated energy densities according to \cref{eq:P_e}, i.e.
\begin{equation}
	\hat{\vect{e}}_0:=\frac{\partial \hat{e}(\vect{F},\hat{\vect{d}}_0)}{\partial\hat{\vect{d}}_0}\,,\qquad 
		{\vect{e}}_0:=\frac{\partial {e}(\vect{F},\vect{d}_0)}{\partial{\vect{d}}_0}\,,
\end{equation}
which, making use of \cref{eqn:scaled_ti}, permits to establish the relationship between both fields as
\begin{equation}
	\hat{\vect{e}}_0=\sqrt{\frac{\mu_1}{\varepsilon_1}}\vect{e}_0\,.
\end{equation}
Similarly, the scaled (dimensionless) first Piola-Kirchhoff stress tensor $\hat{\vect{P}}$ and the true counterpart $\vect{P}$ can be obtained from their respective associated energy densities according to \cref{eq:P_e}, i.e.
\begin{equation}
	\hat{\vect{P}}:=\frac{\partial \hat{e}(\vect{F},\hat{\vect{d}}_0)}{\partial{\vect{F}}}\,,\qquad 
		{\vect{P}}:=\frac{\partial {e}(\vect{F},\vect{d}_0)}{\partial{\vect{F}}}\,,
\end{equation}
which, making use of \cref{eqn:scaled_ti}, permits to establish the relationship between both fields as
\begin{equation}
	\hat{\vect{P}}={\frac{1}{\mu_1}}\vect{P}\,.
\end{equation}

\paragraph{Rank-one laminate.}
With regard to the internal energies of each of the phases $a$ and $b$ of the rank-one laminate composite described in \cref{sec:R1}, c.f.~\cref{eqn:the model for ROL}, we used the following scaled internal energies with respect to the material parameters $\mu_1$ and $\varepsilon$ of the phase $a$, namely
%
%
\begin{equation}\label{eqn:the model for ROL scaled}
\begin{aligned}
\hat{\widetilde{\mathcal{P}}}^{a}(\hat{\vect{\mathcal{I}}}^{a}) &=
\frac{1}{2}{{I}}^{a}_{1} + \frac{\mu_2}{2\mu_1}{{I}}_2^{a} + \frac{\hat{I}^a_5}{2 J^a}\,,
\qquad & 
\hat{\mathcal{P}}_{\text{vol}}^{a}(J^{a}) &=
\frac{\lambda}{2\mu_1}(J^a-1)^2\,,
\\
\hat{\widetilde{\mathcal{P}}}^{b}(\hat{\vect{\mathcal{I}}}^{b}) &=		f_m\Big(\frac{1}{2}{{I}}^b_{1} + \frac{\mu_2}{2\mu_1}{{I}}^b_{2}\Big) + \frac{\hat{I}^b_{5}}{2 f_e J^b}\,,
\qquad & 
\hat{\mathcal{P}_{\text{vol}}^{b}}(J^{b}) &=
f_m\frac{\lambda}{2\mu_1}(J^b-1)^2\,,
\end{aligned}	
\end{equation}
with 
\begin{equation} \label{eqn:the model for ROL scaled invar}
    \hat{I}_5^{a,b}=\big\|\hat{\vect{d}}^{a,b}\big\|^2\,,\qquad 
    \hat{\vect{d}}^{a,b}=\vect{F}^{a,b}\,\hat{\vect{d}}_0^{a,b}\,, \qquad
    \hat{\vect{d}}_0^{a,b}=\frac{\vect{d}_0^{a,b}}{\sqrt{\mu_1\varepsilon}}\,.
\end{equation}


\paragraph{Cubic metamaterial.}
An identical scaling procedure in terms of the parameters $\mu_1,\varepsilon$ of the matrix, analogous to \cref{eqn:the model for ROL scaled,eqn:the model for ROL scaled invar}, is applied to the internal energies of the matrix and inclusion material phases within the RVE of the cubic metamaterial in Section \ref{sec:cub}.

\section{Concentric sampling strategy} \label{sec:F_sampling}

In order to create data of the global behavior of the analytical potential in \cref{sec:ana_pot}, c.f.~\cref{p_ti}, the sampling of the input space $\bF\in\text{GL}^+(3)$ and $\bd\in\bbR^3$ in a wide range is now described.
Uniform sampling of $\text{GL}^+(3)$ and $\bbR^3$ is anything but practical, as it would quickly exceed a reasonable dataset size while possibly containing large deformations, potentially outside a relevant range. Therefore, we choose the sampling strategy proposed by \textcite{kunc2019}, which is now shortly described. 

At first, the polar decomposition $\F=\R\,\U$
is applied to the deformation gradient, where $\R\in\SO(3)$ is the rotation tensor and $\U$ is the symmetric positive definite (s.p.d.) right stretch tensor \cite[Section 2.6]{Holzapfel2000}. The objectivity condition implies that constitutive models must be independent of $\R$.
Therefore, it is sufficient to sample $\U$, on which the isochoric-volumetric decomposition $\U=J^{\nicefrac{1}{3}}\,\Bar{\U}$ is applied \cite[Section 6.4]{Holzapfel2000}. In doing so, the volume ratio $J$ of the deformation and the isochoric
part of the deformation $\Bar{\U}$ can be sampled independent of each other. This makes it easier to generate stretch tensors $\U$ in a practically relevant range. 

As $J\in\bbR^+$, it is straightforward to sample the volume ratio by choosing sensible values around $J=1$, depending on the compressibility of the considered material. As we assume nearly incompressible material behavior, we set $J=1$.
For the isochoric part, the unimodular
s.p.d.\ tensor space in which $\Bar{\U}$ lies has to be sampled.
After some Lie group theory, this can be formulated equivalently to sampling a five-dimensional linear space.
Altogether, the strategy proposed by \textcite{kunc2019} provides a straightforward generation of uniformly distributed stretch tensors $\U$. Still, it should be noted that uniformly distributed stretch tensors $\U$ may not be optimal, as this approach does not take the material symmetry into account and may yield redundant information. However, as the optimal sampling of deformation spaces for specific symmetry groups is not a straightforward task, we pragmatically use this algorithm. 
Here, for the data generation in \cref{sec:ana_pot}, we employ 30 deviatoric directions and 50 deviatoric amplitudes between 0.1 and 1 for the sampling of $\U$.

\medskip

The sampling of the scaled electric displacement field $\hat{\bd}_0$ in the Euclidian vector space $\bbR^3$ is much more straightforward. 
Here, we sample by using (nearly) equidistant points on a unit sphere and scaling these unit vectors with different amplitudes.
In particular, we use 20 equidistant vectors on a unit sphere, which are scaled by 50 amplitudes between 0 and 4.

\medskip

Altogether, this results in a fairly large dataset with 1.5 million points. 
As it is shown in \cref{sec:ana_pot}, such a large dataset is not necessary to calibrate (physics-augmented) machine learning models, however, it is useful to examine some specific model properties.

\section{Analytical homogenization of rank-one laminated composites}\label{sec:ROL_appendix}

\subsection{Preliminaries}

Let us consider an electro-active material whose microstructure corresponds to a rank-one laminate composite as introduced in \cref{sec:R1} and illustrated in \cref{fig: normal vector}, c.f.~\cite{marin2021}.
Now, the macroscopic deformation gradient $\vect{F}$ and electric displacement field $\vect{d}_0$ are defined as the weighted sum of those in each phase, namely
\begin{equation}\label{eqn:homogenisation}
\vect{F} = c^{a}\vect{F}^{a}+c^{b}\vect{F}^{b}\,,
\qquad
\vect{d}_0 = c^{a}\vect{d}_0^{a}+c^{b}\vect{d}_0^{b}\,,
\end{equation}
where indices $a$ and $b$ are used to indicate the respective microscale phases of the laminate. Strong continuity of the tangential and normal components of $\vect{F}$ and $\vect{d}_0$ across the laminate interface entails
\begin{equation}\label{eqn:interface conditions}
\begin{aligned}
\llbracket\vect{F}\rrbracket\Cross \vect{l}_0&=\vect{0}\,,&\qquad
\llbracket\vect{d}_0\rrbracket\cdot\vect{l}_0&=0\qquad \text{with}\qquad \llbracket\bullet\rrbracket =(\bullet)^a-(\bullet)^b\,.
\end{aligned}
\end{equation}
These conditions can alternatively be written in a more convenient form as
\begin{equation}\label{eqn:interface conditions II}
\begin{aligned}
\llbracket\vect{F}\rrbracket&=\vect{\alpha}\otimes\vect{l}_0\,,&\qquad
\llbracket\vect{d}_0\rrbracket&=\vect{T}_{\vect{l}_0}\vect{\beta}\,,\\
\end{aligned}
\end{equation}
where $\vect{\alpha} \in \mathbb{R}^3$ denotes the microscale deformation gradient amplitude vector and ${\vect{\beta}} \in \mathbb{R}^2$ the microscale electric displacement amplitude.  $\vect{T}_{\vect{l}_0}\in\bbR^{3\times 2}$ is defined as 
$\vect{T}_{\vect{l}_0}=\vect{t}_{0_1}\otimes \vect{e}^{\star}_1 + \vect{t}_{0_2}\otimes \vect{e}^{\star}_2$,
with $\vect{t}_{0_1}$ and $\vect{t}_{0_2}$ being two vectors orthonormal to $\vect{l}_0$ and to each other, and with the 2D unit vectors $\vect{e}^{\star}_1=\begin{pmatrix}
 1&0
 \end{pmatrix}^T$ and $\vect{e}^{\star}_2=\begin{pmatrix}
 0&1
 \end{pmatrix}^T$. 
Combination of \cref{eqn:homogenisation} and \cref{eqn:interface conditions II} permits to obtain  $\vect{F}^{a,b}$ and  $\vect{d}_0^{a,b}$ as
\begin{equation}\label{eqn:homogenisation rules}
	\begin{aligned}
		\vect{F}^{a}(\vect{F},\vect{\alpha})&=\vect{F}+c^{b}\vect{\alpha}\otimes\vect{l}_0\,,&\qquad 
		\vect{d}_0^{a}(\vect{d}_0,\vect{\beta})&=\vect{d}_0+c^{b}\vect{T}_{\vect{l}_0}\vect{\beta}\,,\\
		\vect{F}^{b}(\vect{F},\vect{\alpha})&=\vect{F}-c^{a}\vect{\alpha}\otimes\vect{l}_0\,,&\qquad 
		\vect{d}_0^{b}(\vect{d}_0,\vect{\beta})&=\vect{d}_0-c^{a}\vect{T}_{\vect{l}_0}\vect{\beta}\,.
	\end{aligned}
\end{equation}

The homogenized or effective internal energy  of the composite $e(\vect{F},\vect{d}_0)$ can be postulated as
\begin{equation}\label{eqn:effective internal energy}
e\left(\vect{F},\vect{d}_0\right)=\min_{\vect{\alpha},\vect{\beta}}\left\{{e}^{\star}(\vect{F},\vect{d}_0,\vect{\alpha},\vect{\beta})\right\}\,,
\end{equation}
with 
\begin{equation}\label{eqn:effective contributions}
\begin{aligned}
{e}^{\star}(\vect{F},\vect{d}_0,\vect{\alpha},\vect{\beta})=c^{a} e^{a}\left(\vect{F}^{a}\left(\vect{F},\vect{\alpha}\right),\vect{d}_0^{a}\left(\vect{d}_0,\vect{\beta}\right)\right)
+c^{b} e^{b}\left(\vect{F}^{b}\left(\vect{F},\vect{\alpha}\right),\vect{d}_0^{b}\left(\vect{d}_0,\vect{\beta}\right)\right)\,,
\end{aligned}
\end{equation}
where $e^a(\vect{F}^a,\vect{d}_0^a)$ and $e^b(\vect{F}^b,\vect{d}_0^b)$ are the microscale internal energy functions expressed in terms of their respective microscale fields, c.f.~\cref{eqn:the model for ROL}. 
The stationary conditions of \cref{eqn:effective contributions} with respect to $\vect{\alpha},\vect{\beta}$ yield
\begin{equation}\label{eqn:jump_P_E}
\begin{aligned}
D{e}^{\star}[\delta\vect{\alpha}]&=0 \qquad\Rightarrow&\qquad 	\llbracket\vect{P}\rrbracket \vect{l}_0&=\vect{0}\,, \\
D{e}^{\star}[\delta\vect{\beta}]&=0 \qquad\Rightarrow&\qquad \vect{T}_{\vect{l}_0}^T \llbracket\vect{e}_0\rrbracket &=\vect{0}
\qquad\Leftrightarrow\qquad \llbracket\vect{e}_0  \rrbracket \times \vect{l}_0 =\vect{0}\,,
\end{aligned}
\end{equation}
which represent the strong enforcement of the normal and tangential components of the traction vector and $\vect{e}_0$, respectively.
Thus, the homogenized energy functional $e(\vect{F},\vect{d}_0)$ can be re-written as
\begin{equation}\label{eqn:equivalenc energies ROL}
e(\vect{F},\vect{d}_0)=
{e}^{\star}(\vect{F},\vect{d}_0,\vect{\alpha},\vect{\beta})\big|_{\text{s.t. }\{\llbracket\vect{P}\rrbracket \vect{l}_0=\vect{0},\; \llbracket\vect{e}_0\rrbracket \times \vect{l}_0 =\vect{0}\}}\,.
\end{equation}

\subsection{Solution of amplitude vectors
}\label{sec:positive definitess ROL}

The stationary conditions \cref{eqn:jump_P_E} represent a system of nonlinear equations, where the microscale amplitude vectors $\vect{\alpha},\vect{\beta}$ can be resolved in terms of the macroscale homogenized fields $\vect{F},\vect{d}_0$. 
The computation of $\vect{\alpha},\vect{\beta}$ can be carried out with a $k$-iterative Newton-Raphson algorithm, namely
\begin{equation}\label{eqn:NR ROL}
\begin{aligned}
&\text{Solve} \quad & &
D{e}^{\star}[\delta \vect{\alpha},\delta \vect{\beta}]\Big|^k
+ D^2{e}^{\star}[\delta \vect{\alpha},\delta \vect{\beta};\Delta \vect{\alpha},\Delta \vect{\beta}]\Big|^k=\boldsymbol{0} \quad\text{for}\quad \Delta \vect{\alpha}\,,\Delta \vect{\beta}\,,\\
&\text{Update} \quad & &
\vect{\alpha}^{k+1}=\vect{\alpha}^{k}+\Delta \vect{\alpha},\qquad
\vect{\beta}^{k+1}=\vect{\beta}^{k}+\Delta \vect{\beta}\,.
\end{aligned}
\end{equation}
In \cite{marin2021} it was shown that \cref{eqn:NR ROL} can be written as
\begin{equation}\label{eqn:system of equations ROL}
\left[\begin{matrix}
\Delta \vect{\alpha}\\ \Delta \vect{\beta}
\end{matrix}\right]=-\left[\hat{\mathbb{H}}_{e}^{\vect{l}_0}\right]^{-1}_k
\left[\begin{matrix}
\llbracket\vect{P}\rrbracket \vect{l}_0 \\
\vect{{T}}_{\vect{l}_0}^T\llbracket\vect{e}_0\rrbracket 
\end{matrix}\right]_k\,,
\end{equation}
where the second order tensor $\left[\hat{\mathbb{H}}_{e}^{\vect{l}_0}\right]$ is defined as 
\begin{equation}\label{eqn:He}
\left[\hat{\mathbb{H}}_{e}^{\vect{l}_0}\right]=c^b\begin{bmatrix}
\vect{K}^a_1 &  \vect{K}^a_2\\
\left(\vect{K}^a_2\right)^T&
\vect{K}^a_3
\end{bmatrix} + 
c^a\begin{bmatrix}
\vect{K}^b_1 &  \vect{K}^b_2\\
\left(\vect{K}^b_2\right)^T&
\vect{K}^b_3
\end{bmatrix}\,,
\end{equation}
where the individual sub-matrices are defined as
\begin{equation}\label{eqn:matrices K}
\begin{aligned}
\left(\vect{K}^{a,b}_1\right)_{ij}&=\left(\frac{\partial^2 e^{a,b}}{\partial\vect{F}^{a,b}\partial\vect{F}^{a,b}}\right)_{iPjQ}(\vect{l}_0)_P(\vect{l}_0)_{Q}\,,\qquad &
\vect{K}^{a,b}_3&=\vect{T}_{\vect{l}_0}^T\frac{\partial^2 e^{a,b}}{\partial\vect{F}^{a,b}\partial\vect{d}_0^{a,b}}\vect{T}_{\vect{l}_0},\\    
\left(\vect{K}^{a,b}_2\right)_{iJ}&=\left(\vect{l}_0\right)_P\left(\frac{\partial^2 e^{a,b}}{\partial\vect{F}^{a,b}\partial\vect{d}_0^{a,b}}\right)_{iPQ}\left(\vect{T}_{\vect{l}_0}\right)_{QJ}\,.
\end{aligned}
\end{equation}

To ensure the existence of solutions of \cref{eqn:system of equations ROL}, and hence for $\vect{\alpha},\vect{\beta}$, the second directional derivative of ${e}^{\star}$ with respect to $\{\vect{\alpha},\vect{\beta}\}$ must be (strictly) positive definite, namely
\begin{equation}\label{eqn:convexity}
D^2 {e}^{\star}[\delta \vect{\alpha},\delta \vect{\beta};\delta \vect{\alpha},\delta \vect{\beta}] >0\qquad \forall \delta \vect{\alpha},\delta \vect{\beta}\,,
\end{equation}  
\textcite{marin2021} showed that $D^2{e}^{\star}[\delta \vect{\alpha},\delta \vect{\beta};\delta \vect{\alpha},\delta \vect{\beta}]$ is positive provided that both $e^a$ and $e^b$ are  elliptic, c.f.~see \cref{eqn:ellipticity condition}. 	
Thus, provided that $e^a$and $e^b$ are both polyconvex, above inequality \cref{eqn:convexity} is fulfilled, and thus existence of the solution $\vect{\alpha},\vect{\beta}$ is always guaranteed here.

\subsection{Effective stress and electric field}

From the internal energy of each phase, namely $e^a(\vect{F}^a,\vect{d}_0^a)$ or $e^b(\vect{F}^b,\vect{d}_0^b)$, it is possible to obtain the respective first Piola-Kirchhoff stress tensor and electric field according to \cref{eq:P_e} as
\begin{equation}\label{eqn:constitutive relationships ROL}
\begin{aligned}	
	\vect{P}^a=\frac{\partial e^a(\vect{\mathcal{U}}^a)}{\partial\vect{F}^a}\,,\qquad
	\vect{P}^b=\frac{\partial e^b(\vect{\mathcal{U}}^b)}{\partial\vect{F}^b}\,,\qquad
	\vect{e}_0^a=\frac{\partial e^a(\vect{\mathcal{U}}^a)}{\partial\vect{d}_0^a}\,,\qquad
	\vect{e}_0^b=\frac{\partial e^b(\vect{\mathcal{U}}^b)}{\partial\vect{d}_0^b}\,.
\end{aligned}	
\end{equation}
Differentiation of \cref{eqn:equivalenc energies ROL} with respect to time  yields the balance of power between macro and micro scales, i.e.
\begin{equation}\label{eqn:balance power ROL}
\dot{e}\left(\vect{F},\vect{d}_0\right)=\dot{e}^{\star}(\vect{F},\vect{d}_0,\vect{\alpha},\vect{\beta}) 
\qquad \Rightarrow \qquad
\vect{P}:\dot{\vect{F}} + \vect{e}_0\cdot\dot{\vect{d}}_0=De^{\star}[\dot{\vect{F}}]+De^{\star}[\dot{\vect{d}}_0]\,,
\end{equation}
where the directional derivatives $De^{\star}[\dot{\vect{F}}]$ and $De^{\star}[\dot{\vect{d}}_0]$ are
\begin{equation}\label{first directional derivative of effective energy with respect to F}
\begin{aligned}
De^{\star}[\dot{\vect{F}}]&=\left(c^a\vect{P}^a+c^b\vect{P}^b\right):\dot{\vect{F}} 
+\underbrace{c^ac^b
	D\tilde{\vect{\alpha}}[\dot{\vect{F}}]\cdot\left(\llbracket\vect{P}\rrbracket\vect{l}_0\right)}_{=0}
+\underbrace{c^ac^b
	D\tilde{\vect{\beta}}[\dot{\vect{F}}]\cdot\left(\vect{{T}}_{{\vect{l}_0}}^T\llbracket\vect{e}_0\rrbracket\right)}_{=0}\,,\\
De^{\star}[\dot{\vect{d}}_0]&=\left(c^a\vect{e}_0^a+c^b\vect{e}_0^b\right)\cdot \dot{\vect{d}}_0+ \underbrace{c^ac^b
	D\tilde{\vect{\alpha}}[\dot{\vect{d}}_0]\cdot\left(\llbracket\vect{P}\rrbracket\vect{l}_0\right)}_{=0}
+\underbrace{c^ac^b
	D\tilde{\vect{\beta}}[\dot{\vect{d}}_0]\cdot\left(\vect{\mathcal{T}}_{\vect{l}_0}^T\llbracket\vect{e}_0\rrbracket\right)}_{=0}\,,
\end{aligned}
\end{equation}
where use of  \cref{eqn:constitutive relationships ROL} has been made. Use of  \cref{eqn:jump_P_E}  enables the last two contributions in \cref{first directional derivative of effective energy with respect to F} to vanish. 
Then, introduction of \cref{first directional derivative of effective energy with respect to F} into  \cref{eqn:balance power ROL} permits to obtain the Hill-Mandel principle in the context of electro-mechanics, particularized to the case of rank-one laminate composites, i.e.
\begin{equation}\label{eqn:Hill-Mandel ROL}
\vect{P}:\dot{\vect{F}}+\vect{e}_0\cdot\dot{\vect{d}}_0=\Big(c^a\vect{P}^a+c^b\vect{P}^b\Big):\dot{\vect{F}} + \Big(c^a\vect{e}_0^a+c^b\vect{e}_0^b\Big)\cdot\dot{\vect{d}}_0\,.
\end{equation}
From this Hill-Mandel principle, it is possible to obtain the relationship between the macroscopic fields $\vect{P},\vect{e}_0$ and their microscopic counterparts $\vect{P}^a,\vect{P}^b$ and $\vect{e}_0^a,\vect{e}_0^b$, as
\begin{equation}
\vect{P}=c^a\vect{P}^a + c^b\vect{P}^b\,,\qquad
\vect{e}_0=c^a\vect{e}_0^a + c^b\vect{e}_0^b\,.
\end{equation}

\section{Numerical homogenization of the cubic metamaterial}\label{sec:RVE homogenization}

\subsection{Preliminaries}

For the case of an electro-active material whose microstructure corresponds to the cubic metamaterial as introduced in \cref{sec:cub} and illustrated in \cref{fig:cubic cell}, the microscopic fields $\vect{F}_{\mu}$ and $\vect{e}_{0_{\mu}}$ are defined in terms of the microscopic current position $\vect{x}_{\mu}$ and electric potential $\varphi_{\mu}$ fields as
\begin{equation}\label{eqn_:micro fields_appendix}
	\vect{F}_{\mu}=\frac{\partial \vect{x}_{\mu}}{\partial\vect{X}_{\mu}},\qquad 
	\vect{e}_{0_{\mu}}=-\frac{\partial\varphi_{\mu}}{\partial\vect{X}_{\mu}}.
\end{equation}
In the context of periodic first-order homogenization, c.f.~\cite{miehe2016},  $\vect{x}_{\mu}$ and  $\varphi_{\mu}$ can be decomposed  into macroscopic and microscopic contributions
\begin{equation}\label{eqn:def x and phi micro}
	\begin{aligned}
		\vect{x}_{\mu}&=\widetilde{\vect{x}}_{\mu}+\vect{F}\vect{X}_{\mu},&\qquad
		\varphi_{{{\mu}}}&=\widetilde{\varphi}_{\mu}-\vect{e}_{0}\cdot\vect{X}_{\mu},\\
	\end{aligned}
\end{equation}
where $\widetilde{\vect{x}}_{\mu}:\mathcal{B}_{0_{\mu}}\rightarrow \mathbb{R}^3$ and $\widetilde{\varphi}_{\mu}:\mathcal{B}_{0_{\mu}}\rightarrow \mathbb{R}$ represent the microscopic fluctuations. The latter are subjected to periodic boundary conditions of the form  $\llbracket\widetilde{\vect{x}}_{\mu}\rrbracket=\vect{0}$ and $\llbracket\widetilde{{\varphi}}_{\mu}\rrbracket={0}$, where $\llbracket\left({\cdot}\right)\rrbracket=\left({\cdot}\right)^+ - \left({\cdot}\right)^-$ represents the jump of the field $\left({\cdot}\right)$ across opposite boundaries of $\mathcal{B}_{0_{\mu}}$.
The microscopic fields $\vect{F}_{\mu}$ and $\vect{e}_{0_{\mu}}$ can be related with their macroscopic counterparts $\vect{F}$ and $\vect{e}_0$  as 
\begin{equation}\label{eqn:homogenisation of F_v2}
\begin{aligned}	
\vect{F} &= \frac{1}{V_{\mu}}\int_{\mathcal{B}_{0_{\mu}}}\vect{F}_{\mu}({\vect{X}_{{\mu}}})\,dV_{\mu}=\frac{1}{V_{\mu}}\int_{\partial\mathcal{B}_{0_{\mu}}}\vect{x}_{\mu}({\vect{X}_{{\mu}}})\otimes \vect{n}_{0_{\mu}}({\vect{X}_{{\mu}}})\,dV_{\mu},
\\
\vect{e}_0& = \frac{1}{V_{\mu}}\int_{\mathcal{B}_{0_{\mu}}}\vect{e}_{0_{\mu}}({\vect{X}_{{\mu}}})\,dV_{\mu}=\frac{1}{V_{\mu}}\int_{\partial\mathcal{B}_{0_{\mu}}}\varphi_{\mu}({\vect{X}_{{\mu}}})\,\vect{n}_{0_{\mu}}({\vect{X}_{{\mu}}})\,dV_{\mu}.
\end{aligned}
\end{equation}

We assume the existence of a Helmholtz type energy density at both macro and microscales, denoted as $\Psi(\vect{F},\vect{e}_0)$ and $\Psi_{\mu}\left({\vect{F}_{\mu}},\vect{e}_{0_{\mu}}\right)$, respectively.  Then, the effective or macroscopic Helmholtz free energy density is postulated as 
\begin{equation}\label{eqn:homogenisation cubic cell}
\underbrace{\Psi(\vect{F},\vect{e}_0)V_{\mu}}_{\Pi(\widetilde{\vect{x}}_{\mu},\widetilde{\varphi}_{\mu})}=\inf_{\widetilde{\vect{x}}_{\mu}\in\,\mathbb{V}^{x}}\sup_{\widetilde{\varphi}_{\mu}\in\mathbb{V}^{\varphi}}\Bigg\{\int_{\mathcal{B}_{0_{\mu}}}\Psi_{\mu}\left(\vect{F}_{\mu},\vect{e}_{0_{\mu}}\right)\,dV_{\mu}\Bigg\},
\end{equation}
where the relevant functional spaces $\mathbb{V}^x$ and $\mathbb{V}^{\varphi}$ are 
\begin{equation}
\begin{aligned}
\mathbb{V}^{x}&=\Big\{\widetilde{\vect{x}}_{\mu}:\left(\widetilde{\vect{x}}_{\mu}\right)_i\in H^1_0(\mathcal{B}_{0_{\mu}}),\,\,&\llbracket\widetilde{\vect{x}}_{\mu}\rrbracket=\vect{0},\,\,&\widetilde{\vect{x}}_{\mu}=\vect{0}\,\,\,\text{on}\,\,\,\partial\mathcal{B}^x_{0_{\mu}}\Big\},\\
\mathbb{V}^{\varphi}&=\Big\{\widetilde{{\varphi}}_{\mu}:\widetilde{\varphi}_{\mu}\in H^1_0(\mathcal{B}_{0_{\mu}}),\,\,&\llbracket\widetilde{{\varphi}}_{\mu}\rrbracket={0},\,\,&\widetilde{{\varphi}}_{\mu}={0}\,\,\,\text{on}\,\,\,\partial\mathcal{B}^{\varphi}_{0_{\mu}}\Big\},
\end{aligned}
\end{equation}
where $\partial\mathcal{B}^x_{0_{\mu}}\subset \partial\mathcal{B}_{0_{\mu}}$
is the region where zero Dirichlet boundary conditions are prescribed for $\widetilde{\vect{x}}_{\mu}$, preventing rigid body motions. Furthermore, $\partial\mathcal{B}^{\varphi}_{0_{\mu}}\subset \partial\mathcal{B}_{0_{\mu}}$  is the region where zero Dirichlet boundary conditions are prescribed for $\widetilde{{\varphi}}_{\mu}$, with excluding non-unique solutions of the form $\widetilde{\varphi}_{\mu}+C$, $\forall C\in \mathbb{R}$.

\subsection{Solution of micro-fluctuations}

The stationary conditions of the homogenized energy in \cref{eqn:homogenisation cubic cell} yields
\begin{equation}\label{eqn:stationary conditions unit cell}
\begin{aligned}
D\Pi(\widetilde{\vect{x}}_{\mu},\widetilde{\varphi}_{\mu})[\vect{v}_x]&=\int_{\mathcal{B}_{0_{\mu}}}\vect{P}_{\mu}:\frac{\partial \vect{v}_x}{\partial\vect{X}_{\mu}}\,dV_{\mu}={0},\qquad 
D\Pi(\widetilde{\vect{x}}_{\mu},\widetilde{\varphi}_{\mu})[v_{\varphi}]&=-\int_{\mathcal{B}_{0_{\mu}}}\vect{d}_{0_{\mu}}\cdot\frac{\partial {v}_{\varphi}}{\partial\vect{X}_{\mu}}\,dV_{\mu}={0},	
\end{aligned}
\end{equation}
which permit to obtain the micro-fluctuations $\widetilde{\vect{x}}_{\mu}$ and $\widetilde{\varphi}_{\mu}$. 

In \cref{eqn:stationary conditions unit cell}, both  $\vect{P}_{\mu}$ and $\vect{d}_{0_{\mu}}$ could in principle be obtained through the following relationships
\begin{equation}\label{eqn:micro relations I}
\vect{P}_{\mu}=\frac{\partial\Psi_{\mu}(\vect{F}_{\mu},\vect{e}_{0_{\mu}})}{\partial\vect{F}_{\mu}},\qquad 
\vect{d}_{0_{\mu}}=-\frac{\partial\Psi_{\mu}(\vect{F}_{\mu},\vect{e}_{0_{\mu}})}{\partial\vect{e}_{0_{\mu}}}.
\end{equation}
However, we advocate for constitutive models which are defined in terms of the microscopic internal energy $e_{\mu}(\vect{F}_{\mu},\vect{d}_{0_{\mu}})$ instead of the Helmholtz free energy $\Psi_{\mu}(\vect{F}_{\mu},\vect{e}_{0_{\mu}})$. Specifically, we consider internal energy densities $e_{\mu}(\vect{F}_{\mu},\vect{d}_{0_{\mu}})$ complying with the polyconvexity condition in \cref{eq:mvc,eqn:polyconvexity in second derivatives}, namely
\begin{equation}
e_{\mu}(\vect{F}_{\mu},\vect{d}_{0_{\mu}})=\cP_{\mu}(\vect{\mathcal{V}}_{\mu})\qquad\text{with}\qquad
\vect{\mathcal{V}}_{\mu}=\{\vect{F}_{\mu},\vect{H}_{\mu},J_{\mu},\vect{d}_{0_{\mu}},\vect{d}_{\mu}\},\quad \vect{d}_{\mu}=\vect{F}_{\mu}\vect{d}_{0_{\mu}},
\end{equation}
where the microscopic fields $\vect{H}_{\mu}$ and $J_{\mu}$ are defined in terms of $\vect{F}_{\mu}$ as in \cref{eqn:J and H}. 
As described in \cref{remark legendre}, polyconvexity of the microscopic internal energy density $e_{\mu}\left({\vect{F}_{\mu},\vect{d}_{0_{\mu}}}\right)$
guarantees
the definition of the microscopic Helmholtz energy density $\Psi_{\mu}(\vect{F}_{\mu},\vect{e}_{0_{\mu}})$ according to the Legendre transformation 
\begin{equation}\label{eqn:legendre micro}
\Psi_{\mu}(\vect{F}_{\mu},\vect{e}_{0_{\mu}})=-\sup_{\vect{d}_{0_{\mu}}}\Big\{\vect{e}_{0_{\mu}}\cdot\vect{d}_{0_{\mu}} - e_{\mu}(\vect{F}_{\mu},\vect{d}_{0_{\mu}})\Big\}.
\end{equation}
Given the microscopic fields $\vect{F}_{\mu},\vect{e}_{0_{\mu}}$, it is possible to obtain $\vect{P}_{\mu}$ and $\vect{d}_{0_{\mu}}$ from the (nonlinear) stationary condition of the Legendre transformation in \cref{eqn:legendre micro} as
\begin{equation}
\vect{P}_{\mu}=\frac{\partial e_{\mu}(\vect{F}_{\mu},\vect{d}_{0_{\mu}})}{\partial\vect{F}_{\mu}}, \qquad
\vect{e}_{0_{\mu}}=\frac{\partial e_{\mu}(\vect{F}_{\mu},\vect{d}_{0_{\mu}})}{\partial\vect{d}_{0_{\mu}}}. 
\end{equation}

Generally, the boundary value problem defined by \cref{eqn:stationary conditions unit cell} can only be solved numerically for given macro $\bF,\bd_0$, e.g., by discretizing the microscale RVE $\cB_{0_\mu}$ and the sought fields $\widetilde{\vect{x}}_{\mu}$ and $\widetilde{\varphi}_{\mu}$ using finite element methods \cite{Gil_electro_partII_2016,Daniel_Magneto_2021}. As mentioned in \cref{sec:cub:data} and illustrated in \cref{fig:cubic cell:fe}, here we employ tetrahedral elements with quadratic shape functions for both $\widetilde{\vect{x}}_{\mu}$ and $\widetilde{\varphi}_{\mu}$.
Since the problem is nonlinear, similar to \cref{sec:positive definitess ROL}, an iterative solution procedure in terms of a Netwon-Raphson method is required.

\subsection{Effective stress and electric displacement fields}

Differentiation with respect to time in above \cref{eqn:homogenisation cubic cell} and making use of \cref{eqn:def x and phi micro,eqn:micro relations I} yields the following power balance between macro and micro scales,
\begin{equation}\label{eqn:Hill-Mandel I}
\Big(\vect{P}:\dot{\vect{F}}-\vect{d}_0\cdot\dot{\vect{e}}_0\Big)V_{\mu}=\int_{\mathcal{B}_{0_{\mu}}}\vect{P}_{\mu}:\Big(\frac{\partial\dot{\widetilde{\vect{x}}}_{\mu}}{\partial\vect{X}_{0_{\mu}}}+\dot{\vect{F}}\Big)-\vect{d}_{0_{\mu}}\cdot\Big(-\frac{\partial\dot{\widetilde{\varphi}}_{\mu}}{\partial\vect{X}_{\mu}} + \dot{\vect{e}}_{0}\Big)\,dV_{\mu}.
\end{equation}
For $\frac{\partial\dot{\widetilde{\vect{x}}}_{\mu}}{\partial\vect{X}_{0_{\mu}}}\in\mathbb{V}^x$
and $\frac{\partial\dot{\widetilde{\varphi}}_{\mu}}{\partial\vect{X}_{\mu}}\in\mathbb{V}^\varphi$, substitution of \cref{eqn:stationary conditions unit cell} into \eqref{eqn:Hill-Mandel I} permits to obtain the Hill-Mandel principle in the context of electro-mechanics, namely
\begin{equation}\label{eqn:Hill-Mandel II}
\Big(\vect{P}:\dot{\vect{F}}-\vect{d}_0\cdot\dot{\vect{e}}_0\Big)V_{\mu}=\int_{\mathcal{B}_{0_{\mu}}}\Big(\vect{P}_{\mu}:\dot{\vect{F}}-\vect{d}_{0_{\mu}}\cdot{\dot{\vect{e}}}_{0}\Big)\,dV_{\mu}.
\end{equation}
As a consequence, the following relationships between the macroscopic fields $\vect{P},\vect{d}_0$ and their microscopic counterparts $\vect{P}_{\mu},\vect{d}_{0_{\mu}}$ can be established,
\begin{equation}\label{eqn:macro const relationships RVE}
\vect{P}=\frac{1}{V_{\mu}}\int_{\mathcal{B}_{0_{\mu}}}\vect{P}_{\mu}\,dV_{\mu},\qquad
\vect{d}_0=\frac{1}{V_{\mu}}\int_{\mathcal{B}_{0_{\mu}}}\vect{d}_{0_{\mu}}\,dV_{\mu}.
\end{equation}

\renewcommand*{\bibfont}{\footnotesize}
\printbibliography

\end{document}